\documentclass[a4paper,11pt]{article}
\pdfoutput=1
\usepackage{geometry}
\usepackage{amssymb,mathrsfs,amsmath}
\usepackage{color,xcolor}
\usepackage{slashed,soul}
\usepackage{graphicx}
\usepackage{amsfonts}
\usepackage{lscape}
\def\linkcolor{cyan!70!black}

\usepackage[
colorlinks=true
,urlcolor=\linkcolor
,anchorcolor=\linkcolor
,citecolor=\linkcolor
,filecolor=\linkcolor
,linkcolor=\linkcolor
,menucolor=\linkcolor
,linktocpage=true
,pdfproducer=medialab
,pdfa=true
]{hyperref}
\usepackage{amsthm}
\usepackage{booktabs}
\usepackage{array}
\usepackage{rotating}
\usepackage{bbm}

\usepackage[numbers, sort&compress]{natbib}
\newlength{\bibitemsep}
\newlength{\bibparskip}
\let\oldthebibliography\thebibliography
\renewcommand\thebibliography[1]{%
  \oldthebibliography{#1}%
  \setlength{\parskip}{\bibitemsep}%
  \setlength{\itemsep}{\bibparskip}%
}

\usepackage{multirow}
\usepackage{float}
\usepackage[utf8]{inputenc}
\usepackage[T1]{fontenc}
\usepackage{appendix}
\usepackage{cancel}
\usepackage{comment}

\usepackage[caption=false]{subfig} % loads caption3
\usepackage{ragged2e} % for the \justifying macro
\DeclareCaptionJustification{justified}{\justifying}

\graphicspath{{figs/}}

\newcommand{\braket}[1]{\ensuremath{\left\langle#1\right\rangle}}

\newcommand{\calO}{{\mathcal{O}}}

\renewcommand{\Re}{{\rm Re}}

\def\mesonP{\mathcal P}

\newcommand{\vp}{\varphi}
\newcommand{\vpj }{\mbox{${\vp^\dag\, \raisebox{1.5mm}{${}^\leftrightarrow$}\hspace{-4mm} D_\mu\,\vp}$}}

\newcommand{\cw}{\cos\theta_W}
\newcommand{\sw}{\sin\theta_W}

\newcommand\gVLV[1][\mesonV]{g_{VL}^{\tau\ell#1}} %Vector current L for vector meson #1
\newcommand\gVRV[1][\mesonV]{ g_{VR}^{\tau\ell#1}} %Vector current R for vector meson #1
\newcommand\gTLV[1][\mesonV]{ g_{TL}^{\tau\ell#1}} %Tensor current L for vector meson #1
\newcommand\gTRV[1][\mesonV]{ g_{TR}^{\tau\ell#1}} %Tensor current r for vector meson #1
\newcommand\gTLVd[1][\mesonV]{ \tilde g_{TL}^{\tau\ell#1}} %Tensor current L for vector meson #1
\newcommand\gTRVd[1][\mesonV]{ \tilde g_{TR}^{\tau\ell#1}} %Tensor current r for vector meson #1

\newcommand\gLP[1][\mesonP]{g_{L}^{\tau\ell#1}} %Eff coupling L for pseudoscalar meson #1
\newcommand\gRP[1][\mesonP]{g_{R}^{\tau\ell#1}} %Eff coupling R for pseudoscalar meson #1
\newcommand\gVLP[1][\mesonP]{g_{VL}^{\tau\ell#1}} %Vector current L for pseudoscalar meson #1
\newcommand\gVRP[1][\mesonP]{g_{VR}^{\tau\ell#1}} %Vector current R for pseudoscalar meson #1
\newcommand\gSLP[1][\mesonP]{g_{SL}^{\tau\ell#1}} %Scalar current L for pseudoscalar meson #1
\newcommand\gSRP[1][\mesonP]{g_{SR}^{\tau\ell#1}}

\newcommand{\fref}[1]{Figure~\ref{#1}} 
\newcommand{\eref}[1]{Eq.\,(\ref{#1})}

\newcommand{\aref}[1]{Appendix~\ref{#1}}
\newcommand{\sref}[1]{Section~\ref{#1}}
\newcommand{\tref}[1]{Table~\ref{#1}}
\allowdisplaybreaks

\begin{document}

\vspace{1cm}

\begin{titlepage}

\vspace*{-1.0truecm}
\begin{flushright}
CPPC-2022-08 \\
IFT-UAM/CSIC-22-82
 \end{flushright}
\vspace{0.8truecm}

\begin{center}
\renewcommand{\baselinestretch}{1.8}\normalsize
\boldmath
{\LARGE\textbf{
Indirect constraints on lepton-flavour-violating quarkonium decays
}}
\unboldmath
\end{center}

\vspace{0.4truecm}

\renewcommand*{\thefootnote}{\fnsymbol{footnote}}

\begin{center}

{\bf Lorenzo  Calibbi$\,^a$\footnote{\href{mailto:calibbi@nankai.edu.cn}{calibbi@nankai.edu.cn}}, 
Tong Li$\,^a$\footnote{\href{mailto:litong@nankai.edu.cn}{litong@nankai.edu.cn}},
Xabier Marcano$\,^b$\footnote{\href{mailto:xabier.marcano@uam.es}{xabier.marcano@uam.es}}
and Michael A.~Schmidt$\,^{c}$\footnote{\href{mailto:m.schmidt@unsw.edu.au}{m.schmidt@unsw.edu.au}}}
\vspace{0.5truecm}

{\footnotesize

$^a${\sl School of Physics, Nankai University, Tianjin 300071, China \vspace{0.15truecm}}

$^b${\sl  Departamento de F\'{\i}sica Te\'orica and Instituto de F\'{\i}sica Te\'orica UAM/CSIC,\\
Universidad Aut\'onoma de Madrid, Cantoblanco, 28049 Madrid, Spain \vspace{0.15truecm}}

$^c${\sl Sydney Consortium for Particle Physics and Cosmology,\\
School of Physics, The University of New South Wales,\\
Sydney, New South Wales 2052, Australia \vspace{0.15truecm}}

}

\vspace*{5mm}
\end{center}

\renewcommand*{\thefootnote}{\arabic{footnote}}
\setcounter{footnote}{0}

\vspace{0.4cm}
\begin{abstract}
 \noindent Within an effective-field-theory framework, we present a model-independent analysis of the potential of discovering new physics by searching for lepton flavour violation in heavy quarkonium decays and, more in general, we study the phenomenology of lepton-flavour-violating (LFV) 2~quark\,-\,2~lepton~($2q2\ell$) operators with two charm or bottom fields. We compute the constraints from LFV muon and tau decays on the new-physics operators that can induce LFV processes involving $c\bar c$ and $b\bar b$ systems, thus providing a comprehensive list of indirect upper limits on processes such as $J/\psi \to \ell\ell^\prime$, $\Upsilon(nS) \to \ell\ell^\prime$, $\Upsilon(nS) \to \ell\ell^\prime \gamma$ etc., which can be sought at BESIII, Belle~II, and the proposed super tau-charm factory.
  We show that such indirect constraints are so stringent that they prevent the detection of quarkonium decays into $e\mu$. 
  In the case of decays of quarkonia into $\ell\tau$ ($\ell=e,\mu$), we find that an improvement by 2-3 orders of magnitude on the current sensitivities is in general required in order to discover or further constrain new physics. However, we show that cancellations among different contributions to the LFV tau decay rates are possible, such that \mbox{$\Upsilon(nS)\to \ell\tau$} can saturate the present experimental bounds.
  We also find that, interestingly, searches for LFV $Z$ decays, $Z\to\ell\tau$, at future $e^+e^-$ colliders are complementary probes of $2q2\ell$ operators with third generation quarks.
\end{abstract}

\end{titlepage}

\tableofcontents

%%%%%%%%%%%%%%%%%%%%%%%%%%%%%%%%%%%%%%%%%%%%%%%
\section{Introduction}
\label{sec:Intro}
%%%%%%%%%%%%%%%%%%%%%%%%%%%%%%%%%%%%%%%%%%%%%%%

The lack of conclusive evidence for new physics~(NP) at the Large Hadron Collider (LHC) makes it crucial to pursue a diversified experimental programme in search for
Nature's next fundamental energy scale beyond the electroweak~(EW) one. With this respect, high-intensity frontier experiments, in particular searches for charged lepton flavour violation~(LFV), represent an ideal laboratory capable to test scales above $10^3$-$10^4$~TeV, way beyond the reach of any foreseeable high-energy collider~\cite{Calibbi:2017uvl}. 

Neutrino oscillations have provided evidence that lepton family numbers are not
conserved and one can expect non-standard contributions to LFV processes in the context of any extension 
of the Standard Model (SM) involving new fields that couple to leptons. 
On the other hand, the physics case for LFV searches has been recently reinforced by the first results of the FNAL Muon~{g-2} experiment~\cite{Abi:2021gix} and the persistent hints for violation of lepton flavour universality~(LFU) in semileptonic $B$ meson decays~\cite{Bifani:2018zmi,London:2021lfn}, especially those of the kind $b\to s \mu\mu$. Both anomalies seem to point to a new-physics sector, coupled preferably with muons, at a scale below $100$~TeV~\cite{Allwicher:2021jkr,DiLuzio:2017chi}.
Moreover, any new physics interacting with muons is not in general expected to exhibit a flavour structure aligned to the SM one, that is, LFV effects induced by the fields possibly behind the muon $g-2$ and $b\to s \mu\mu$ anomalies are difficult to avoid unless very peculiar flavour symmetries are imposed~\cite{Glashow:2014iga,Calibbi:2021qto,Isidori:2021gqe,Bigaran:2022kkv}. 
Therefore, LFV rates at observable level are likely if these experimental anomalies will be confirmed to be signal of new physics.

The hints for LFU violation in $B$ decays require a new-physics sector that couple to both quarks and leptons\,---\,the typical example being scalar or vector leptoquarks~\cite{Dorsner:2016wpm}. 
Such new physics can be described in a model-independent way within an effective field theory~(EFT) in terms of 2~quarks-2~leptons~($2q2\ell$) operators, as long as its scale is much larger than the typical energy scales of the processes under study.
It has been shown that the $B$ anomalies can be addressed by operators involving 3rd-generation fermions only, the couplings to lighter generations being induced by field rotations from the interaction basis to the mass basis~\mbox{\cite{Bhattacharya:2014wla,Calibbi:2015kma,Feruglio:2016gvd,Buttazzo:2017ixm}}.

\begin{table}[t!]
\begin{center}
{\small
\renewcommand{\arraystretch}{1.2}
\begin{tabular}{lclc}
\hline
\multicolumn{1}{c}{LFVQD} & \multicolumn{3}{c}{Present bounds on BR $(90\%\,\text{CL})$} \\
%%%%%%%%%%%%%
\hline 
$J/\psi\to e\mu$ &  $4.5\times10^{-9}$ & BESIII (2022) & \cite{BESIIInew}
\\
$\Upsilon(1S)\to e\mu$ &  $3.6\times10^{-7}$ & Belle (2022) & \cite{Belle:2022cce} \\
$\Upsilon(1S)\to e\mu\gamma$ &  $4.2\times10^{-7}$ & Belle (2022) & \cite{Belle:2022cce} \\
%%%%%%%%%%%%%
\hline 
$J/\psi\to e\tau$ &  $7.5\times10^{-8}$ & BESIII (2021) & \cite{BESIII:2021slj} \\
$\Upsilon(1S)\to e\tau$ &  $2.4\times10^{-6}$ & Belle (2022) & \cite{Belle:2022cce} \\
$\Upsilon(1S)\to e\tau\gamma$ &  $6.5\times10^{-6}$ & Belle (2022) & \cite{Belle:2022cce} \\
$\Upsilon(2S)\to e\tau$ &  $3.2\times10^{-6}$ & BaBar (2010) & \cite{BaBar:2010vxb} \\
$\Upsilon(3S)\to e\tau$ &  $4.2\times10^{-6}$ & BaBar (2010) & \cite{BaBar:2010vxb} \\
%%%%%%%%%%%%
\hline 
$J/\psi\to \mu\tau$ &  $2.0\times10^{-6}$ & BES (2004) & \cite{BES:2004jiw} \\
$\Upsilon(1S)\to \mu\tau$ &  $2.6\times10^{-6}$ & Belle (2022) & \cite{Belle:2022cce} \\
$\Upsilon(1S)\to \mu\tau\gamma$ &  $6.1\times10^{-6}$ & Belle (2022) & \cite{Belle:2022cce} \\
$\Upsilon(2S)\to \mu\tau$ &  $3.3\times10^{-6}$ & BaBar (2010) & \cite{BaBar:2010vxb} \\
$\Upsilon(3S)\to \mu\tau$ &  $3.1\times10^{-6}$ & BaBar (2010) & \cite{BaBar:2010vxb} \\
\hline
\end{tabular}
\caption{Present $90\%\,\text{CL}$ upper limits on vector quarkonium  LFV decays. 
No limit is currently available for LFV decays of (pseudo)scalar or other vector resonances. 
\label{tab:quarkonia}}
}
\end{center}
\end{table}

The above considerations prompt us to address the experimental prospects of LFV processes involving heavy quark flavours, either flavour-violating or flavour-conserving in the quark sector. In this paper, we focus on the latter case, in particular on new physics that can induce LFV decays of heavy quarkonia, that is, $c\bar c$ and $b\bar b$ bound states. The existing limits on  LFV quarkonium decays (LFVQD), concerning vector resonances only, are listed in \tref{tab:quarkonia}. 
We note the recent results by BESIII and Belle, which improved previous bounds notably and even searched for new channels such as $\Upsilon(1S)\to\ell\ell'\gamma$.
The experimental prospects of these processes are even more interesting:
the extended run of BESIII~\cite{BESIII:2020nme} and the proposed Super-Tau-Charm Factory~(STCF)~\cite{Barnyakov:2020vob,Zhou:2021rgi,Lyu:2021tlb} could increase the sensitivity on the $J/\psi\to \ell_i \ell_j$ decays by several orders of magnitude and, for the first time, search for LFV decays of (pseudo)scalar charmonium states. 
Similarly, Belle~II~\cite{Kou:2018nap} is expected to reach an integrated luminosity about two orders of magnitude larger than the previous B factories, hence it should improve the limits on the $\Upsilon(nS)$ modes by at least one order of magnitude.

However, any new physics giving rise to this kind of decays would also induce other LFV processes, in particular LFV muon or tau decays~\cite{Nussinov:2000nm}, as well as other high-energy LFV processes such as LFV $Z$ decays, which will give competitive limits at future high-energy $e^+e^-$ colliders\,---\,see Ref.~\cite{Calibbi:2021pyh}.
The obvious question is then whether the stringent constraints on the latter processes (see \tref{tab:LFVlimits}) still allow sizeable effects for LFV quarkonia decay. 
In other words, is it possible to discover new physics searching for quarkonium LFV? The aim of this paper is to give a precise quantitative answer to this question, providing model-independent indirect upper limits on the LFV decay rates of quarkonia, in a similar way to what was done in Ref.~\cite{Calibbi:2021pyh} for LFV $Z$ decays.
\begin{table}[t]
\begin{center}
{\small
\renewcommand{\arraystretch}{1.2}
\begin{tabular}{lrlrl}
\hline
\multicolumn{1}{c}{LFV observable} & \multicolumn{2}{c}{Present bounds} & \multicolumn{2}{c}{Expected future limits} \\
\hline
BR$(\mu\to e\gamma)$ &  $4.2\times10^{-13}$ &MEG (2016)~\cite{TheMEG:2016wtm} & $6\times 10^{-14} $& MEG~II~\cite{Baldini:2018nnn}\\
BR$(\mu\to eee)$ &   $1.0\times10^{-12}$ & SINDRUM (1988)~\cite{Bellgardt:1987du} & $10^{-16}$ &Mu3e~\cite{Blondel:2013ia} \\
CR$(\mu\to e,{\rm Au})$ & $7.0\times10^{-13}$ &SINDRUM~II (2006)~\cite{Bertl:2006up}&  \multicolumn{2}{c}{--} \\
CR$(\mu\to e,{\rm Al})$ & \multicolumn{2}{c}{--} & $6\times10^{-17}$ &COMET/Mu2e~\cite{Kuno:2013mha,Bartoszek:2014mya} \\
BR$(Z\to e\mu)$& $ 2.62\times 10^{-7}$ & ATLAS (2022)~\cite{ATLAS:2022uhq} &
$10^{-8}$\,--\,$10^{-10}$ & FCC-ee/CEPC~\cite{Dam:2018rfz}  \\
\hline
BR$(\tau\to e\gamma)$ & $3.3\times10^{-8}$ &BaBar (2010)~\cite{Aubert:2009ag} & $ 9\times 10^{-9\phantom{0}}$ &Belle~II~\cite{Kou:2018nap,Banerjee:2022xuw}\\
BR$(\tau\to eee)$ & $2.7\times10^{-8}$ &Belle (2010)~\cite{Hayasaka:2010np} & $4.7\times 10^{-10}$ &Belle~II~\cite{Kou:2018nap,Banerjee:2022xuw}\\
BR$(\tau\to e\mu\mu)$ & $2.7\times10^{-8}$ &Belle (2010)~\cite{Hayasaka:2010np} & $4.5\times 10^{-10}$ &Belle~II~\cite{Kou:2018nap,Banerjee:2022xuw}\\
BR$(\tau\to \pi e)$ & $8.0\times10^{-8}$  &  Belle (2007)~\cite{Miyazaki:2007jp} & $7.3\times 10^{-10}$ &Belle~II~\cite{Kou:2018nap,Banerjee:2022xuw}\\
BR$(\tau\to \rho e)$ & $1.8\times10^{-8}$ &Belle (2011)~\cite{Miyazaki:2011xe} & $3.8\times 10^{-10}$ &Belle~II~\cite{Kou:2018nap,Banerjee:2022xuw}\\
BR$(Z\to e\tau)$ &$ 5.0\times 10^{-6}$ & ATLAS (2021)~\cite{newATLAS} & $10^{-9}$ & FCC-ee/CEPC~\cite{Dam:2018rfz}  \\
\hline
BR$(\tau\to \mu\gamma)$ & $4.2\times10^{-8}$ &Belle (2021)~\cite{Belle:2021ysv} & $6.9\times 10^{-9\phantom{0}}$ &Belle~II~\cite{Kou:2018nap,Banerjee:2022xuw}\\
BR$(\tau\to \mu\mu\mu)$ & $2.1\times10^{-8}$ &Belle (2010)~\cite{Hayasaka:2010np} & $3.6\times 10^{-10}$ &Belle~II~\cite{Kou:2018nap,Banerjee:2022xuw}\\
BR$(\tau\to \mu ee)$ & $1.8\times10^{-8}$ &Belle (2010)~\cite{Hayasaka:2010np} & $2.9\times 10^{-10}$ &Belle~II~\cite{Kou:2018nap,Banerjee:2022xuw}\\
BR$(\tau\to \pi\mu)$ &  $1.1\times10^{-7}$ & Babar (2006)~\cite{Aubert:2006cz} & $7.1\times 10^{-10}$ &Belle~II~\cite{Kou:2018nap,Banerjee:2022xuw}\\
BR$(\tau\to \rho\mu)$ & $1.2\times10^{-8}$ &Belle (2011)~\cite{Miyazaki:2011xe} & $5.5\times 10^{-10}$ &Belle~II~\cite{Kou:2018nap,Banerjee:2022xuw}\\
BR$(Z\to \mu\tau)$ &$ 6.5\times 10^{-6}$ & ATLAS (2021)~\cite{newATLAS} & $10^{-9}$ & FCC-ee/CEPC~\cite{Dam:2018rfz}  \\
\hline
\end{tabular}
\caption{Present $90\%\,\text{CL}$ upper limits ($95\%\,\text{CL}$ for the $Z$ decays) and future expected sensitivities for the set of LFV transitions relevant for our analysis.
\label{tab:LFVlimits}}
}
\end{center}
\end{table}

\begin{figure}[t!]
    \centering
    \includegraphics[width=.8\textwidth]{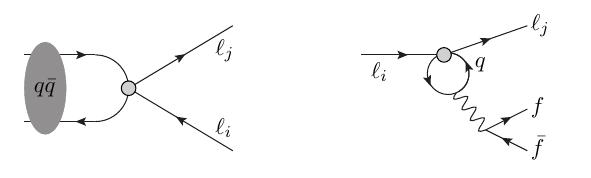}
    \caption{Diagrammatic example of how the same EFT vertex (grey circle) generating quarkonia LFV decays can induce other LFV processes at loop level.}
    \label{LFVdiag}
\end{figure}

To be agnostic about the new dynamics that can give rise to these effects, we employ an effective-field-theory approach, working within both the so-called Low-Energy Effective Field Theory~(LEFT)~\cite{Jenkins:2017jig}, which involves QED$\times$QCD invariant operators of fields below the EW scale, and the Standard Model Effective Field Theory (SMEFT) where invariance under the full SM gauge group and also heavy fields are considered~\cite{Buchmuller:1985jz,Grzadkowski:2010es}\,---\,for a review cf.~Ref.~\cite{Brivio:2017vri}.
In this context, new physics contributions to the quarkonium decays we are interested in are described by $2q2\ell$ operators of the schematic form $\bar c\,c\,\bar\ell_i\ell_j$ and $\bar b\,b\,\bar\ell_i\ell_j$ ($\ell_{i,j}=e,\mu,\tau$, $i\neq j$). 
On the other hand, diagrams obtained by closing the quark loop will induce ({\it e.g.}~via a virtual photon exchange, as illustrated in \fref{LFVdiag}) other LFV operators involving lighter quarks as well as purely leptonic LFV operators~\cite{Crivellin:2017rmk}. 
These radiative effects\,---\,that can be summarised by the operator mixing induced by the one-loop renormalisation group equations (RGEs) of the operator coefficients~\cite{Jenkins:2013zja,Jenkins:2013wua,Alonso:2013hga,Jenkins:2017dyc}\,---\,will provide contributions to LFV $\mu$ and $\tau$ decays, from which we can then obtain the above-mentioned indirect constraints on the coefficients of the $\bar c\,c\,\bar\ell_i\ell_j$ and $\bar b\,b\,\bar\ell_i\ell_j$ operators and, thus, on the rates of LFV quarkonium decays.

Earlier works focusing on or including constraints on LFV $2q2\ell$ operators can be found in Refs.~\cite{Gutsche:2009vp,Carpentier:2010ue,Crivellin:2013hpa,Cai:2015poa,Hazard:2016fnc,Hazard:2017udp,Crivellin:2017rmk,Davidson:2018rqt,Dib:2018rpy,Angelescu:2020uug,Gonzalez:2021tqc,Cirigliano:2021img,Hoferichter:2022mna,Arun:2022uqn}. 
The authors of Ref.~\cite{Hazard:2016fnc}, in particular, calculated quarkonium LFV decay rates and obtained bounds on the associated operators. However, we have found no systematic comparison with the constraints from other LFV processes in the literature, nor an assessment of the largest possible effects for quarkonia compatible with such bounds\,---\,see however \cite{Gutsche:2009vp,Gonzalez:2021tqc} for works focusing on a limited number of quarkonium processes and indirect constraints.
In this paper, we extend beyond the existing literature and  answer in a systematic way the above question about the sensitivity to new physics of future searches for LFV quarkonium decays, including heavy (pseudo)scalar quarkonia and radiative LFV decays of vector quarkonia. Interestingly, our approach based on high-intensity frontier observables is complementary to that of Ref.~\cite{Angelescu:2020uug}, where indirect constraints on low-energy processes induced by $2q2\ell$ operators were set based on high-energy measurements of di-lepton distributions from $pp\to \ell_i\ell_j$ at the LHC. 

The rest of the paper is organised as follows. In \sref{sec:eft} we introduce the EFT framework we employ and our conventions.
Our calculations for the quarkonium LFV decay rates in terms of the coefficients of LEFT operators are presented in \sref{sec:obs}. The running of LEFT operators is employed in \sref{sec:LEFT} in order to estimate the indirect constraints on quarkonium LFV, while the effects of operator mixing above the EW scale, the sensitivity of LFV experiments to high-scale NP, and possible cancellations due to the interference of multiple operators are discussed in \sref{sec:SMEFT} adopting the SMEFT framework. We summarise our results and conclude in \sref{sec:Con}. In the Appendices, more technical results and useful formulae are collected.

%%%%%%%%%%%%%%%%%%%%%%%%%%%%%%%%
\section{EFT framework}
%%%%%%%%%%%%%%%%%%%%%%%%%%%%%%%%
\label{sec:eft}

As discussed in the introduction, we parameterise the effects of LFV new physics in terms of non-renormalisable operators. 
Throughout this work, we assume that the new particles related to the NP scale $\Lambda$ responsible for LFV are much heavier than the EW scale, $\Lambda \gg m_W$. In such a scenario, in order to assess the NP effects across different scales, it is then convenient to work within the SMEFT framework, whose Lagrangian consists of that of the SM extended with a tower of higher-dimensional operators constructed by gauge-invariant combinations of the SM fields only and suppressed by inverse powers of the scale $\Lambda$:
\begin{align}
    \mathcal{L}_\textsc{smeft} = \mathcal{L}_\textsc{sm} + \sum_{d>4} \sum_{a} \,\frac{C_a^{(d)}}{\Lambda^{d-4}}\,\calO_a^{(d)} \,,
\end{align}
where $\calO_a^{(d)}$ are the effective operators of dimension-$d$ and the $C_a^{(d)}$ represent the corresponding Wilson Coefficients (WCs), whose values depend on the renormalisation scale $\mu$. Notice that we are working with dimensionless SMEFT WCs. 
In the rest of the paper, we will focus on dimension-6 operators\,---\,that are expected to provide the dominant contributions to LFV processes\,---\,and adopt the conventions of the Warsaw basis~\cite{Grzadkowski:2010es}. All dimension-6 SMEFT operators that can induce LFV effects~\cite{Crivellin:2013hpa} are listed in \tref{Tab:dim6}.

\begin{table}[t!] 
\centering
\renewcommand{\arraystretch}{1.2}
\setlength{\tabcolsep}{15pt}
\begin{tabular}{cccc} 
\hline
\multicolumn{4}{c}{$2q2\ell$ operators} \\
\hline \\[-2.6ex]
$\calO_{\ell q,prst}^{(1)}$ & $(\bar L_p \gamma_\mu L_r)(\bar Q_s \gamma^\mu Q_t)$ & 
$\calO_{\ell q,prst}^{(3)}$ &  $(\bar L _p \gamma_\mu \tau^I L_r) (\bar Q_s \gamma^\mu \tau^I Q_t)$ \\
$\calO_{\ell u,prst}$ & $(\bar L_{p} \gamma_\mu L_{r}) (\bar u_s \gamma^\mu u_t)$  &
$\calO_{\ell d,prst}$ &   $(\bar L_{p} \gamma_\mu L_{r}) (\bar d_s \gamma^\mu d_t)$ \\
$\calO_{eu,prst}$ &  $(\bar e_{p} \gamma_\mu e_{r}) (\bar u_s \gamma^\mu u_t)$ &
$\calO_{ed,prst}$ &   $(\bar e_{p} \gamma_\mu e_{r}) (\bar d_s \gamma^\mu d_t)$ \\
$\calO_{qe,prst}$ &   $(\bar Q_p \gamma^\mu Q_r) (\bar e_{s} \gamma_\mu e_{t})$ &
$\calO_{\ell edq,prst}$ &   $(\bar L_p e_{r}) (\bar d_s Q_t)$  \\
$\calO_{\ell equ,prst}^{(1)}$ &   $(\bar L_p^a e_{r})\epsilon_{ab} (\bar Q_s^b u_t)$ &
$\calO_{\ell equ,prst}^{(3)}$ &   $(\bar L_p^a \sigma_{\mu\nu} e_{r})  \epsilon_{ab} (\bar Q_s^b \sigma^{\mu\nu} u_t )$ \\[1ex]
\hline
\multicolumn{2}{c}{4$\ell$ operators} &
\multicolumn{2}{c}{Dipole operators} \\
\hline \\[-2.6ex]
$\calO_{\ell\ell,prst}$ & $ (\bar L _p \gamma_\mu L_r) (\bar L_s \gamma^\mu L_t)$ & 
$\calO_{eW,pr}$ & $(\bar L_p \sigma^{\mu\nu} e_{r}) \tau^I \varphi W_{\mu\nu}^I$  \\
$\calO_{ee,prst}$ & $(\bar e _p \gamma_\mu e_r) (\bar e_s \gamma^\mu e_t)$ &
$\calO_{eB,pr}$ & $(\bar L_p \sigma^{\mu\nu} e_{r}) \varphi B_{\mu\nu}$ \\
$\calO_{\ell e,prst}$ & $(\bar L _p \gamma_\mu L_r) (\bar e_s \gamma^\mu e_t)$ & & \\[.6ex]
\hline
\multicolumn{4}{c}{Lepton-Higgs operators}\\ 
\hline\\[-2.6ex]
$\calO_{\varphi \ell,pr}^{(1)}$ &  $(\varphi^\dagger i \overleftrightarrow{D}_\mu \varphi) ( \bar L_p \gamma^\mu L_r)$ &
$\calO_{\varphi \ell,pr}^{(3)}$ &  $(\varphi^\dagger i \overleftrightarrow{D}^I_\mu \varphi) ( \bar L_p \gamma^\mu \tau^I L_r)$ \\
$\calO_{\varphi e,pr}$ &  $(\varphi^\dagger i \overleftrightarrow{D}_\mu \varphi) ( \bar e_p \gamma^\mu e_r)$ &
$\calO_{e\varphi3,pr}$ &  $( \bar L_p  e_r \varphi) (\varphi^\dagger \varphi)$ \\[.6ex]
\hline
\end{tabular}
\caption{ Complete list of the dimension-$6$ SMEFT operators relevant to LFV processes.
$Q$ and $L$ respectively denote quark and lepton $SU(2)_L$ doublets ($a,b=1,2$ and $I=1,2,3$ are $SU(2)_L$ indices, $\tau^I$  are the Pauli matrices). $u$, $d$ and $e$ are (up and down) quark and lepton singlets. $\vp$ represents the Higgs doublet with $\vpj\equiv \vp^\dag (D_\mu \vp)-(D_\mu \vp)^\dag \vp$. $B_{\mu\nu}$ and $W^I_{\mu\nu}$ are the $U(1)_Y$ and $SU(2)_L$ field strengths, respectively. $p,r,s,t = 1,2,3$ denote flavour indices.
\label{Tab:dim6}}
\end{table}

In a specific UV-complete model, the WCs at the scale $\Lambda$ can be determined by integrating out the heavy NP fields. 
In the spirit of our model-independent approach, we will instead consider the WC of the $\calO_a^{(d)}$ at $\mu=\Lambda$ as independent free parameters.
However, at lower energies, the coefficients of different operators will mix as an effect of the RGEs. In particular, multiple operators will be induced at the EW scale even if the UV physics is assumed to match dominantly to a single operator (or just a few of them) at the scale $\Lambda$.

Below the EW scale, we work within the LEFT employing the basis introduced by Ref.~\cite{Jenkins:2017jig}.
As we will see in the next section, the observables that we focus on\,---\,the LFV quarkonium decays\,---\,and the LFV decays of muons and taus that will set indirect constraints on them can be induced by dimension-5 photon dipole operators\footnote{We adopt the following convention for the fermionic QED couplings: $\mathcal{L}_\text{QED} = - e Q_f \bar f \cancel{A} f$.}
\begin{align}
	\label{eq:LEFTdip}
		\mathcal{L}_\text{dipole} = ~& C_{e\gamma,pr}\, (\bar{\ell}_p \sigma^{\mu\nu}P_R \ell_r  )\, F_{\mu\nu} + h.c.\,,
\end{align}
by dimension-6 $2q2\ell$ operators
\begin{align}
      \label{eq:LEFT2q2l}
		\mathcal{L}_{2q2\ell} &= C_{eq,prst}^{V,LL}\, (\bar{\ell}_p \gamma^\mu P_L \ell_r)( \bar{q}_s\gamma_\mu P_L q_t) 
		+ C_{eq,prst}^{V,RR} \,(\bar{\ell}_p \gamma^\mu P_R \ell_r)( \bar{q}_s\gamma_\mu P_R q_t)  \nonumber \\
		&+  C_{eq,prst}^{V,LR} \,(\bar{\ell}_p \gamma^\mu P_L \ell_r )(\bar{q}_s\gamma_\mu P_R q_t) 
		   + C_{qe,prst}^{V,LR} \,(\bar{q}_p\gamma_\mu P_L q_r)(\bar{\ell}_s \gamma^\mu P_R \ell_t )   \nonumber \\
		&+ \Big[ C_{eq,prst}^{S,RL} \,(\bar{\ell}_p P_R \ell_r )(\bar{q}_s P_L q_t) + C_{eq,prst}^{S,RR}\, (\bar{\ell}_p P_R \ell_r )(\bar{q}_s P_R q_t) \nonumber \\
		& +  C_{eq,prst}^{T,RR} \, (\bar{\ell}_p \sigma_{\mu\nu} P_R \ell_r)( \bar{q}_s \sigma^{\mu\nu} P_R q_t)+h.c.\Big]\,,
\end{align}
and by dimension-6 4-lepton ($4\ell$) operators
	\begin{align}
	      \label{eq:LEFT4l}
		\mathcal{L}_{4\ell} &= C_{ee,prst}^{V,LL}\, (\bar{\ell}_p \gamma^\mu P_L \ell_r )(\bar{\ell}_s\gamma_\mu P_L \ell_t) + C_{ee,prst}^{V,RR}\, (\bar{\ell}_p \gamma^\mu P_R \ell_r )(\bar{\ell}_s\gamma_\mu P_R \ell_t)  \nonumber \\
		&+ C_{ee,prst}^{V,LR} \,(\bar{\ell}_p \gamma^\mu P_L \ell_r )(\bar{\ell}_s\gamma_\mu P_R \ell_t) +\Big[C_{ee,prst}^{S,RR}\, (\bar{\ell}_p P_R \ell_r )(\bar{\ell}_s P_R \ell_t) +h.c. \Big]\,,
	\end{align}
where $\ell$ denotes leptons, $q=u,d$, that is, up-type or down-type quarks, and $p,r,s,t$ are flavour indices. All fields are defined in the physical mass basis. Notice that, in contrast to the SMEFT case, the WCs of the above LEFT interactions are dimensionful parameters.

As we will show in the next section, besides the effects induced by the above LEFT operators, certain quarkonia processes are also sensitive to dimension-7 lepton-gluon and lepton-photon operators that read~\cite{Liao:2020zyx}
	\begin{align}
	\label{eq:L7GG}
		\mathcal{L}_{\ell\ell GG} =~& C_{eGG,pr}\, (\bar{\ell}_p P_R \ell_r )\, G^a_{\mu\nu}G^{a\mu\nu} + C_{eG\tilde{G},pr}\, (\bar{\ell}_p P_R \ell_r )\,G^a_{\mu\nu}\tilde{G}^{a\mu\nu} +h.c. \,, \\
		\mathcal{L}_{\ell\ell FF} =~& C_{eFF,pr}\, (\bar{\ell}_p P_R \ell_r )\, F_{\mu\nu}F^{\mu\nu} ~~+ C_{eF\tilde{F},pr}\, (\bar{\ell}_p P_R \ell_r )\,F_{\mu\nu}\tilde{F}^{\mu\nu} ~\,+h.c. \,,
		\label{eq:L7FF}
	\end{align}
where the dual field strength tensors are defined by $\tilde F^{\mu\nu}=\frac12 \epsilon^{\mu\nu\alpha\beta} F_{\alpha\beta}$. 

The tree-level matching at the EW scale of the dimension-6 SMEFT operators of \tref{Tab:dim6} to the above presented LEFT basis was computed in Ref.~\cite{Jenkins:2017jig}. For completeness, we collect the matching formulae in \aref{app:matching}.
The dimension-7 lepton-gluon/photon operators are obtained from tree-level matching to dimension-8 SMEFT operators and from 1-loop matching to dimension-6 scalar operators with quarks, see e.g.~Ref.~\cite{Liao:2020zyx}. 
Moreover, in order to obtain phenomenological predictions in terms of the WCs, the latter need to be evaluated at the energy scale relevant for the process of interest. As usual, this can be done by solving the RGEs of the WCs that, for the LEFT framework, can be found in~Ref.~\cite{Jenkins:2017dyc},
while for SMEFT operators the running of the WCs is given by the RGEs calculated in Refs.~\cite{Jenkins:2013zja,Jenkins:2013wua,Alonso:2013hga}.

%%%%%%%%%%%%%%%%%%%%%%%%%%%%%%%%
\section{Decay rates for LFV quarkonium decays}
%%%%%%%%%%%%%%%%%%%%%%%%%%%%%%%%
\label{sec:obs}

In this section we present our calculation for the LFV decay rates of quarkonia in terms of the LEFT operators defined in the previous section.
We follow the calculation in Ref.~\cite{Hazard:2016fnc} (see also Ref.~\cite{Abada:2015zea}).
Due to the C parity conservation in the decay of vector quarkonia $V$ with $J^C = 1^-$, $V\to \ell_i\ell_j$
and $V\to \ell_i\ell_j \gamma$ decays are induced by C-odd and C-even operators, respectively, and are thus complementary.
The expressions for the other LFV processes relevant to our analysis are collected in \aref{app:obs}.

\subsection{LFV leptonic vector quarkonium decay: \texorpdfstring{$V \to \ell_i^- \ell_j^+$}{V->li- lj+}}
\label{sec:Vllp}
We parameterise the quarkonium decay amplitude by
\begin{align}
	\mathcal{M} & = 
	\frac12 \bar u_i \,{\cancel\epsilon}_V (V_L P_L + V_R P_R) v_j 
	+ \frac{2i}{m_V}\bar u_i \,\epsilon_V^\mu \sigma_{\mu\nu} P^\nu  ( T_L P_L + T_R P_R ) v_j \,,
\end{align}
where $P^\nu$, $m_V$, and $\epsilon_V$ are respectively the momentum, mass, and polarisation vector of the vector quarkonium. The coefficients parametrising vector and tensor interactions can be expressed in terms of the LEFT Wilson coefficients and are given by
	\begin{align}
		V_L & = f_V m_V \left(C_{eq,ijqq}^{V,LL} + C_{eq,ijqq}^{V,LR} + \frac{2 e^2 Q_q Q_\ell \delta_{ij}}{m_V^2}\right)\,,
		    &
		T_L & = m_V f_V^T C_{eq,jiqq}^{T,RR*} -eQ_qf_V C_{e\gamma,ji}^{*}\,,
		\label{eq:VTL} 
		\\
		V_R & = f_V m_V \left(C_{eq,ijqq}^{V,RR} + C_{qe,qqij}^{V,LR} + \frac{2 e^2 Q_q Q_\ell \delta_{ij}}{m_V^2}\right) \,,
		    &
		T_R & = m_V f_V^T C_{eq,ijqq}^{T,RR} -eQ_qf_V C_{e\gamma,ij}\,,
		\label{eq:VTR} 
	\end{align}
where $Q_\ell=-1$ and $Q_q$ are the electric charges of leptons and quarks. The lepton flavour conserving contribution is dominated by tree-level photon exchange which enters the coefficients parametrising the vector interactions $V_{L,R}$.
The two form factors $f_V$ and $f_V^T$ parameterise the hadronic vector and tensor matrix elements
\begin{align}
	\braket{0\left|\bar q \gamma^\mu q \right| V(P)} & = f_V m_V \epsilon^\mu_V \,,
						   &
    \braket{0\left|\bar q \sigma^{\mu\nu} q \right| V(P)} & = if_V^T (\epsilon^\mu_V P^\nu - \epsilon^\nu_V P^\mu)\;. 
\end{align}
The resulting branching ratio for $V\to \ell_i^-\ell_j^+$ is
\begin{align}\nonumber
		\mathrm{BR}(V\to \ell_i^-\ell_j^+)  = \frac{m_V}{\Gamma_V} \frac{\lambda^{1/2}(1,y_i^2,y_j^2) }{16\pi}\Bigg[&
		\frac{|V_L|^2+|V_R|^2}{12} \Big(2-y_i^2-y_j^2 -(y_i^2-y_j^2)^2\Big)
		\\\nonumber&
		+\frac43 \left(|T_L|^2+|T_R|^2\right)\Big(1+y_i^2+y_j^2-2(y_i^2 -y_j^2)^2\Big)
		\\\nonumber&
	+y_i y_j \Big(\mathrm{Re}\big(V_L V_R^*\big) + 16\, \mathrm{Re}\big(T_R T_L^*\big)\Big)
		\\\nonumber&
		+2y_i \left(1+y_j^2-y_i^2\right)\, \mathrm{Re}\big(V_R T_R^* + V_L T_L^*\big)
		\\&
		+2y_j \left(1+y_i^2-y_j^2\right)\, \mathrm{Re}\big(V_L T_R^* + V_R T_L^*\big)
		\Bigg]\,,
	\end{align}
where $y_i=m_i/m_V$, for a lepton of mass $m_i$, and  $\lambda(x,y,z)=x^2+y^2+z^2-2xy -2xz-2yz$ denotes the K\"all\'en function. We used \texttt{FeynCalc}~\cite{Mertig:1990an,Shtabovenko:2016sxi,Shtabovenko:2020gxv} to obtain the squared matrix element. Our result agrees with Ref.~\cite{Hazard:2016fnc} in the limit $y_i\to 0$ except for an additional factor $(1+y_j^2/2)$ in the first line for the vector operator contribution.

\subsection{Radiative LFV leptonic vector quarkonium decay: \texorpdfstring{$V \to \ell_i^- \ell_j^+\gamma$}{V->li- lj+ gamma}}
\label{sec:Vllgamma}
In light of the recent analysis of the radiative LFV $\Upsilon(1S)$ decays performed by Belle~\cite{Belle:2022cce}, we calculate the radiative LFV leptonic vector quarkonium decay using the non-relativistic colour singlet model, following~Refs.~\cite{Li:2021phq} and~\cite{Barger:1987nn}. The final state photon can originate from one of the initial state quarks, one of the final state leptons or result from the effective vertex in Eq.~\eqref{eq:L7FF}. The operators contributing to final state radiation are strongly constrained by the LFV vector quarkonium decay $V\to \ell_i^- \ell_j^+$. We thus neglect contributions from final state radiation in the analysis of the radiative decay $V\to \ell_i^- \ell_j^+ \gamma$, see \aref{app:Vllgamma} for full details.
Taking the different Lorentz and polarisation structures into account, the quarkonium decay amplitude for $V(P,\epsilon_V)\to \ell_i^-(p_i)\ell_j^+(p_j)\gamma(q,\epsilon)$ is given by
\begin{align}
	\mathcal{M}& = 
	\frac{Q_q e}{x_\gamma m_V^3}  \, \Big[(P\cdot q)\, (\epsilon_V\cdot \epsilon^*) - (P \cdot \epsilon^*) \,(q\cdot \epsilon_V)\Big]\,\bar u_i \left[(S_R+\tilde S_Rx_\gamma) P_R + (S_L+\tilde S_Lx_\gamma) P_L\right] v_j
	\nonumber\\&
	+\frac{Q_q e}{ x_\gamma m_V^3}  \,i \epsilon_{\alpha\beta\mu\nu} P^\alpha q^\beta \epsilon_V^\mu \epsilon^{*\nu} \, \bar u_i \left[(P_R^\prime+i\tilde P_R^\prime x_\gamma) P_R + (P_L^\prime+i\tilde P_L^\prime x_\gamma) P_L\right] v_j 
	\nonumber\\&
	+\frac{Q_q e}{x_\gamma m_V^2} i \epsilon_{\alpha\beta\mu\nu} q^\beta \epsilon_J^\mu \epsilon^{*\nu}  \bar u_i \gamma^\alpha (A_R P_R + A_L P_L ) v_j\,,
\end{align}
and depends on the following combinations of LEFT WCs
\begin{align}
	S_R &=  2 m_V f_V \big( C^{S,RR}_{eq,ijqq}+  C^{S,RL}_{eq,ijqq} \big)\,,
	&
	\tilde S_R &= 4 m_V^2 f_V C_{eFF,ij} \,,
	\nonumber\\
	S_L &=   2 m_V f_V \big(C^{S,RL}_{eq,jiqq} +C^{S,RR}_{eq,jiqq}\big)^*\,,
	&
	\tilde S_L &= 4 m_V^2 f_V C_{eFF,ji}^* \,,
	\nonumber\\
	P_R^\prime &=   2 m_V f_V \big( C^{S,RR}_{eq,ijqq}-  C^{S,RL}_{eq,ijqq} \big)\,,
	&
	\nonumber
	\tilde P_R^\prime &= 4  m_V^2 f_V C_{eF\tilde F,ij}\,,
	\\\nonumber
	P_L^\prime &=  2 m_V f_V \big(C^{S,RL}_{eq,jiqq} -C^{S,RR}_{eq,jiqq}\big)^*\,,
	&
	\tilde P_L^\prime &= 4  m_V^2 f_V C_{eF\tilde F,ji}^* \,,
	\\
	A_R & = 2 m_V f_V \big(C_{qe,qqij}^{V,LR}-C_{eq,ijqq}^{V,RR}\big)\,,
	\,
	&
	A_L & = 2 m_V f_V \big(C_{eq,ijqq}^{V,LL}-C_{eq,ijqq}^{V,LR}\big)
	\;. \label{eq:SPA}
\end{align}
Here, $A_{L,R}$ denote axial-vector contributions, $S_{L,R}$ scalar contributions and $P^\prime_{L,R}$ pseudoscalar contributions. Finally terms with an overtilde correspond to contributions from the dimension-7 operators with two photon field strength tensors, \eref{eq:L7FF}. Note that the contributions proportional to $C_{eFF}$ and $C_{eF\tilde F}$ are proportional to an additional factor $x_\gamma =2 E_\gamma/m_V$ compared to the pseudoscalar contributions. 

In the limit of one massless final state lepton,
the phase space integration can be carried out analytically and we obtain for the branching ratio
\begin{align}
	\textrm{BR}(V\to \ell_i^-\ell_j^+\gamma)  &= \frac{\alpha Q_q^2 m_V}{192\pi^2 \Gamma_V} \bigg[
		\left(|A_L|^2 + |A_R|^2\right)\, G_A(y)
		+\left(|S_L|^2 + |P_L^\prime|^2 + |S_R|^2+|P_R^\prime|^2\right)\, G_S(y) 
		\nonumber\\&
		+\left(|\tilde S_L|^2 + |\tilde P_L^\prime|^2  + |\tilde S_R|^2  + |\tilde P_R^\prime|^2\right)\, \tilde G_S(y)
		+ I_{PA} \, G_{PA}(y) + \tilde I_{PA} \,\tilde G_{PA}(y)
		\nonumber\\
		&
		+\mathrm{Re}(S_L\tilde S_L^* +S_R \tilde S_R^*)\, \hat G_S(y)
	+\mathrm{Im}( P_L^\prime \tilde P_L^{\prime*} + P_R^\prime \tilde P_R^{\prime*} ) 	\, \left(\hat G_{S}(y)-\frac{1}{12}\right)
		\bigg]\,,
\end{align}
where $y$ denotes the non-zero mass of the charged (anti-)lepton normalised to the vector quarkonium mass, i.e.~$y=y_i$ ($y=y_j$). $I_{PA}$ and $\tilde I_{PA}$ denote the interference terms which differ for the two cases
\begin{align}\label{Interference}
     I_{PA} &= \begin{cases}
	+\mathrm{Re}(A_L P_L^{\prime*} + A_R P_R^{\prime*}) & \text{for}\, y=y_i\neq0,\, y_j=0\,,\\
   -\mathrm{Re}(A_L P_R^{\prime*} + A_R P_L^{\prime*}) & \text{for}\, y_i=0,\, y=y_j\neq 0\,,\\
   \end{cases}
   \nonumber\\
     \tilde I_{P A} &= \begin{cases}
	+\mathrm{Im}( A_L \tilde P_L^{\prime*} + A_R \tilde P_R^{\prime*} ) & \text{for}\, y=y_i\neq0,\, y_j=0\,,\\
  -\mathrm{Im}(  A_L \tilde P_R^{\prime*}+   A_R \tilde P_L^{\prime*}) & \text{for}\, y_i=0,\, y=y_j\neq 0\,,\\
  \end{cases}\;
\end{align}
because for a massless final state lepton the different chiralities do not interfere.
The kinematic functions entering the branching ratio are given by
\begin{align}
	G_A(y) & = \frac{1}{36}\left(8-45y^2+36y^4+y^6+12(y^2-6)y^4\ln y\right),
	\nonumber\\
	G_S(y)&=\frac{1}{12}\left(1-6y^2+3y^4+2y^6-12y^4\ln y\right),
	\nonumber\\
	\tilde G_S(y) & = \frac{1}{120}\left(3-30y^2-20y^4+60y^6-15 y^8+2 y^{10}-120y^4\ln y\right),
	\nonumber\\
	\hat G_S(y) & = \frac{1}{12}\left(1-8y^2+8y^6-y^8-24y^4\ln y\right),
	\nonumber\\
	G_{PA}(y) & = \frac{y}{2}\left(1+4y^2-5y^4+4(2+y^2)y^2\ln y\right),
	\nonumber\\
	\tilde G_{PA}(y) & =  \frac{y}{3} \left(1 +9 y^2 -9y^4 -y^6 + 12 (1+y^2) y^2 \ln y\right).
\end{align}

\subsection{LFV leptonic pseudoscalar quarkonium decay: \texorpdfstring{$P \to \ell_i^- \ell_j^+$}{P->li- lj+}}
Using the equations of motion for the final state spinors, the pseudoscalar decay amplitude 
\begin{align}
	i \mathcal{M} =  \bar u_i \left( S_R  P_R + S_L P_L\right) v_j\;,
\end{align}
can be parameterised in terms of two coefficients %
\begin{align}
		S_R & = \frac{h_P}{4m_q} \left(C_{eq,ijqq}^{S,RR} - C_{eq,ijqq}^{S,RL}\right) 
		- \frac{f_P}{2} \left[m_j \left(C_{eq,ijqq}^{V,LR} - C_{eq,ijqq}^{V,LL}\right) + m_i \left(C_{eq,ijqq}^{V,RR} - C_{qe,qqij}^{V,LR}\right) \right]
		\nonumber\\ &
		+i\frac{4\pi}{\alpha_s} a_P C_{eG\tilde G,ij}\;,
		\\
		S_L & = \frac{h_P}{4m_q} \left(C_{eq,jiqq}^{S,RL} - C_{eq,jiqq}^{S,RR}\right)^* 
		-  \frac{f_P }{2} \left[m_i \left(C_{eq,ijqq}^{V,LR} - C_{eq,ijqq}^{V,LL}\right) + m_j \left(C_{eq,ijqq}^{V,RR} - C_{qe,qqij}^{V,LR}\right) \right]
		\nonumber\\ &
		+i\frac{4\pi}{\alpha_s} a_P C_{eG\tilde G,ji}^*
		\;.
	\end{align}
	Given the proportionality to the final state lepton masses, the pseudoscalar quarkonium decay is mostly sensitive to pseudoscalar WCs.
	The form factors $f_P$, $h_P$ and $a_P$ parameterise the hadronic axialvector, pseudoscalar, and anomaly matrix elements
\begin{align}
	\braket{0\left|\bar q \gamma^\mu\gamma_5 q \right| P(p)} & = i f_P p^\mu \,,
						   &
	\braket{0\left|\bar q i  \gamma_5 q \right| P(p)} & = \frac{h_{P}}{2m_q} \,,
						      &
	\braket{0\left|\frac{\alpha_s}{4\pi} G \tilde G \right| P(p)} & = a_P \,,
\end{align}
which satisfy the relation $h_P=m_P^2 f_P -a_P$ from axialvector current conservation. 
The gluonic matrix elements are expected to be small for $\eta_{b,c}$ and thus we take $a_P=0$. 
The resulting branching ratio for $P\to \ell_i^-\ell_j^+$ is given by
\begin{equation}
		\mathrm{BR}(P\to \ell_i^-\ell_j^+)  = \frac{m_P}{\Gamma_P} \frac{\lambda^{1/2}(1,y_i^2,y_j^2) }{16\pi}
		\Big[
			 \big(|S_L|^2+|S_R|^2\big)\, \left(1-y_i^2-y_j^2\right)
		-4 y_i y_j\, \mathrm{Re}(S_L S_R^*)
		\Big],
	\end{equation}
where $y_i=m_i/m_P$ and we used \texttt{FeynCalc}~\cite{Mertig:1990an,Shtabovenko:2016sxi,Shtabovenko:2020gxv} to obtain the squared matrix element.
Our result agrees with Ref.~\cite{Hazard:2016fnc} in the limit of $m_i\to 0$ for the pseudoscalar and axial-vector contributions and we also find agreement for the anomaly contribution, if we disregard the superfluous $+h.c.$ for the dimension-7 terms with the field strength tensors in~Ref.~\cite{Hazard:2016fnc}.

\subsection{LFV leptonic scalar quarkonium decay: \texorpdfstring{$S \to \ell_i^- \ell_j^+$}{S->li- lj+}}
Using the fact that the vector current form factor vanishes for scalar quarkonia, the scalar decay amplitude 
\begin{align}
	 \mathcal{M} =  \bar u_i \left( S_R  P_R + S_L P_L\right) v_j\,,
\end{align}
can be parameterised in terms of two coefficients 
\begin{align}
		S_R & = \frac{m_S f_S}{2} \left(C_{eq,ijqq}^{S,RR} + C_{eq,ijqq}^{S,RL}\right) 
		+\frac{4\pi}{\alpha_s} a_S C_{eGG,ij}\,,
		\\
		S_L & = \frac{m_S f_S}{2} \left(C_{eq,jiqq}^{S,RL} + C_{eq,jiqq}^{S,RR}\right)^* 
		+\frac{4\pi}{\alpha_s} a_S C_{eGG,ji}^*
		\;.
	\end{align}
	The form factors $f_S$ and $a_S$ are defined as~\cite{Cheng:2013fba}
\begin{align}
	\braket{0\left|\bar q q \right| S} & = m_S f_S \,,
						   &
	\braket{0\left|\frac{\alpha_s}{4\pi} G G \right| S} & = a_S \;.
\end{align}
The gluonic matrix elements are expected to be small and thus we take $a_S=0$ in the analysis.
The branching ratio for $S\to \ell_i^-\ell_j^+$ is given by
\begin{equation}
		\mathrm{BR}(S\to \ell_i^-\ell_j^+)  = \frac{m_S}{\Gamma_S} \frac{\lambda^{1/2}(1,y_i^2,y_j^2) }{16\pi}
		\Big[
			 \big(|S_L|^2+|S_R|^2\big)\, \left(1-y_i^2-y_j^2\right)
		-4 y_i y_j\, \mathrm{Re}(S_L S_R^*)
		\Big],
	\end{equation}
where $y_i=m_i/m_S$
and we used \texttt{FeynCalc}~\cite{Mertig:1990an,Shtabovenko:2016sxi,Shtabovenko:2020gxv} to obtain the squared matrix element.
Our result agrees with Ref.~\cite{Hazard:2016fnc} in the limit of $m_i\to 0$.

%%%%%%%%%%%%%%%%%%%%%%%%%%%%%%%%%%%%%%%%%%%%%%%

\section{Numerical results}
\label{sec:Results}

In this section we analyse the LFV decays of quarkonium states, focusing mostly on the vector quarkonium $c\bar c$ states $J/\Psi$ and $\Psi(2S)$, and the $b\bar b$ states $\Upsilon(1S)$, $\Upsilon(2S)$ and $\Upsilon(3S)$. 
We also provide indirect upper limits on the LFV decay rates of the lightest (pseudo)scalar $c\bar c$ and $b\bar b$ resonances, as well as on the radiative decays of vector quarkonia. In contrast to the  2-body decays of vector quarkonia $V\to\ell\ell^\prime$, these latter processes are all sensitive  to scalar operators, as shown by the formulae presented in \sref{sec:obs}. Hence they potentially provide complementary information.

In the previous section we computed the 
contributions to the LFV quarkonia decays (LFVQD) to leading order. 
Nevertheless, in order to reduce the hadronic uncertainties, we will compute LFVQD as a double ratio, normalising the LFV channel to the experimentally measured lepton flavour conserving decay to electrons:
\begin{equation}
{\rm BR}(V\to \ell\ell') = 
\frac{{\rm BR}(V\to ee)_{\rm exp}}{{\rm BR}(V\to ee)_{\rm LO}}\,
{\rm BR}(V\to \ell\ell')_{\rm LO}\,,
\end{equation}
where the subscript LO refers to the leading order expressions derived in \sref{sec:obs} and the subscript exp to the corresponding experimental value~\cite{Workman:2022ynf}. We checked that this introduces a small correction, in general a 2-4\% reduction of the rates, with the only exception of $\Upsilon(3S)$, whose rates increase by about 8\%.
The (pseudo)scalar quarkonium decays to an electron-positron pair have not been measured yet, therefore we just consider the LO predictions for those decays. 
For all quarkonium decays we include both lepton flavour combinations as final states, $\ell^+ \ell^{\prime-}$ and $\ell^- \ell^{\prime +}$.

For the numerical analysis, we implemented the expressions of \sref{sec:obs} for LFVQD in the \texttt{flavio}~\cite{Straub:2018kue} python code.
This allows us to use the range of flavour observables already included in the routine, as well as the renormalisation group evolution implemented by means of the \texttt{wilson}~\cite{Aebischer:2018bkb} package. 
The latter also includes the full tree-level matching between SMEFT and LEFT (cf.~\aref{app:matching}), which we use in the following to explore both EFT frameworks. When evaluating quarkonium processes, we set the renormalisation scales for decays of bottomonium resonances to their respective masses, while the renormalisation scales for charmonium decays are fixed at $\mu=2$~GeV.

The numerical values for the masses, decay widths and decay constants of the quarkonia we consider are collected in \tref{Tab:input}.
Notice that the total widths of the scalar bottomonium states $\chi_{b0}$ have not been measured yet. Following Ref.~\cite{Godfrey:2015vda}, we evaluate it using the theoretically calculated partial decay width of the radiative decays $\chi_{b0}(nP)\to \Upsilon(mS) + \gamma$ and the experimentally measured branching ratio BR($\chi_{b0}(nP) \to \Upsilon(mS) + \gamma$) in order to obtain
\begin{equation}
    \Gamma_{\chi_{b0}(nP) } = \frac{\Gamma(\chi_{b0}(nP)\to \Upsilon(mS) +\gamma)_{\rm th}}{\mathrm{BR}(\chi_{b0}(nP) \to \Upsilon(mS) +\gamma)_{\rm exp}} \;.
\end{equation}
For $\chi_{b0}(1P)$ the only available decay is $\chi_{b0}(1P) \to \Upsilon(1S) + \gamma$ with $\Gamma(\chi_{b0}\to\Upsilon(1S)+ \gamma) = 23.8$~keV~\cite{Godfrey:2015dia} and BR($\chi_{b0}(1P)\to\Upsilon(1S) + \gamma$)=$(1.94\pm 0.27)$\%~\cite{ParticleDataGroup:2020ssz}. For $\chi_{b0}(2P)$, we take the simple weighted average 
of the total widths obtained from the decay rates to $\Upsilon(1S)+\gamma$ and $\Upsilon(2S)+\gamma$, which have partial widths of 2.5~keV and 10.9~keV~\cite{Godfrey:2015dia}, and branching ratios of $(3.8\pm1.7)\times 10^{-3}$ and $(1.38\pm0.30)\%$~\cite{ParticleDataGroup:2020ssz}, respectively, with the errors added in quadrature.

\begin{table}[htb!]
\begin{center}
{\small
\renewcommand{\arraystretch}{1.3}
\begin{tabular}{
>{\centering}p{2.5cm}
>{\centering}p{2.5cm}
>{\centering}p{2.5cm}
>{\centering}p{2.5cm}
c
}
\hline
Quarkonium & Mass (MeV) & $\Gamma$ (keV) & $f_{V}$ (GeV) & $f_{V}^T$ (GeV)  \\
%%%%%%%%%%%%%
\hline 
$J/\psi$ &  $3096.900\pm0.006$ & $92.6\pm1.7$ & 0.4104(17)~\cite{Hatton:2020qhk} & 0.3927(27)~\cite{Hatton:2020vzp}\\
$\psi(2S)$ &  $3686.10\pm0.06$ & $294\pm8$ & 0.2926(12)~\cite{Abada:2015zea}& -- \\
$\Upsilon(1S)$ &  $9460.30\pm0.26$ & $54.02\pm1.25$ &0.6772(97)~\cite{Hatton:2021dvg} & -- \\
$\Upsilon(2S)$ &  $10023.26\pm0.31$ & $31.98\pm2.63$ & 0.481(39)~\cite{Colquhoun:2014ica}& -- \\
$\Upsilon(3S)$ &  $10355.2\pm0.5$ & $20.32\pm1.85$ & 0.395(25)~\cite{Chung:2020zqc}& --\\
\hline
\end{tabular}

\vspace{2ex}

\begin{tabular}{
>{\centering}p{2.5cm}
>{\centering}p{3cm}
>{\centering}p{3.5cm}
c
}
\hline
Quarkonium & Mass (MeV) & $\Gamma$ (MeV) & $f_{M}$ (GeV)   \\
%%%%%%%%%%%%%
\hline
$\eta_c(1S)$ &  $2983.9\pm0.4$ & $32.0\pm0.7$ & 0.387(7)~\cite{Becirevic:2013bsa}\\
$\eta_b(1S)$ &  $9398.7\pm2.0$ & $10\pm5$ & 0.724(12)~\cite{Hatton:2021dvg}\\
\hline
$\chi_{c0}(1P)$ &  $3414.71\pm0.30$ & $10.8\pm0.6$ & $-i$\,0.887~\cite{Godfrey:2015vda}\\
$\chi_{b0}(1P)$ &  $9859.44\pm0.52$ & $1.23\pm0.17$ & $-i$\,0.423~\cite{Godfrey:2015vda} \\
$\chi_{b0}(2P)$ &  $10232.5\pm0.6$ & $0.76\pm0.15$ & $-i$\,0.421~\cite{Godfrey:2015vda} \\
\hline
\end{tabular}
\caption{Quarkonium masses, widths, and decay constants, taken from the PDG~\cite{ParticleDataGroup:2020ssz} with the exception of $\chi_{b0}(nP)$ which have not been measured yet.
They 
have been obtained following Ref.~\cite{Godfrey:2015vda} from the calculated decay width of the radiative decay $\chi_{b0}(nP)\to  \Upsilon(mS)+\gamma$ and its measured branching ratio as discussed in the text.
When the transverse form factor is missing, we assume $f_{V}^T\equiv f_{V}$, following~Ref.~\cite{Khodjamirian:2015dda}, which is motivated by the observation that vector and tensor decay constants of light vector mesons are of a similar order of magnitude. This is also consistent with the non-relativistic colour singlet model~\cite{Appelquist:1974zd,DeRujula:1974rkb,Kuhn:1979bb,Keung:1980ev,Berger:1980ni,Clavelli:1982hp,Clavelli:2001zi,Clavelli:2001gb}. Following Ref.~\cite{Hazard:2016fnc}, we also use the scalar decay constants obtained in~Ref.~\cite{Godfrey:2015vda} using the mock meson approach in the quark model. 
\label{Tab:input}}
}
\end{center}
\end{table}

\subsection{LEFT analysis}
\label{sec:LEFT}

\newcolumntype{R}[1]{>{\flushright\arraybackslash}m{#1}}
\newcolumntype{L}[1]{>{\flushleft\arraybackslash}m{#1}}
\newcolumntype{C}[1]{>{\centering\arraybackslash}m{#1}}

\begin{table}[pt!]\centering
\renewcommand{\arraystretch}{1.3}
\subfloat[
\small Vector and tensor operators.
The operators  $C^{V,RR}_{eu, ijcc}$,  $C^{V,LR}_{ue, ccij}$,  $C^{T,RR}_{eu, ji cc}$ and $C_{e\gamma, ji}$ lead, respectively, to the same results as $C^{V,LL}_{eu, ijcc}$,  $C^{V,LR}_{eu, ijcc}$,  $C^{T,RR}_{eu, ijcc}$ and $C_{e\gamma, ij}$.\label{Table:LEFTccVTD}]{
\begin{tabular}{C{2cm} C{4cm} C{3.5cm} C{3.5cm}}
\hline
\multirow{2}{*}{Operator} & \multirow{2}{*}{Strongest constraint} & \multicolumn{2}{c}{Indirect upper limits on BR}\\\cline{3-4}\\[-4ex]
&& $J/\psi\to\ell\ell' \hspace{0.6cm}$ & $\psi(2S)\to\ell\ell'  \hspace{0.6cm}$ \\
\hline\\[-3ex]
 $C^{V,LL}_{eu, \mu ecc}$  &  $\mu  \to e,\,\text{Au}$ & [1.6 - 0.07] $\times\,10^{-15} $  & [2.8 - 0.2] $\times\,10^{-16} $     \\
 $C^{V,LR}_{eu, \mu ecc}$   &  $\mu  \to e,\,\text{Au}$& [1.5 - 0.07] $\times\,10^{-15} $  & [2.8 - 0.2] $\times\,10^{-16} $     \\
  $C^{T,RR}_{eu, \mu ecc}$   &  $\mu  \to e\gamma$  & [3.4 - 0.5] $\times\,10^{-21}$  &[7.8 - 1.4] $\times\,10^{-22} $    \\
  $C_{e\gamma,\mu e}$  &  $\mu  \to e\gamma$  & [2.6 - 2.5] $\times 10^{-26} $  & [6.3 - 0.5] $\times 10^{-27} $    \\
  %%%
  [1ex]\hline\\[-3ex]
 $C^{V,LL}_{eu, \tau ecc}$  &  $\tau  \to \rho e$ & [6.6 - 0.1] $\times\, 10^{-9} $  & [1.2 - 0.05] $\times\,10^{-9} $     \\
  $C^{V,LR}_{eu, \tau ecc}$  &  $\tau  \to \rho e$& [6.5 - 0.1] $\times\, 10^{-9} $  & [1.2 - 0.04] $\times\,10^{-9} $      \\
 $C^{T,RR}_{eu, \tau ecc}$  &  $\tau  \to e\gamma$  & [1.2 - 0.05] $\times\, 10^{-12} $  & [2.3 - 0.2] $\times\,10^{-13} $     \\
  $C_{e\gamma, \tau e}$  &  $\tau  \to e\gamma$  & [1.7 - 1.6] $\times 10^{-18} $  & [4.7 - 3.5] $\times 10^{-19} $    \\
  %%%
 [1ex]\hline\\[-3ex]
 $C^{V,LL}_{eu, \tau \mu cc}$  &  $\tau  \to \rho \mu$ & [4.5 - 0.09] $\times\, 10^{-9} $  & [7.9 - 0.3] $\times\,10^{-10} $     \\
  $C^{V,LR}_{eu,\tau\mu cc}$  &  $\tau  \to \rho \mu$& [4.4 - 0.09] $\times\, 10^{-9} $  & [7.9 - 0.3] $\times\,10^{-10} $      \\
 $C^{T,RR}_{eu, \tau \mu cc}$  &  $\tau  \to \mu\gamma$  & [1.6 - 0.07] $\times\, 10^{-12} $  & [2.9 - 0.3] $\times\,10^{-13} $     \\
  $C_{e\gamma, \tau \mu}$  &  $\tau  \to \mu\gamma$  & [2.2 - 2.1] $\times 10^{-18} $  & [6.1 - 4.5] $\times 10^{-19} $    \\[1ex]
\hline
\end{tabular}}
\vspace{2ex}

\subfloat[
\small Scalar operators. We find similar limits for $\psi(2S) \to \ell \ell'\gamma$, about a factor of 4\,(2) stronger for the $\mu e\, (\tau\ell)$ channels. 
See text for details on how the indirect upper limits have been estimated.
\label{Table:LEFTccScalars}
]{
\begin{tabular}{C{1.9cm} C{1.9cm} C{3.4cm} C{3.2cm} C{3.2cm} }
\hline
\multirow{2}{*}{Operator} & \multirow{2}{*}{Str.~const.} & \multicolumn{3}{c}{Indirect upper limits on BR}\\\cline{3-5}\\[-4ex]
&& $J/\psi\to\ell\ell' \gamma \hspace{0.4cm}$ 
& $\eta_c\to\ell\ell' \hspace{0.4cm}$ & $\chi_{c0}(1P)\to\ell\ell' \hspace{0.4cm}$ \\
 \hline\\[-3ex]
  $C^{S,RR}_{eu, \mu e cc}$  &  $\mu  \to e$, Au & [1.5 - 1.4] $\times\,10^{-21}  $  & [2.0 - 1.9] $\times\,10^{-20} $  & [3.4 - 3.2] $\times\,10^{-19} $     \\
  $C^{S,RL}_{eu, \mu e cc}$  &  $\mu  \to e$, Au & [1.5 - 1.4] $\times\,10^{-21}  $  & [2.0 - 1.9] $\times\,10^{-20} $  & [3.4 - 3.2] $\times\,10^{-19} $     \\
  [1ex]\hline\\[-3ex]
  $C^{S,RR}_{eu, \tau e cc}$  &  $\tau  \to  e\gamma$ & [1.7 - 0.003] $\times\,10^{-10} $  & [6.8 - 0.01] $\times\,10^{-9} $ & [1.5 - 0.003] $\times\,10^{-7} $     \\
  $C^{S,RL}_{eu, \tau e cc}$  & $\tau  \to  e\gamma$    & [2.0 - 0.09] $\times\,10^{-10}$  & [9.2 - 0.4] $\times\,10^{-9}$   & [1.3 - 0.08] $\times\,10^{-7}$     \\
  [1ex]\hline\\[-3ex]
 $C^{S,RR}_{eu, \tau \mu cc}$  &  $\tau  \to \mu\gamma$ & [2.2 - 0.004] $\times\,10^{-10} $  & [8.7 - 0.02] $\times\,10^{-9} $   & [1.9 - 0.003] $\times\,10^{-7} $   \\
  $C^{S,RL}_{eu, \tau \mu cc}$  & $\tau  \to \mu\gamma$   & [2.6 - 0.1] $\times\,10^{-10}$  & [1.2 - 0.05] $\times\,10^{-8}$   & [1.7 - 0.1] $\times\,10^{-7}$       \\[1ex]
\hline
\end{tabular}}
\caption{Indirect upper limits on the branching ratio of LFV charmonium decays
considering a single non-vanishing LEFT operator at a scale $\mu\in(m_{q\bar q},m_Z)$.
The intervals show how the indirect limits become stronger as $\mu$ increases. 
The second column displays the low-energy observable that gives the strongest constraint.
}\label{Table:LEFTcc}
\end{table}

\begin{table}[tp!]
\centering
\renewcommand{\arraystretch}{1.3}
\subfloat[
\small Vector and tensor operators. The operators $C^{V,RR}_{ed,ijbb}$, $C^{V,LR}_{de,bbij}$, $C^{T,RR}_{ed,jibb}$ and $C_{e\gamma,ji}$ lead, respectively to the same results as $C^{V,LL}_{ed,ijbb}$, $C^{V,LR}_{ed,ijbb}$, $C^{T,RR}_{ed,ijbb}$ and $C_{e\gamma,ij}$.\label{Table:LEFTbbVTD}]{
\begin{tabular}{C{1.9cm} C{1.9cm} C{3.2cm} C{3.2cm} C{3.2cm}}
\hline
\multirow{2}{*}{Operator} & \multirow{2}{*}{Str.~const.} & \multicolumn{3}{c}{Indirect upper limits on BR}\\\cline{3-5}\\[-4ex]
&& $\Upsilon(1S)\to\ell\ell' \hspace{0.4cm}$ & $\Upsilon(2S)\to\ell\ell' \hspace{0.4cm}$ & $\Upsilon(3S)\to\ell\ell' \hspace{0.4cm}$ \\ 
  \hline\\[-3ex]
  $C^{V,LL}_{ed, \mu e bb}$  &  $\mu  \to e,\,\text{Au}$ & [1.1 - 0.08] $\times\,10^{-12}  $  & [9.9 - 0.8] $\times\,10^{-13} $  & [1.1 - 0.1] $\times\,10^{-12} $     \\
  $C^{V,LR}_{ed, \mu e bb}$ &  $\mu  \to e,\,\text{Au}$& [1.1 - 0.08] $\times\,10^{-12} $  & [9.9 - 0.8] $\times\,10^{-13} $   & [1.1 - 0.1] $\times\,10^{-12} $     \\
  $C^{T,RR}_{ed, \mu e bb}$  &  $\mu  \to e\gamma$  & [4.7 - 0.7] $\times\,10^{-19} $ & [4.3 - 0.7] $\times\,10^{-19} $   & [4.8 - 0.9] $\times\,10^{-19} $   \\
  $C_{e\gamma, \mu e}$  &  $\mu  \to e\gamma$  & $1.6\times 10^{-25} $  & $1.5\times 10^{-25} $  & $1.6\times 10^{-25} $   \\
  [1ex]\hline\\[-3ex]
 $C^{V,LL}_{ed, \tau e bb}$  &  $\tau  \to \rho e$ & [3.1 - 0.2] $\times\,10^{-6} $  & [2.8 - 0.2] $\times\,10^{-6} $ & [3.0 - 0.3] $\times\,10^{-6} $     \\
  $C^{V,LR}_{ed, \tau e bb}$  &  $\tau  \to \rho e$& [3.1 - 0.2] $\times\,10^{-6} $  & [2.8 - 0.2] $\times\,10^{-6} $   & [3.0 - 0.3] $\times\,10^{-6} $   \\
  $C^{T,RR}_{ed, \tau e bb}$  &  $\tau  \to e\gamma$  & [4.0 - 0.6] $\times\,10^{-11} $  & [3.7 - 0.6] $\times\,10^{-11} $  & [4.1 - 0.8] $\times\,10^{-11} $   \\
  $C_{e\gamma, \tau e}$  &  $\tau  \to e\gamma$  & $1.4\times 10^{-17} $  & $1.3\times 10^{-17} $  & $1.4\times 10^{-17} $   \\
  [1ex]\hline\\[-3ex]
 $C^{V,LL}_{ed, \tau \mu bb}$  &  $\tau  \to \rho \mu$ & [2.1 - 0.2] $\times\,10^{-6} $  & [1.9 - 0.2] $\times\,10^{-6} $   & [2.1 - 0.2] $\times\,10^{-6} $   \\
  $C^{V,LR}_{ed, \tau \mu bb}$ &  $\tau  \to \rho \mu$& [2.1 - 0.2] $\times\,10^{-6} $  & [1.9 - 0.3] $\times\,10^{-6} $  & [2.1 - 0.2] $\times\,10^{-6} $     \\
  $C^{T,RR}_{ed, \tau \mu bb}$  &  $\tau  \to \mu\gamma$  & [5.2 - 0.7] $\times\,10^{-11} $  & [4.8 - 0.7] $\times\,10^{-11} $  & [5.3 - 0.9] $\times\,10^{-11} $   \\
  $C_{e\gamma, \tau \mu}$  &  $\tau  \to \mu\gamma$  & $1.8\times 10^{-17} $  & $1.6\times 10^{-17} $  & $1.8\times 10^{-17} $   \\[1ex]
\hline
\end{tabular}}
\vspace{2ex}

\subfloat[
\small Scalar operators. The results for $\Upsilon(2S)$ are similar in size and the ones for $\Upsilon(3S)$ are slightly less constrained. 
See text for details on how the indirect upper limits have been estimated.
\label{Table:bbScalars}]{
\begin{tabular}{C{1.9cm} C{1.9cm} C{3.2cm} C{3.2cm} C{3.2cm} }
\hline
\multirow{2}{*}{Operator} & \multirow{2}{*}{Str.~const.} & \multicolumn{3}{c}{Indirect upper limits on BR}\\\cline{3-5}\\[-4ex]
&& $\Upsilon(1S)\to\ell\ell' \gamma \hspace{0.4cm}$ 
& $\eta_b\to\ell\ell' \hspace{0.4cm}$ & $\chi_{b0}(1P)\to\ell\ell' \hspace{0.4cm}$ \\
 \hline\\[-3ex]
 $C^{S,RR}_{ed, \mu e bb}$  &  $\mu  \to e$, Au & [9.2 - 5.6] $\times\,10^{-19}  $  & [1.2 - 0.73] $\times\,10^{-16} $  & [3.0 - 1.9] $\times\,10^{-16} $     \\
 $C^{S,RL}_{ed, \mu e bb}$  &  $\mu  \to e$, Au & [9.2 - 5.6] $\times\,10^{-19}  $  & [1.2 - 0.73] $\times\,10^{-16} $  & [3.0 - 1.9] $\times\,10^{-16} $     \\
  [1ex]\hline\\[-3ex]
 $C^{S,RR}_{ed, \tau e bb}$  &  $\tau  \to  e\gamma$ & [7.6 - 0.1] $\times\,10^{-9} $  & [1.1 - 0.02] $\times\,10^{-6} $ & [2.8 - 0.05] $\times\,10^{-6} $     \\
  $C^{S,RL}_{ed, \tau e bb}$  & $\tau  \to  e\gamma$    & [3.5 - 0.3] $\times\,10^{-8}$  & [5.3 - 0.4] $\times\,10^{-6}$   & [1.2 - 0.09] $\times\,10^{-5}$     \\
  [1ex]\hline\\[-3ex]
 $C^{S,RR}_{ed, \tau \mu bb}$  &  $\tau  \to \mu\gamma$ & [9.8 - 0.2] $\times\,10^{-9} $  & [1.4 - 0.03] $\times\,10^{-6} $   & [3.7 - 0.07] $\times\,10^{-6} $   \\
  $C^{S,RL}_{ed, \tau \mu bb}$  & $\tau  \to \mu\gamma$   & [4.5 - 0.3] $\times\,10^{-8}$  & [6.8 - 0.5] $\times\,10^{-6}$   & [1.5 - 0.1] $\times\,10^{-5}$       \\[1ex]
\hline
\end{tabular}}
\caption{Same as Table~\ref{Table:LEFTcc}, but for $b\bar b$ states. 
}
\label{Table:LEFTbb}
\end{table}

We are interested in assessing how large LFVQD are allowed to be given the current constraints on any other LFV process.
We can already get a good feeling about the answer to this question by working in the LEFT framework, valid below the EW scale, and switching on only those WCs that contribute directly to the LFVQD, which is the a priori most favourable scenario for these processes.
Due to RGE effects these WCs will still generate other LFV processes, including in particular strongly constrained leptonic decays, which will actually tell us how large the LFVQD could be without violating any existing bound.

We start showing our numerical results obtained by switching on a single LEFT operator at a time.
While certainly being a simplified scenario, and probably unrealistic within many UV theories, we regard this as a useful first exercise in order to assess the most relevant WCs for the processes we are interested in. 
We discuss deviations from this simplified assumption in the next subsection, where we show the results of our analysis within the SMEFT framework. 
The main results of our LEFT study are collected in Tables~\ref{Table:LEFTcc} and \ref{Table:LEFTbb}.

In Table~\ref{Table:LEFTccVTD} we summarise the results for vector and tensor operators with two charm quarks. The first two columns list the WCs (following the notation presented in~\sref{sec:eft}) and the most constraining process for a given operator. In the last two columns, we quote the resulting indirect upper limits on the branching ratios of \mbox{$J/\psi\to \ell \ell^\prime$} and \mbox{$\psi(2S)\to \ell\ell^\prime$} considering a single non-vanishing LEFT operator at a scale $\mu$ that we choose to be either the quarkonium mass scale $m_{q\bar q}$ or the $Z$ boson mass scale $m_Z$. 
As one can see, the indirect upper limits become stronger as the scale $\mu$ increases, due to the larger separation of scales (and thus a larger logarithm from the RGEs). Notice that the choice $\mu = m_{q\bar q}$
corresponds to the arguably unrealistic case that, right at the quarkonium mass scale, the single non-vanishing operator is the one the induces LFVQD, thus enhancing the latter process compared to other LFV observables. From an UV point of view this situation\,---\,if possible at all\,---\,may require very unlikely cancellations or correlations among the parameters. We still show this possibility in order to encompass even tuned scenarios favourable to LFVQD, although the case $\mu = m_Z$ leading to stronger bounds should be regarded as a more realistic situation.

For tensor and dipole operators, the strongest constraint arises from muon and tau LFV radiative decays. While the dipole operator directly contributes to the radiative LFV decay $\ell\to\ell^\prime \gamma$, the tensor WC $C^{T,RR}_{eq, \ell\ell^\prime qq}$  contributes to $C_{e\gamma,\ell\ell^\prime}$ via RG running. 
Instead, the vector operators contribute to the dipole operator only at 2-loop in the RG running, while the relevant vector operators for $\tau\to \rho \ell$, $\ell=e,\mu$, and $\mu\to e$ conversion in nuclei are generated at 1-loop, as illustrated by \fref{LFVdiag}. Thus we find that $\tau\to\rho \ell$ and $\mu\to e$ conversion in gold provide the most stringent constraints. 
We also find the same upper limits for operators with exchanged chiralities, $L\leftrightarrow R$.

In Table~\ref{Table:LEFTccScalars} we present the results for the scalar operators. 
Scalar operators with heavy quarks contribute to $\mu\to e$ conversion via gluon operators after integrating out the heavy quark, see Appendix~\ref{app:meconv}. The dimension-7 gluon operators are not implemented in \texttt{flavio}, 
but we estimate the contribution of $2q2\ell$ scalar operators with heavy quarks to gluon operators following Ref.~\cite{Crivellin:2017rmk,Kitano:2002mt,Cirigliano:2009bz,Crivellin:2014cta}.
Neglecting other loop-induced operators, we find for the $\mu\to e$ conversion rate from the operator $C^{S,RR}_{eq,\mu e qq}$ or $C^{S,RL}_{eq,\mu e qq}$
\begin{align}
    \mathrm{CR}(\mu N \to e N) & = \frac{m_\mu^5}{36\pi^2 \Gamma_{\rm capt}}\left| m_p S^{(p)} f_{Gp} +  m_n S^{(n)} f_{Gn}\right|^2 \frac{\left| C^{S,RX}_{eq,\mu e qq}(\mu=m_q) \right|^2}{m_q^2} \;,
\end{align}
where $m_q$ denotes the quark mass $m_{b,c}$, $X=R,L$, $N=p,n$ the nucleon, $f_{GN}$ is the gluon form factor, $S^{(N)}$ the scalar overlap integral, and $\Gamma_{\rm capt}$ the muon capture rate. As the ratio $C_{eq,\mu e qq}^{S,RX}/m_q$ does not run in QCD, the Wilson coefficient at scale $\mu$ can be obtained by multiplying with the running quark mass at $\mu$. 
The expressions for the other two $2q2\ell$ scalar operators are obtained by replacing the Wilson coefficients $C^{S,RX}_{eq,\mu e qq}$ by $C^{S,RX*}_{eq,e\mu qq}$.

Furthermore, scalar operators with same chirality contribute to the RG running of the dipole operator at 1-loop order and thus are strongly constrained by the non-observation of radiative LFV lepton decays. On the contrary, scalar operators with mixed chirality do not contribute the dipole operator at 1-loop order in the RG evolution. However, starting from 2-loop order, there are contributions which we estimate  in the leading-log approximation as
\begin{align}
 C_{e\gamma,ij}(m_\ell) &
 =  \frac{e^3}{(4\pi)^4} \frac{16}{3} m_c \,\ln\left(\frac{\bar\mu}{\mathrm{max}(m_\ell,m_c)}\right) \,  C_{eq,ijcc}^{S,RL}(\bar\mu)  \,,
 \end{align}
where we employed the 2-loop anomalous dimensions calculated in \cite{Crivellin:2017rmk}. 
Using this equation and setting $\bar \mu = m_Z$ or the quarkonium mass scale $m_{q\bar q}$, we estimate the indirect bounds for the scalar operators with mixed chiralities.

We find similar results for LEFT operators with $b$ quarks which are presented in Table~\ref{Table:LEFTbb}. The strongest constraints also originate from radiative LFV lepton decays for all operators with the exception of vector operators where $\tau\to\rho \ell$ and $\mu\to e$ conversion in gold provide the most stringent upper bounds. The 2-loop contribution of mixed chirality scalar operators to the dipole operator can be estimated as
 \begin{align}
C_{e\gamma,ij}(m_\ell) &
=  \frac{e^3}{(4\pi)^4} \frac{4}{3} m_b \,\ln\left(\frac{\bar\mu}{m_b}\right) \,  C_{eq,ijbb}^{S,RL}(\bar\mu) 
\;.
\end{align}
Looking at the results in Tables~\ref{Table:LEFTcc} and \ref{Table:LEFTbb}, one can already get a good feeling about the most promising WCs and decay channels. 
Firstly, we see that there is no hope to study LFV dipole operators via LFVQD.
This should not be surprising, since these operators generate the severely constrained processes $\ell^\prime\to\ell\gamma$ already at the tree level.
Tensor operators share the same fate, as large RGE effects mix them to the dipole operators. 
Secondly, we note that the $e\mu$ LFVQD modes, if induced by vector operators, are less suppressed but still far from the current experimental sensitivities both for $c\bar c$ and $b\bar b$ states, cf.~Table~\ref{tab:quarkonia}.
In this case, the most relevant constraints arises from $\mu \to e$ conversion in nuclei, whose bound is expected to improve impressively in the next few years (see \tref{tab:LFVlimits}) thus suppressing even more our hope to observe $q\bar q\to e\mu $ decays.

\begin{figure}[t!]
\begin{center}
\includegraphics[width=.9\textwidth]{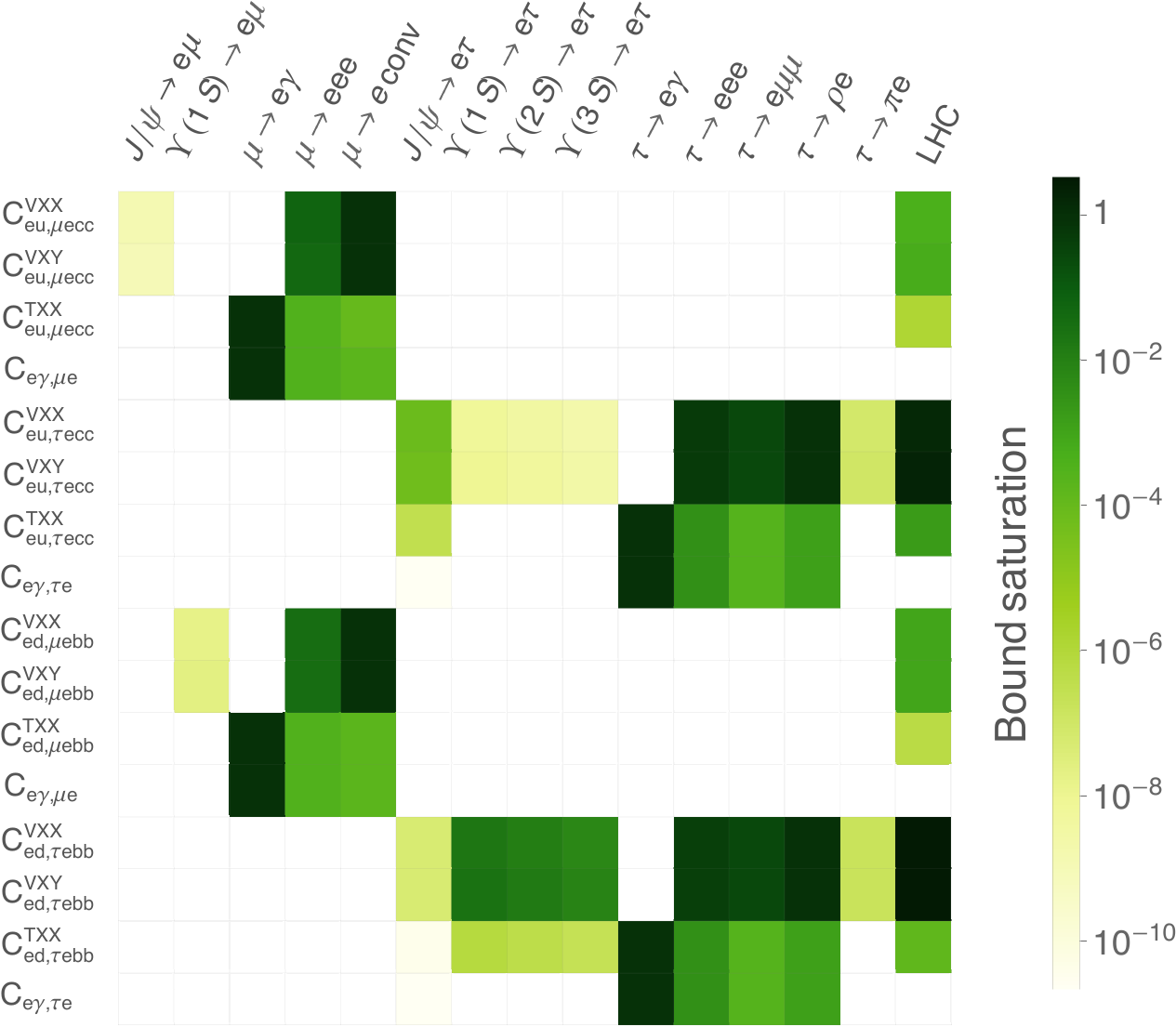}
\caption{
Matrix plot showing how far the predictions for different LFV observables are from saturating their current bounds when choosing maximum allowed values for each individual WC at $\mu=m_Z$. $X, Y \in \{L,R\}$ with $X\neq Y$.
For the $\tau\mu$ sector, we find results similar to those shown for the $\tau e$ processes, with the exception of $J/\psi\to\mu\tau$, whose limit has not been updated by BESIII yet, cf.~\tref{tab:quarkonia}.
}
\label{fig:matrixplot}
\end{center}
\end{figure}

Finally, the results in the $\tau \ell$ sector seem more optimistic for future LFVQD searches. 
In the case of $c\bar c\to\ell\tau$ decays, we find maximum allowed rates at the level of $10^{-9}$-$10^{-10}$,\footnote{The effectiveness of indirect constraints from tau decays such as $\tau \to \ell \rho$ stems from the fact that the width of the $J/\psi$ resonance is about 7 orders of magnitude larger than the tau width. This obviously contributes to suppress the branching ratios of the LFV $J/\psi$ decays compared to the tau ones.} which are about one-two orders or magnitude below the latest BESIII results for $J/\psi\to e \tau $,\footnote{Although BESIII has not provided results for $J/\psi\to \mu\tau$ yet, we assume they can set a limit at the same level as the $J/\psi\to e\tau$ one, thus improving the current bound by almost two orders of magnitude, cf.~\tref{tab:quarkonia}.} and may be partly within the sensitivity of a future super tau-charm factory (STCF).\footnote{In a $\sim$3-year run, the STCF  could produce $\sim 10^{13}\,J/\psi$ decays~\cite{Lyu:2021tlb}, that is, 1000 more than those employed by BESIII to set the present constraint~\cite{BESIIInew}.}
On the other hand, we find larger allowed rates for $b\bar b\to\ell\tau$ decays, of the order of $10^{-6}$-$10^{-7}$.
This is a consequence of a combination of phase space, narrower widths and smaller QED-induced RGE effects, since $b$ quarks carry half the electric charge of $c$ quarks.
Interestingly, the resulting rates for $b\bar b\to\ell\tau$ decays are at the level of current sensitivities, implying that Belle~II can probe these LEFT vector operators beyond the reach of any other experiment. Notice that the results in \tref{Table:LEFTbbVTD} show that the sensitivities to new physics of $\Upsilon(1S)$, $\Upsilon(2S)$ and $\Upsilon(3S)$ are comparable, since the effect of the different widths ($\Gamma[\Upsilon(1S)]>\Gamma[\Upsilon(2S)]>\Gamma[\Upsilon(3S)]$) is largely compensated by the different masses and decay constants, cf.~\tref{Tab:input}. Hence, running the experiment longer at the centre-of-mass energy of only one of these resonances may be a more effective probe of our LFV operators than collecting data in shorter runs for each resonance.\footnote{On the other hand the width of $\Upsilon(4S)$ is about 1000 times larger than that of $\Upsilon(3S)$, thus we do not expect that studying exotic decays of the former resonance would be beneficial to testing new physics.}

The results for scalar operators reported in Tables~\ref{Table:LEFTccScalars} and~\ref{Table:bbScalars} give a quantitative target for future experiments. Indeed, they provide indirect upper limits for a number of processes that have never been searched for, with the exception of the $\Upsilon(1S) \to \ell\ell^\prime \gamma$ modes. As shown in \tref{tab:quarkonia}, the Belle collaboration has recently released the first limits on these processes, which are about 2-3 orders of magnitude above our indirect limits (for the $\ell\tau$ modes).
Certainly, searches for the processes in Tables~\ref{Table:LEFTccScalars} and~\ref{Table:bbScalars} are worth pursuing, since they are sensitive to different LEFT operators\,---\,hence potentially to different kinds of new physics\,---\,compared to the 2-body quarkonia decays.
However, one should point out that the UV completion of some of these operators is not straightforward. For instance, one can see from Eqs.~(\ref{eq:CeuSRL},\,\ref{eq:CedSRR}) in \aref{app:matching} that $C^{S,RL}_{eu}$ and $C^{S,RR}_{ed}$ match to no dimension-6 SMEFT operator at tree level.

Before moving forward, it is important to clarify that, even if Tables~\ref{Table:LEFTcc} and \ref{Table:LEFTbb} indicate the most constraining observable for each operator, these are not the only relevant processes.
In order to illustrate this, we show in~\fref{fig:matrixplot} the relative importance of different LFV processes for each individual LEFT operator. 
For each WC, we take the largest possible value (at $\mu=m_Z$) that is still allowed by all constraints.
The rate of each observable is normalised to its current experimental upper limit, so the closer it is to 1 (darker colour), the more relevant that process is.
We see, for instance, that the three-body leptonic decays are also important for constraining the vectorial operators, even if $\tau \to \rho \ell$ gives the strongest bound at present.
This fact is important in particular when considering more complex scenarios in an attempt to evade some of the bounds and maximise the LFVQD rates, since suppressing the most constraining observable given in Tables~\ref{Table:LEFTcc} and \ref{Table:LEFTbb} might not be enough. 
We will discuss this point in the following subsection within the SMEFT framework.

It is also interesting to compare these results with the limits obtained from high-energy measurements of the tail of di-lepton distributions at the LHC~\cite{Angelescu:2020uug}, although it is important to note that they are valid only if the NP scale is high enough so that the EFT is still valid at the LHC ({\it i.e.},~above a few TeV).
We see that the LHC bounds in the last column of~\fref{fig:matrixplot} are several orders of magnitude weaker than low-energy constraints in the $\mu e$ sector, although they are  similar (even slightly stronger in some cases) for the $\tau\ell$ sector. This nicely shows the complementarity between low- and high-energy LFV searches.

%%%%%%%%%%%%%%%%%%%%%%%%%%%%%%%%%%%%%%%%%%%%%%%

\subsection{SMEFT analysis}
\label{sec:SMEFT}

\begin{figure}[t!]
\begin{center}
\includegraphics[width=.49\textwidth]{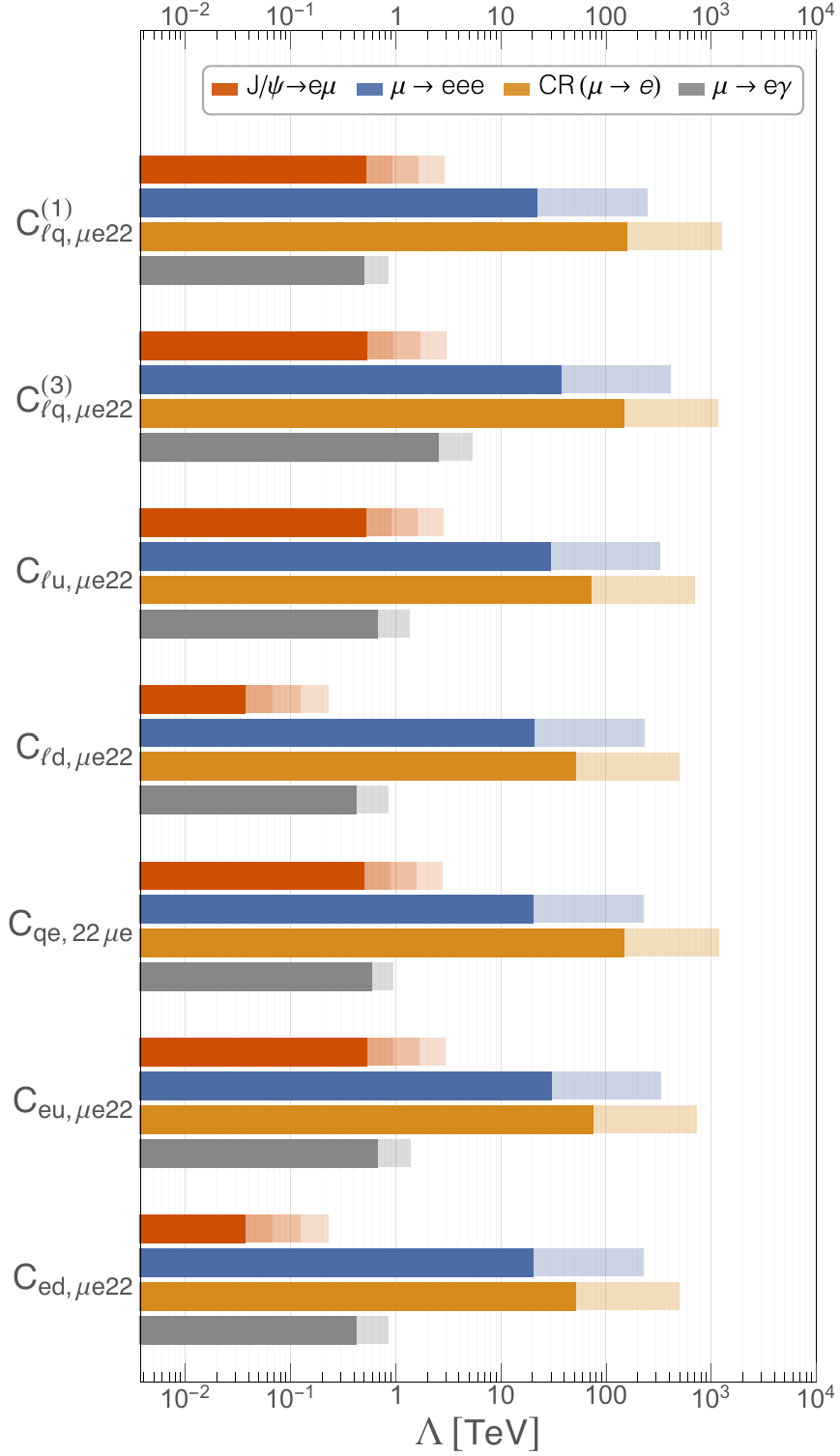}\hfill
\includegraphics[width=.49\textwidth]{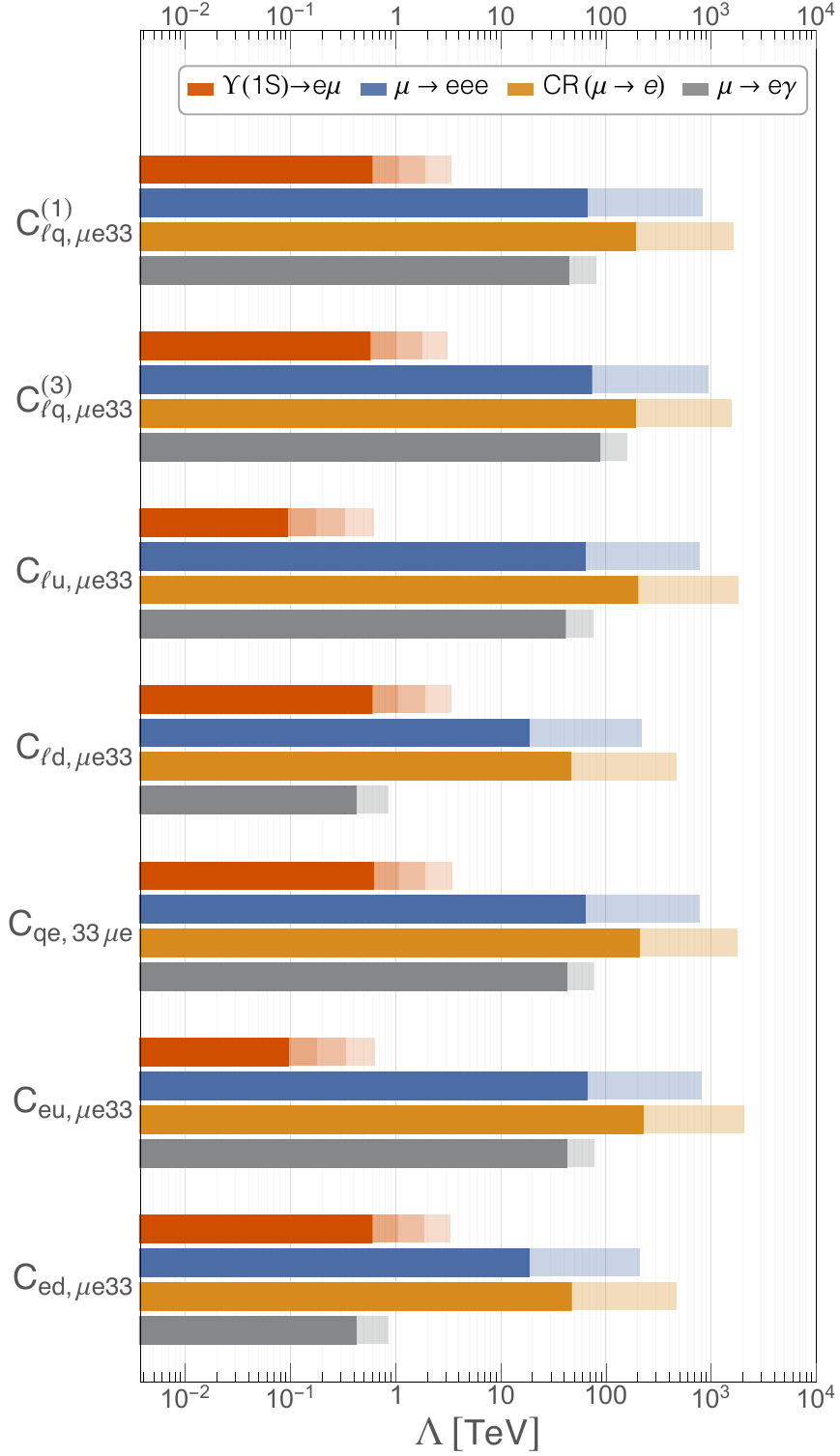}
\caption{Defining a single non-zero SMEFT WC at $\mu=\Lambda$, and assuming a perturbative coefficient $|C(\Lambda)|\leq1$, these bars show the highest NP scale that each $\mu e$ LFV observable can probe. Darker colours are for current bounds, while lighter ones are for future sensitivities. 
For LFV quarkonium decays, we show the prospects assuming a future improvement of 1, 2, 3 orders of magnitude.
}\label{barplot_mue}
\end{center}
\end{figure}

Next, we consider the SMEFT framework.
While it is the natural EFT setup when the new physics scale lies above the EW scale, it does not provide a valid description for low-energy processes such as the LFVQD. 
Therefore, a proper LFV analysis of our observables in terms of the SMEFT operators requires a convolution of SMEFT RGE~\cite{Jenkins:2013zja,Jenkins:2013wua,Alonso:2013hga} down to the EW scale, matching to LEFT~\cite{Jenkins:2017jig}, and LEFT RGE~\cite{Jenkins:2017dyc} to the physical scale of interest, {\it i.e.}, the quarkonium mass.
The first two steps introduce additional contributions that might distort the LEFT results discussed in the previous subsection.

Given the results of the analysis above, here we only focus on LFV vector quarkonium decays, $V\to \ell\ell^\prime$.
Moreover,  due to the strong bounds from $\ell_i\to\ell_j\gamma$, we neglect the dipole operators and consider just the $2q2\ell$ operators in Table~\ref{Tab:dim6},
with the exception of $\calO_{\ell equ}^{(1/3)}$ and $\calO_{\ell edu}$, since they induce large dipole operators through RGE and are thus very tightly constrained. 

In \fref{barplot_mue} we present the results for the single SMEFT operator analysis for the $\mu e$ sector, switching on at the scale $\Lambda$ just a single $2q2\ell$ operator.
We show for each operator the maximum new physics scale that is being probed by each observable, as accessing larger scales would require of non-perturbative WCs, that is $|{\rm C}(\Lambda)|>1$.
The different coloured bars show the current limits (dark colour) and future reach (light colour) for different observables. 
For LFVQD (orange-red) we illustrate with different shadings possible improvements of the sensitivity to the branching ratio by one, two, and three orders of magnitude. 
The left (right) plot shows the results for $2q2\ell$ operators with second (third) generation quarks, motivated by searches for LFV charmonium (bottomonium) decays. 
For all operators, searches for $\mu\to e$ conversion in nuclei (yellow), followed by $\mu\to eee$ (blue) provide the most stringent constraints with an expected improvement of one order of magnitude in the future. Both LFVQD and $\mu\to e\gamma$ (grey) are less sensitive to 4-fermion $2q2\ell$ SMEFT operators. If new physics mainly generates the operators in \fref{barplot_mue} with couplings $C/\Lambda^2 \gtrsim 1/(1000\,\textrm{TeV})^2 - 1/(100\,\textrm{TeV})^2$, we thus expect that both $\mu \to e$ conversion in nuclei and $\mu\to eee$ will be observed at upcoming experiments, while $\mu \to e\gamma$ and LFVQD, such as $J/\psi  \to e\mu$ and $\Upsilon(nS) \to e\mu$, will not.
Hence any observation of LFVQD to $e\mu$ would be a most striking signal that cannot be explained in terms of a single $2q2\ell$ SMEFT operator.

\begin{figure}[t!]
\begin{center}
\includegraphics[width=.49\textwidth]{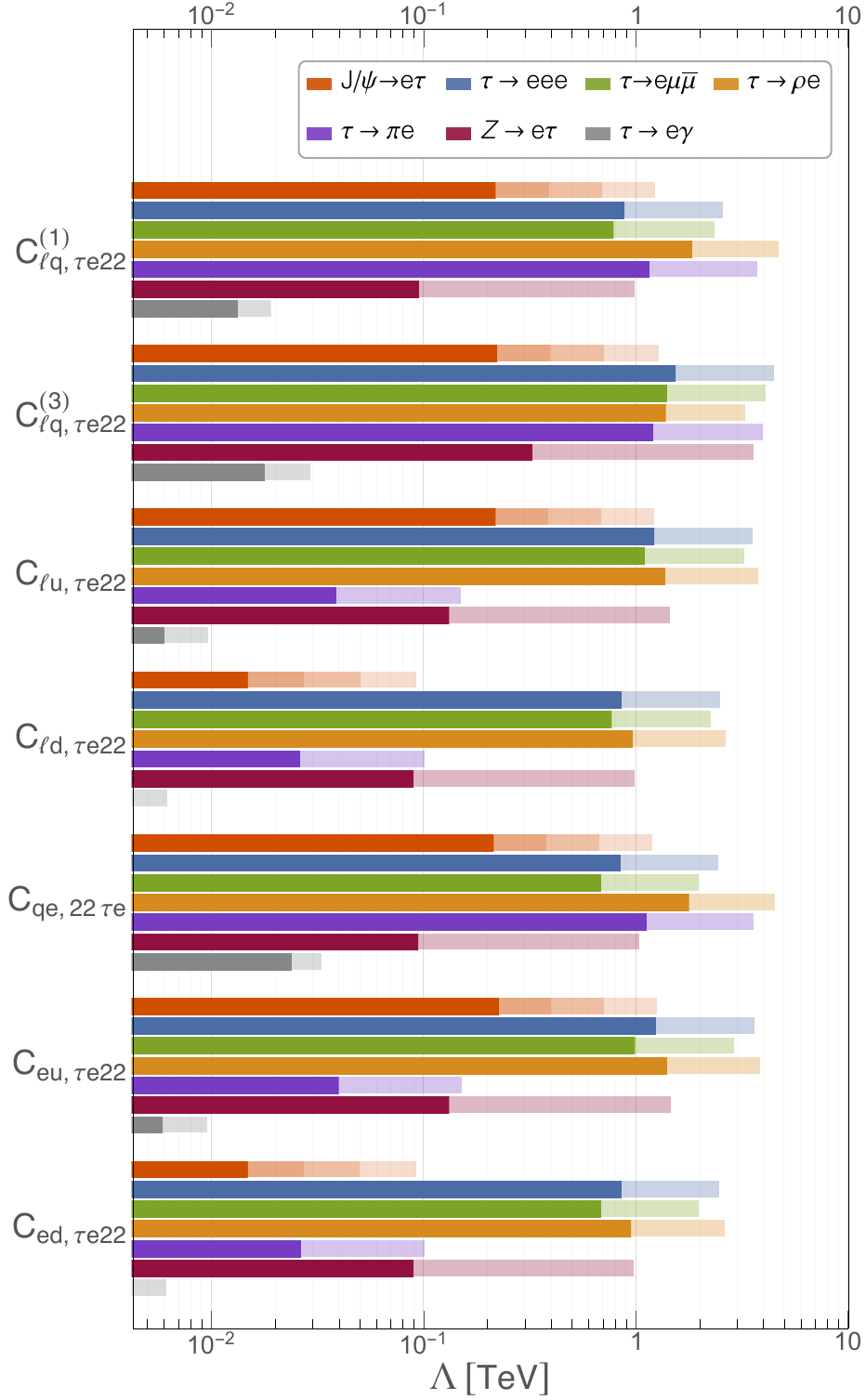}\hfill
\includegraphics[width=.49\textwidth]{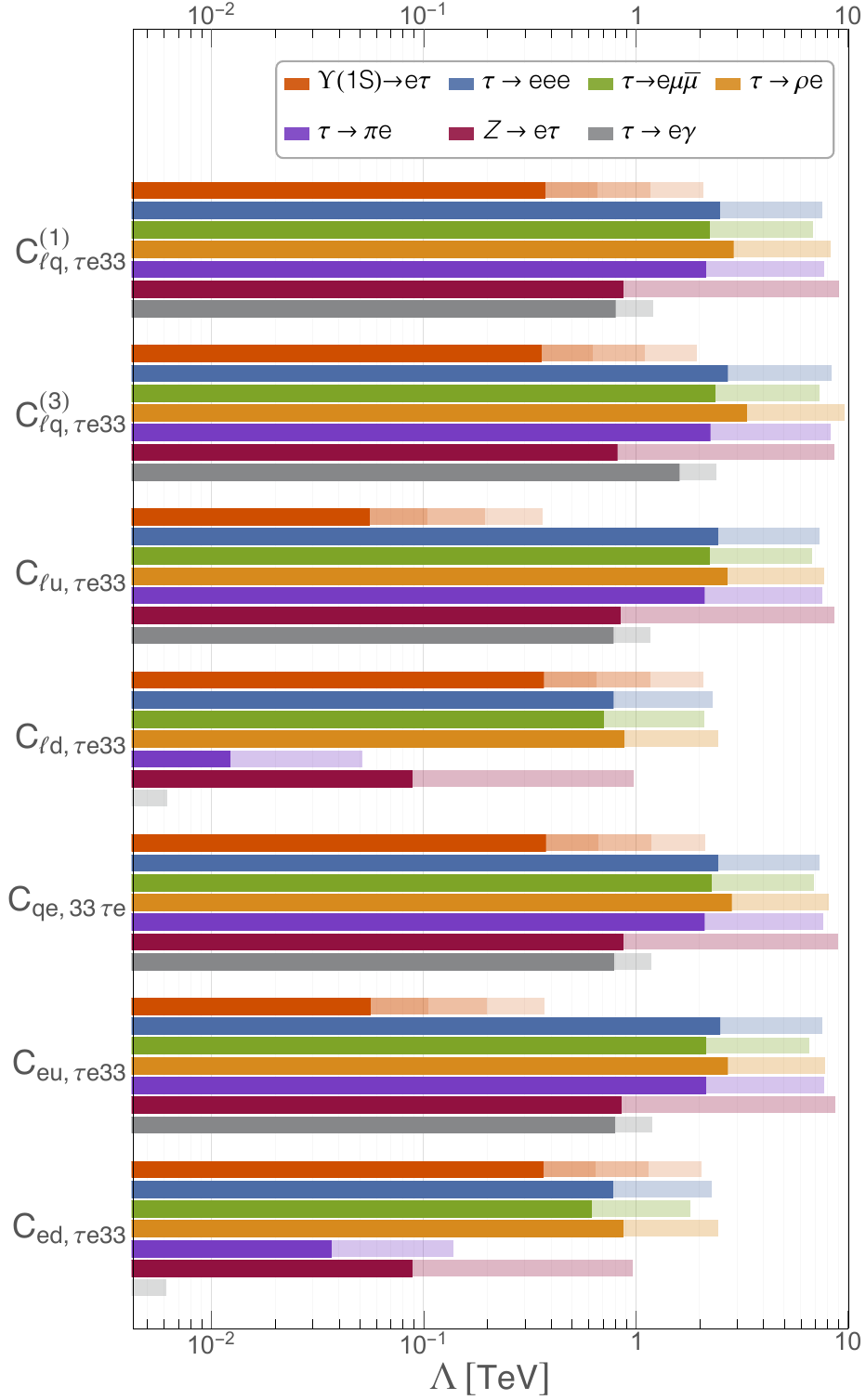}
\caption{Same as \fref{barplot_mue} for the $\tau e$ sector.
}\label{barplot_taue}
\end{center}
\end{figure}

\fref{barplot_taue} displays the analogous results for the single SMEFT operator analysis in the $\tau e$ sector.\footnote{Similar results are obtained for the $\tau \mu$ sector with the only exception being $J/\psi\to \mu\tau$, for which there is no BESIII analysis yet.}
The different coloured bars illustrate now the sensitivity of LFVQD (orange-red), $\tau\to eee$ (blue), $\tau\to e\mu\bar\mu$ (green), $\tau\to \rho e$ (yellow), $\tau\to\pi e$ (purple), $Z\to e \tau$ (dark red) and the radiative decay $\tau\to e\gamma$ (grey). 
We find that current constraints (dark colour) for LFV $\tau$ decays provide the most stringent constraints. 
Nevertheless, if the sensitivity of LFVQD searches is improved by 2-3 orders of magnitude, they may probe currently unexplored new physics scales $\Lambda$ for some of the operators. 
While this observation is in line with the results of the above LEFT analysis for $J/\psi \to e\tau$, the results for $\Upsilon(nS) \to e\tau$ in \fref{barplot_taue} look somewhat less optimistic than those obtained within the LEFT framework, cf.~\tref{Table:LEFTbbVTD}.

The origin of these strong constraints for some of the operators involving $b$ quarks is precisely the above-mentioned additional RGE effects that SMEFT operators are subject to.
In particular, diagrams obtained by closing the quark loop of a $2q2\ell$ operator can contribute to the lepton-Higgs operators displayed in \tref{Tab:dim6}, which induce LFV couplings of the $Z$ boson, see Eqs.~(\ref{eq:ZeL},\,\ref{eq:ZeR}). In turn, such couplings give rise to both LFV $Z$ decays and all kinds of LFV 4-fermion operators ($2q2\ell$ as well as $4\ell$) 
through the matching shown in \aref{app:matching}, see e.g.~Ref.~\cite{Calibbi:2021pyh}. Due to the large coupling to the Higgs field, this effect is particularly pronounced for those operators involving top quarks and it enhances the relative importance of LFV $\tau$ decays and $Z\to e\tau$ compared to LFVQD, as can be seen in the right plot of \fref{barplot_taue}. Interestingly, this plot also shows that, in line to the observations in Ref.~\cite{Calibbi:2021pyh}, a $Z$-pole run of future $e^+e^-$ colliders such as the FCC-ee or the CEPC would probe these operators through $Z$ LFV as well as (or better than) Belle~II will do searching for LFV $\tau$ decays. On the other hand, operators that do not involve top quarks will not generate large $Z$ LFV effects ({\it e.g.}~$C_{\ell d,\tau e bb}$ and $C_{ed,\tau ebb}$) and can be probed better by searches for $\Upsilon(nS)\to e\tau$ (and LFV $\tau$ decays) than $Z \to e\tau$. This provides an interesting example of the complementarity between low-energy and high-energy searches for LFV phenomena.

%%%%%%%%%%%%%%%%%%%%%

As in the previous LEFT analysis, switching on a single WC is a good first approach to analysing the LFVQD. However, it is a somewhat unrealistic scenario for any UV-complete theory.
Unless some additional symmetry is present, we could expect that several of our SMEFT operators are generated at the new physics scale $\Lambda$ where we can integrate out the new degrees of freedom, and this could induce possible interferences and cancellations among different operators, changing the conclusions drawn above. 
Indeed, this is not an unlikely outcome, given the interplay among $2q2\ell$ operators and RGE-induced lepton-Higgs operators that we have just discussed.

\begin{figure}[t!]
\begin{center}
\includegraphics[width=\textwidth]{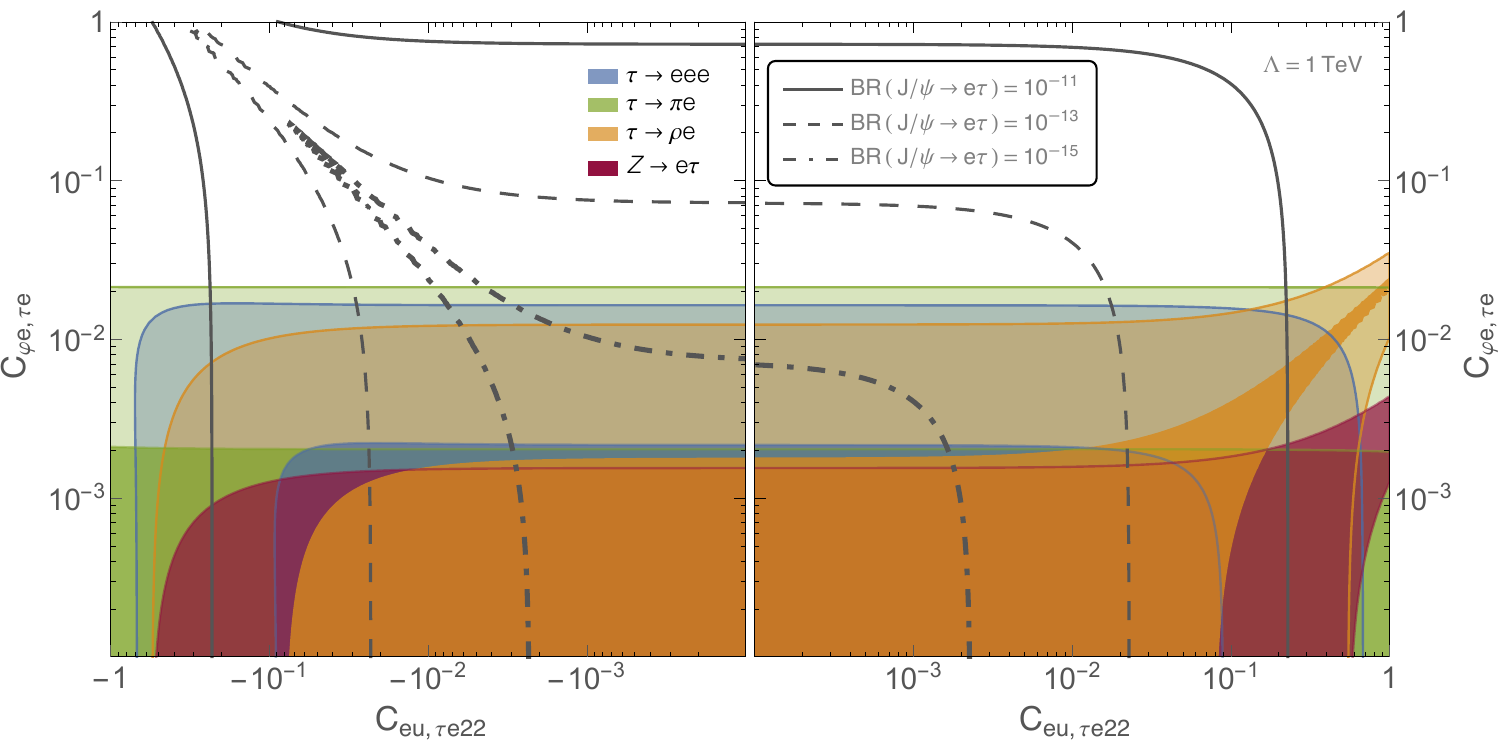}
\includegraphics[width=\textwidth]{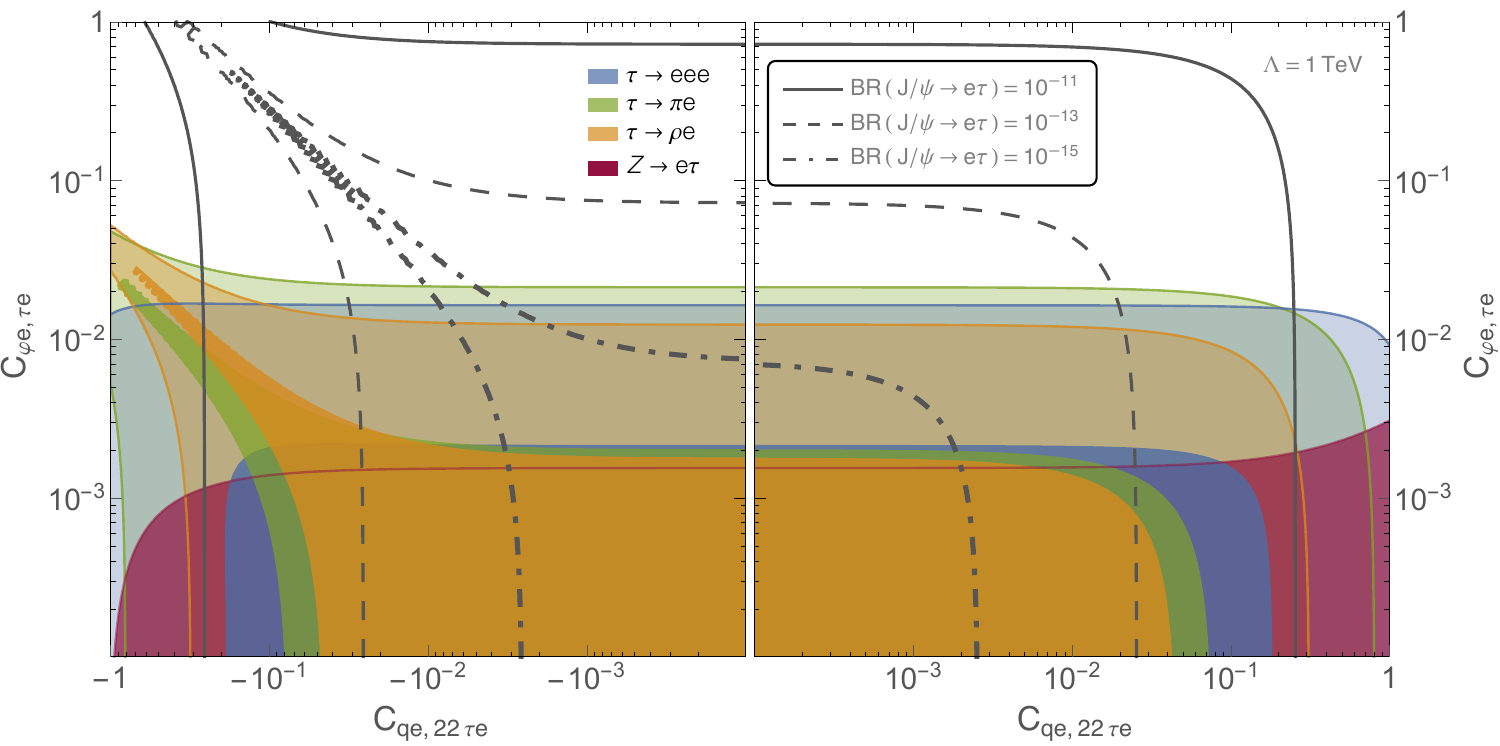}
\end{center}
\caption{Contours of BR($J/\psi \to e\tau $) as a function of the Wilson coefficients $C_{eu,\tau e22}$ and $C_{\varphi e,\tau e}$ (top panel) and $C_{eq,\tau e22}$ and $C_{\varphi e,\tau e}$ (bottom panel) at \mbox{$\Lambda=1$~TeV}. The light coloured regions are allowed by the current constraints on $\tau\to eee$ (blue), $\tau\to\pi e$ (green), $\tau\to\rho e$ (orange). The dark coloured regions show the respective future sensitivities and, in addition, that of $Z\to e\tau$ (red).} 
\label{plots2Dcc}
\end{figure}

\begin{figure}[t!]
\begin{center}
\includegraphics[width=\textwidth]{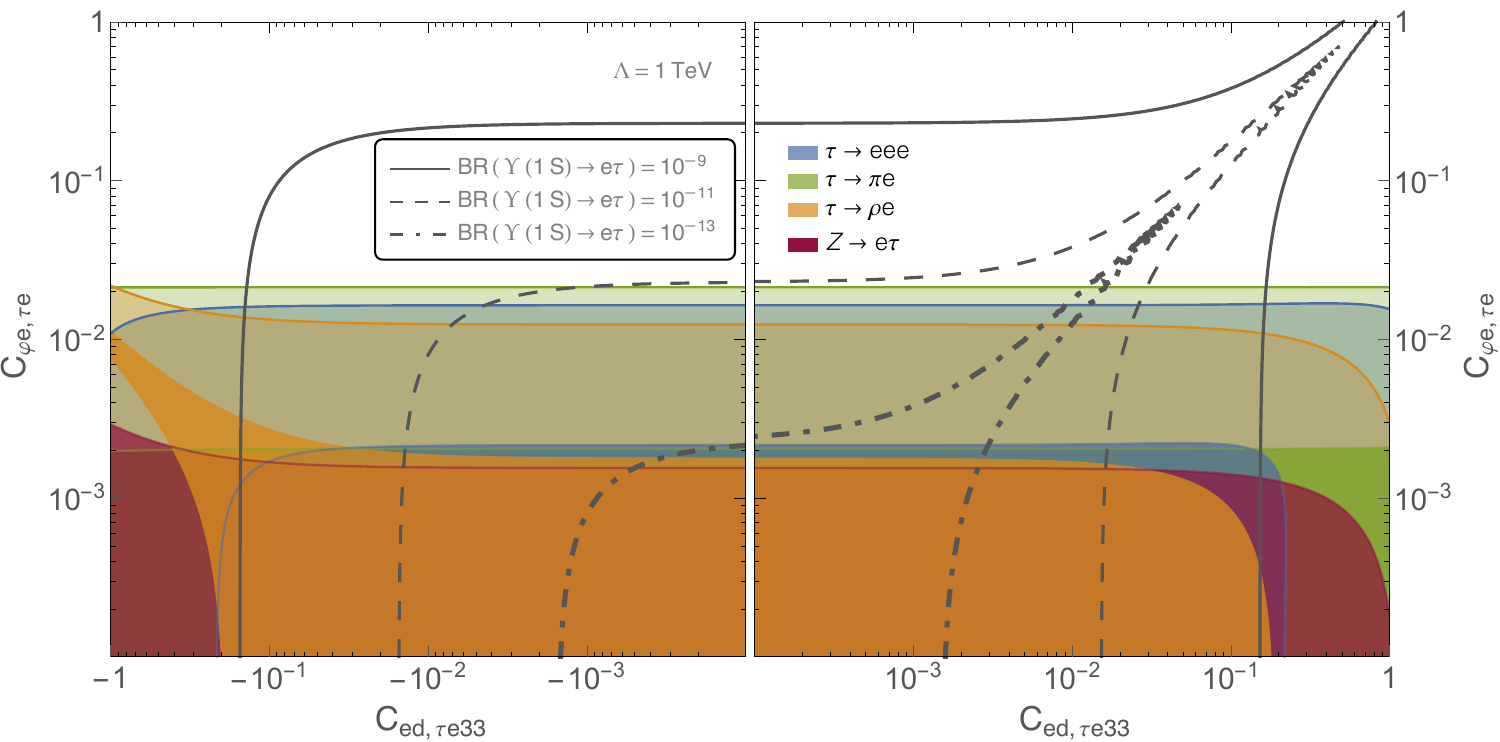}
\includegraphics[width=\textwidth]{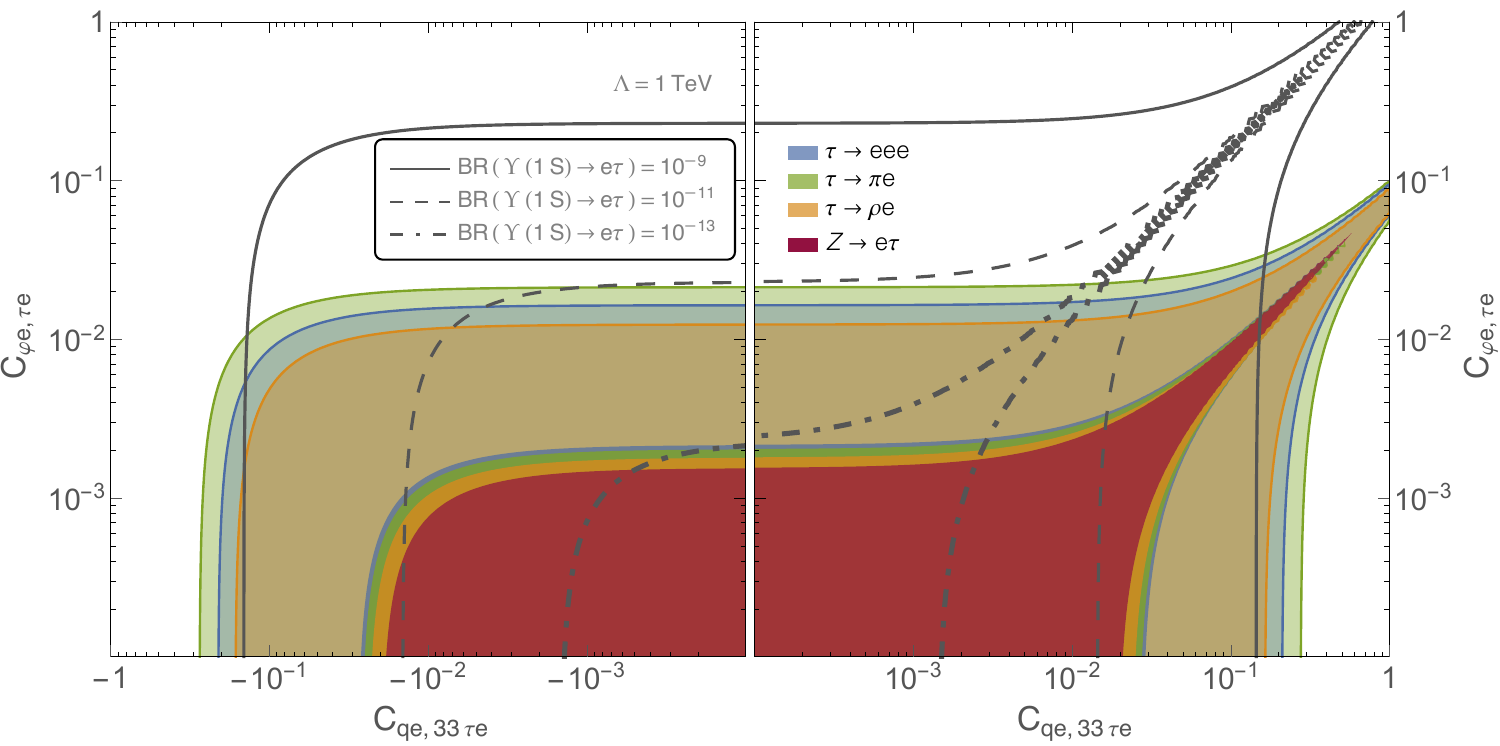}
\end{center}
\caption{Contours of BR($\Upsilon(1S)\to e\tau $) as a function of the Wilson coefficients $C_{ed,\tau e33}$ and $C_{\varphi e,\tau e}$ (top panel) and $C_{eq,\tau e33}$ and $C_{\varphi e, \tau e}$ (bottom panel) at the scale \mbox{$\Lambda=1$~TeV}. Colours as in~\fref{plots2Dcc}.}\label{plots2Dbb}
\end{figure}

In order to explore possible deviations from the single operator analysis, we now turn to a two-operator SMEFT analysis in the $\tau e$ sector.
In Figures~\ref{plots2Dcc} and \ref{plots2Dbb} we show the resulting LFVQD branching ratios as functions of the $2q2\ell$ and lepton-Higgs Wilson coefficients on a logarithmic scale. We choose the lepton-Higgs WC to be positive, hence in the right panels both WCs are positive, while in the left panels the $2q2\ell$ WC is negative. The top panels show results for operators involving right-handed quark currents and the bottom panels for left-handed quark currents. 
For illustration purposes, we only show results for right-handed lepton currents but we find qualitatively similar results for operators built from the corresponding left-handed currents.
Notice that we set \mbox{$\Lambda =1$~TeV} for all plots.

The light-coloured regions in Figures~\ref{plots2Dcc} and \ref{plots2Dbb} are \emph{allowed} by the present bounds on $\tau\to eee$ (blue), $\tau\to \pi e$~(green), $\tau\to \rho e$~(yellow). The corresponding darker colours indicate the future reach of these processes, that is, how negative results of future searches would reduce the allowed parameter space. Besides those three $\tau$ decays, we display the impact of the future sensitivity on $Z\to e\tau$~(red), while we do not show its current bound, as this process is not sensitive enough to constrain the displayed WCs at present.
The plots show that constraints from LFV $Z$ (future) and $\tau$ decays are generally more relevant than LFVQD, in line with the results previously shown in \fref{barplot_taue}. However, there exist non-trivial relations among the Wilson coefficients that can lead to cancellations in one or more of the decay rates. These are visible as \emph{flat directions}, where the contour lines or the shaded regions extend to arbitrarily large values of the Wilson coefficients. The cancellation is generally only possible for a single observable in a given direction, so that the overall bound on the size of the WCs is not much affected. 
In other words, most of the times different LFV $\tau$ decays are complementary and cover each others flat directions.
There is, however, the possibility of an intriguing simultaneous cancellation in all observables with the exception of $\Upsilon(1S)\to e\tau $, as shown in the bottom right panel of \fref{plots2Dbb}. This means that, along that direction, BR$(\Upsilon(1S)\to e\tau $) is not subject to indirect constraints from LFV $\tau$ decays and can in principle be as large as to saturate the present experimental limit.

These flat directions can be understood by looking at the leading order running and matching conditions of our two EFTs.
The LFVQD branching ratio for right-handed charged leptons is proportional to the square of $|V_R|$ in Eq.~\eqref{eq:VTR}, which can be expressed in terms of SMEFT operators at the matching scale $\mu=m_Z$ following the relations in \aref{app:matching}.
Neglecting RGE effects, the amplitude  for $J/\psi\to e\tau$ is then proportional to 
\begin{equation}
    V_{R}  \propto 
    C_{eu, \tau e  22} + C_{qe,22\tau e } + \left(1-\frac{8}{3} s_{\mathrm{w}}^2\right) C_{\varphi e, \tau e } 
    \approx C_{eu, \tau e  22} + C_{qe,22 \tau e } + 0.4 C_{\varphi e, \tau e }\,,
 \end{equation}
where $s_{\mathrm{w}}\equiv\sin \theta_W$ is the sine of the weak mixing angle.
Then, we clearly see that there are flat directions with vanishing $V_R$, which can be observed in the left panels of~\fref{plots2Dcc}.
 Similarly for $\Upsilon(1S)\to e\tau$ the amplitude is proportional to 
 \begin{equation}
    V_{R}  \propto 
    C_{ed,\tau  e 33} + V_{ib}^* V_{jb} C_{qe,ij \tau e } - \left(1-\frac{4}{3} s_{\mathrm{w}}^2\right) C_{\varphi e,\tau e } 
    \approx 
    C_{ed,\tau  e 33} + V_{ib}^* V_{jb} C_{qe,ij \tau  e} - 0.7 C_{\varphi e,\tau e }\,,
\end{equation}
where $V$ is the CKM matrix. 
Notice that the relative sign between the $C_{\varphi e}$ and the $2q2\ell$ operators is now opposite, hence the flat directions for $\Upsilon(1S)\to e\tau$ appear in the right panels of~\fref{plots2Dbb}. 

Understanding the flat directions for the other LFV decays in the figures is more involved. 
The reason is that the $2q2\ell$ operators we are switching on at $\mu=\Lambda$ do not generate directly any of these processes, therefore we need to consider their RGE effects that induce the relevant WCs: $2q2\ell$ operators with $u\bar u$ or $d\bar d$, $4\ell$ operators and lepton-Higgs operators.
In general, the dominant contributions\footnote{As the LEFT $2q2\ell$ vector Wilson coefficients do not receive 1-loop QCD RG corrections, it is enough to consider the SMEFT running.} come from the gauge RGEs~\cite{Alonso:2013hga}, whose coefficients depend on the quantum numbers of all the involved particles. 
This means that the RGE-induced WCs will be different for each observable in each panel, so in general we can expect that the flat directions, if any, will be different for every observable. 
Indeed, by doing this exercise and solving the gauge RGEs in the leading log approximation, it is straightforward to reproduce almost every flat direction in Figures~\ref{plots2Dcc} and \ref{plots2Dbb}.

The only exception is when the third generation of the quark doublet is involved, as in the lower panels of Figure~\ref{plots2Dbb}.
Even if we were interested just in bottom quarks, the same $SU(2)_L$-invariant operator involves the top quark, whose large Yukawa coupling dominates the RGEs over the gauge contributions. 
In particular, this Yukawa term induces a large lepton-Higgs operator~\cite{Jenkins:2013wua}, which in the leading log approach is given by
\begin{align}
    C_{\varphi e,\tau e}(\mu) \approx  \frac{6Y_t^2}{16\pi^2}\, \log\left(\frac\mu\Lambda\right)\, C_{eq,\tau e33}(\Lambda)\,.
\end{align}
Due to this large contribution every observable, but the LFVQD, in Figure~\ref{plots2Dbb} is completely dominated by the lepton-Higgs operator, either by the one we switched on directly at $\mu=\Lambda$ (if $C_{\varphi e}\simeq C_{eq}$) or by the RGE-induced one (if $C_{\varphi e}\ll C_{eq}$).
In between, these two contributions compete and can actually cancel each other.
In other words, along the common flat direction in the right panel of Figure~\ref{plots2Dbb} both effect conspire in order to have $C_{\varphi e,\tau e}(\mu= m_Z)\simeq0$, suppressing all the constraining LFV processes at the same time.

This last result is just an example pointing towards the LFVQD as the only observable to explore this kind of flat directions along which all the other observables vanish.
Notice however that these cancellations do not necessarily exactly hold at higher orders, which we did not include in our analysis. 
Nevertheless, even if these higher order terms would spoil this perfect cancellation, we still expect a strong suppression in all these LFV processes, leading the LFVQD as our best hope to explore these directions of the parameter space.\footnote{At the level of precision of our calculations, analogous flat directions can be observed in the $\mu e$ sector. However, given the strong constraints set by muon processes, they require fine-tuned cancellations that will likely be destabilised by higher-order corrections. For these reasons, we refrain from a detailed discussion of this possibility.}

%%%%%%%%%%%%%%%%%%%%%%%%%%%%%%%%%%%%%%%%%%%%%%%
\section{Summary and conclusions}
\label{sec:Con}
%%%%%%%%%%%%%%%%%%%%%%%%%%%%%%%%%%%%%%%%%%%%%%%

In this paper, we have addressed the prospects of testing new physics in LFV decays of $c\bar c$ and $b\bar b$ bound states and, more in general, studied the low-energy phenomenology of LFV $2q2\ell$ operators with two charm or bottom fields. Within an EFT framework, we could identify in a model-independent way the muon and tau LFV processes that, through radiative effects as illustrated in \fref{LFVdiag}, indirectly limit the rates of processes such as $J/\psi \to \ell\ell^\prime$, $J/\psi \to \ell\ell^\prime \gamma$, $\Upsilon(nS) \to \ell\ell^\prime$ etc.,~which can be sought at BESIII, Belle~II, and the proposed super tau-charm factory (STCF). Our analysis goes beyond previous work by considering both LEFT and SMEFT and in the number of considered processes.
The main results of our work can be summarised as follows.
\begin{itemize}
    \item In \sref{sec:obs}, we recomputed the rates of vector quarkonia LFV decays (with or without the emission of a photon) as well as those of (pseudo)scalar quarkonium states. We found good agreement with analogous calculations published in Ref.~\cite{Hazard:2016fnc} and 
    Ref.~\cite{Li:2021phq} with the appropriate adjustments for neutrinos in the final state, apart from minor differences with the results for vector quarkonium decays in Ref.~\cite{Hazard:2016fnc}, see \sref{sec:Vllp} and \aref{app:Vllgamma} for details. 
    \item Indirect limits, obtained within the LEFT, for a comprehensive list of LFV decays of heavy quarkonia are shown in Tables~\ref{Table:LEFTcc} and~\ref{Table:LEFTbb}. 
    \item For flavour violation in the $\mu e$ sector, $\mu \to e$ conversion in nuclei and $\mu\to e\gamma$ set such strong constraints on the relevant operators that the rates of the processes we considered are bound to be way below the most optimistic future expected sensitivities, e.g.~\mbox{$\textrm{BR}(J/\psi \to e\mu)\lesssim 10^{-15}$}, \mbox{$\textrm{BR}(\Upsilon(nS) \to e\mu)\lesssim 10^{-12}$}. Observing processes of this kind would then be a striking signal of some new physics not captured by our EFT framework. 
    \item In the case of flavour violation in the  $\tau\ell$ sector, the maximum allowed rates for $c\bar c\to \ell\tau$ decays are at the level of $10^{-9}$-$10^{-10}$, about one-two orders or magnitude below the latest BESIII bounds, and may be  within the sensitivity of the STCF. The maximal rates for $b\bar b\to\ell\tau$ decays can be larger, of the order of $10^{-6}$-$10^{-7}$, that is, at the level of the current best limits set by B-factory experiments, cf.~\tref{tab:quarkonia}. Hence Belle~II has the potential to test new physics by searching for LFV bottomonium decays.
    \item Our LEFT analysis did not consider possibly relevant effects of the running of the operators above the EW scale, nor possible cancellations among different operators. Both these effects are analysed within the SMEFT framework in \sref{sec:SMEFT}.
    \item As shown in \fref{barplot_taue}, the SMEFT running tends to increase the relative importance of the constraints from LFV tau decays compared to \mbox{$J/\psi \to \ell\tau$} and \mbox{$\Upsilon(nS) \to \ell\tau$}. As a consequence, for most $2q2\ell$ operators, an improvement of three orders of magnitude on the experimental sensitivity to the latter processes would barely suffice to test new physics scales at the level of LFV tau decays. 
    \item This effect is particularly pronounced in the case of operators contributing to \mbox{$\Upsilon(nS) \to \ell\tau$} that involve top quarks. Interestingly, such operators could be better tested not only through tau decays by Belle~II, but also by searches for LFV $Z$ decays, $Z\to \ell\tau$, at a Tera-Z run of a future $e^+e^-$ collider up to scales $\sim 10$~TeV.  
    \item On the other hand, $\Upsilon(nS)\to \ell \tau$ (and LFV tau decays) are more sensitive than $Z \to \ell \tau$ to operators that do not involve top quarks. This provides a nice  example of the complementarity between low-energy and high-energy searches for LFV phenomena.
   \item If the new physics effects are not dominantly captured by a single SMEFT operator, cancellations (accidental or perhaps induced by symmetries of the UV-complete theory) among different contributions to the LFV decay rates are possible. In particular, we showed in \fref{plots2Dbb} an example of a \emph{flat direction}, along which all tau decays are suppressed and thus \mbox{$\Upsilon(nS)\to \ell\tau$} can saturate the present experimental bounds. A qualitatively similar picture is obtained when considering operators involving left-handed instead of right-handed lepton currents. 
    \item As a by-product of our analysis, we revisited the prospects for $\mu e$ flavour violation induced by $2q2\ell$ operators with heavy quarks, see \fref{barplot_mue}. This can be observable by Mu3e and Mu2e/COMET up to new-physics scales $\sim 1000$~TeV, while no other LFV process (in particular $\mu\to e\gamma$) should be observed if these operators are the main source of flavour violation. Hence, if both $\mu \to eee$ and $\mu \to e$ conversion in nuclei are detected and the orders of magnitude of their rates are comparable, that would be an indication of this kind of operators as the origin of lepton flavour violation. In contrast, new physics dominantly inducing 4-lepton operators would give  $\textrm{BR}(\mu \to eee)\gg \textrm{CR}(\mu\,N \to e\,N)$, while it would be the other way round for $2q2\ell$ operators involving light quarks. This nice interplay of different processes highlights once more the model-discriminating power of the upcoming campaign of searches for LFV muon decays.
\end{itemize}

%%%%%%%%%%%%%%%%%%%%%%%%%%%%%%%%%%%%%%%%%%%%%%%
\section*{Acknowledgements}
We would like to thank Claudia Garc\'ia-Garc\'ia for valuable discussions and tips for designing the figures, and Martin Hoferichter for useful feedback on a previous version of this paper.
LC and TL are partially supported by the National Natural Science Foundation of China (NSFC) under the Grant No.~12035008.
TL is also supported by the NSFC (Grant No.~11975129) and ``the Fundamental Research Funds for the Central Universities'', Nankai University (Grant No.~63196013). MS acknowledges support by the Australian Research Council through the ARC Discovery Project DP200101470. 
XM acknowledges partial support from the European Union's Horizon 2020 research and innovation programme under the Marie Sk{\l}odowska-Curie grant agreement No~860881-HIDDeN and from the Spanish Research Agency (Agencia Estatal de Investigaci\'on) through the Grant IFT Centro de Excelencia Severo Ochoa No CEX2020-001007-S and Grant PID2019-108892RB-I00 funded by MCIN/AEI/10.13039/501100011033.
This research includes computations using the computational cluster Katana~\cite{Katana} supported by Research Technology Services at UNSW Sydney.

%%%%%%%%%%%%%%%%%%%%%%%%%%%%%%%%%%%%%%%%%%%%%%
\appendix

\section{EW-scale LEFT-SMEFT matching}
\label{app:matching}
As discussed in Ref.~\cite{Jenkins:2017jig}, at the EW scale, the SMEFT operators shown in \tref{Tab:dim6} match to the dipole operator in \eref{eq:LEFTdip} as
\begin{align}
\hat C_{e\gamma,pr} & = \frac{v}{\sqrt{2}\Lambda^2}\left({C_{eB,pr}} \cos\theta_W - {C_{eW,pr}} \sin\theta_W  \right)\,.
\end{align}
Here and in the following, $\hat C$ denote the WCs in the flavour basis. The corresponding WCs $C$ in the physical mass basis are obtained by applying the unitary transformation $\hat f_X = X_f f_X$, where hatted fields are in the flavour basis, and unhatted fields in the physical mass basis, $f=u,d,e,\nu$ and $X=L,R$. Throughout this work, we work in the basis where $R_{u,d,e,\nu}=L_{u,e,\nu}\equiv \mathbbm{1}$, the identity matrix, and $L_d=V_{\rm CKM}$, the CKM matrix.

The LEFT-SMEFT matching for the $2q2\ell$ operators in \eref{eq:LEFT2q2l} reads
\begin{align}
	\hat C_{eu,prst}^{V,LL} &= \frac{1}{\Lambda^2}\left(C_{\ell q,prst}^{(1)} -C_{\ell q,prst}^{(3)}\right) - \frac{g_Z^2}{m_Z^2} [Z_{eL}]_{pr} [Z_{uL}]_{st}\,,\\
	\hat C_{ed,prst}^{V,LL} &= \frac{1}{\Lambda^2}\left(C_{\ell q,prst}^{(1)} +C_{\ell q,prst}^{(3)}\right) - \frac{g_Z^2}{m_Z^2} [Z_{eL}]_{pr} [Z_{dL}]_{st}\,,\\
	\hat C_{eu,prst}^{V,RR} &= \frac{C_{eu,prst}}{\Lambda^2}   - \frac{g_Z^2}{m_Z^2} [Z_{eR}]_{pr} [Z_{uR}]_{st}\,,\\
	\hat C_{ed,prst}^{V,RR} &= \frac{C_{ed,prst}}{\Lambda^2}   - \frac{g_Z^2}{m_Z^2} [Z_{eR}]_{pr} [Z_{dR}]_{st}\,,\\
	\hat C_{eu,prst}^{V,LR} &= \frac{C_{\ell u,prst}}{\Lambda^2}  - \frac{g_Z^2}{m_Z^2} [Z_{eL}]_{pr} [Z_{uR}]_{st}\,,\\
	\hat C_{ed,prst}^{V,LR} &= \frac{C_{\ell d,prst}}{\Lambda^2} - \frac{g_Z^2}{m_Z^2} [Z_{eL}]_{pr} [Z_{dR}]_{st}\,,\\
	\hat C_{ue,prst}^{V,LR} &= \frac{C_{qe,prst}}{\Lambda^2}  - \frac{g_Z^2}{m_Z^2} [Z_{eR}]_{st} [Z_{uL}]_{pr}\,,\\
	\hat C_{de,prst}^{V,LR} &= \frac{C_{qe,prst}}{\Lambda^2}  - \frac{g_Z^2}{m_Z^2} [Z_{eR}]_{st} [Z_{dL}]_{pr}\,,\\
	\hat C_{eu,prst}^{S,RL} &= 0\,, \label{eq:CeuSRL}\\
	\hat C_{ed,prst}^{S,RL} &= \frac{C_{\ell edq,stpr}}{\Lambda^2}  \,,\\
	\hat C_{eu,prst}^{S,RR} &= -\frac{C_{\ell equ,prst}^{(1)}}{\Lambda^2} \,,\\
	\hat C_{eu,prst}^{T,RR} &= -\frac{C_{\ell equ,prst}^{(3)}}{\Lambda^2}  \,,\\
	\hat C_{ed,prst}^{S,RR} &= 0\,,\label{eq:CedSRR} \\
	\hat C_{ed,prst}^{T,RR} &= 0\,, 
\end{align}
and that of the $4$-lepton operators in \eref{eq:LEFT4l} is
\begin{align}
	\hat C_{ee,prst}^{V,LL} &= \frac{C_{\ell\ell,prst}}{\Lambda^2}  - \frac{g_Z^2}{4m_Z^2} [Z_{eL}]_{pr} [Z_{eL}]_{st} - \frac{g_Z^2}{4m_Z^2} [Z_{eL}]_{pt} [Z_{eL}]_{sr}\,,\\
		\hat C_{ee,prst}^{V,RR} &= \frac{C_{ee,prst}}{\Lambda^2}  - \frac{g_Z^2}{4m_Z^2} [Z_{eR}]_{pr} [Z_{eR}]_{st} - \frac{g_Z^2}{4m_Z^2} [Z_{eR}]_{pt} [Z_{eR}]_{sr}\,,\\
		\hat C_{ee,prst}^{V,LR} &= \frac{C_{\ell e,prst}}{\Lambda^2}  - \frac{g_Z^2}{m_Z^2} [Z_{eL}]_{pr} [Z_{eR}]_{st} \,.
\end{align}
In the above expressions, the effective interactions of the $Z$ boson read
\begin{align}
&	[Z_{eL}]_{pr}  = \left[ \delta_{pr} \left(-\frac12 +  \sin^2\theta_W \right) - \frac12 \frac{v^2}{\Lambda^2}  \left(C_{\varphi \ell,pr}^{(1)} + C_{\varphi \ell,pr}^{(3)}\right)\right]\,, \label{eq:ZeL}\\
&	[Z_{eR}]_{pr}  = \left[ \delta_{pr} \left( \sin^2\theta_W\right) - \frac12 \frac{v^2}{\Lambda^2}  C_{\varphi e,pr} \right]\,, \label{eq:ZeR}\\
&	[Z_{uL}]_{pr} = \left[ \delta_{pr} \left(\frac12 -\frac23  \sin^2\theta_W\right) - \frac12 \frac{ v^2 }{\Lambda^2} \left(C_{\varphi q,pr}^{(1)} - C_{\varphi q,pr}^{(3)}\right)\right]\,, \\
&	[Z_{uR}]_{pr}  = \left[ \delta_{pr} \left( -\frac23  \sin^2\theta_W\right) - \frac12 \frac{v^2}{\Lambda^2}  C_{\varphi u,pr} \right]\,, \\
&	[Z_{dL}]_{pr}  = \left[ \delta_{pr} \left(-\frac12 + \frac13  \sin^2\theta_W\right) - \frac12 \frac{v^2}{\Lambda^2} \left( C_{\varphi q,pr}^{(1)} + C_{\varphi q,pr}^{(3)} \right) \right]\,, \\
&	[Z_{dR}]_{pr}  = \left[ \delta_{pr} \left( \frac13  \sin^2\theta_W\right) - \frac12 \frac{v^2}{\Lambda^2}  C_{\varphi d,pr}\right]\,.
\end{align}
For simplicity, we set all other SMEFT operators to zero, when defining the theory at the high scale, in particular, those involving Higgs fields together with gauge field strengths or Higgs fields only. While other Higgs operators will be generated by RG corrections, in particular LFV lepton-Higgs operators, the latter ones will not~\cite{Jenkins:2013zja,Jenkins:2013wua,Alonso:2013hga}. Hence, under this assumption, the Higgs vacuum expectation value, the $Z$ coupling and the Weinberg angle receive no corrections and they are simply given by the usual SM definitions. In particular, in the above expressions, one has \mbox{$g_Z  = {e}/{(\sin\theta_W\cos\theta_W)}$}. For these reasons, the above formulae are somewhat simplified compared to the full ones presented in Ref.~\cite{Jenkins:2017jig}.

Given our choice of basis, the only non-trivial relations between WC in the flavour and physical mass basis for neutral-current LFV 4-fermion LEFT operators are
\begin{align}
    C^{V,LL}_{ed,prst} &= \hat C^{V,LL}_{ed,pr ab} V_{a s}^* V_{bt}
    \,, &
    C^{V,LR}_{de,prst} &= \hat C^{V,LR}_{de,abst} V_{a p}^* V_{br}
    \,, 
    \nonumber\\ 
    C^{S,RL}_{ed,prst} &= \hat C^{S,RL}_{ed,prsa}  V_{at}
    \,, &
    C^{S,RR}_{ed,prst} &= \hat C^{S,RR}_{ed,prat}  V_{as}^*
    \,, &
    C^{T,RR}_{ed,prst} &= \hat C^{T,RR}_{ed,prat}  V_{as}^* \;,
\end{align}
where $V_{ij}$ are entries of the CKM matrix. 
For all other Wilson coefficients of LFV LEFT operators, the flavour basis agrees with the mass basis by construction.
%%%%%%%%%%%%%%%%%%%%%%

\section{Calculational details for radiative LFV vector quarkonium decays}
\label{app:Vllgamma}
This section provides additional calculational details for the radiative LFV vector quarkonium decay presented in \sref{sec:Vllgamma}. 
The full matrix element for $V(P,\epsilon_V)\to \ell_i^-(p_i)\ell_j^+(p_j)\gamma(q,\epsilon)$ is given by
\begin{align}
	\mathcal{M}& = 
	\frac{Q_q e}{x_\gamma m_V^3}  \,\Big[(P\cdot q)\, (\epsilon_V\cdot \epsilon^*) - (P \cdot \epsilon^*) \,(q\cdot \epsilon_V)\Big]\,\bar u_i \left[(S_R+\tilde S_Rx_\gamma) P_R + (S_L+\tilde S_Lx_\gamma) P_L\right] v_j
	\nonumber\\&
	+\frac{Q_q e}{ x_\gamma m_V^3}  \,i \epsilon_{\alpha\beta\mu\nu} P^\alpha q^\beta \epsilon_V^\mu \epsilon^{*\nu} \, \bar u_i \left[(P_R^\prime+i\tilde P_R^\prime x_\gamma) P_R + (P_L^\prime+i\tilde P_L^\prime x_\gamma) P_L\right] v_j 
	\nonumber\\&
	+\frac{Q_q e}{x_\gamma m_V^2} i \epsilon_{\alpha\beta\mu\nu} q^\beta \epsilon_J^\mu \epsilon^{*\nu}  \bar u_i \gamma^\alpha (A_R P_R + A_L P_L ) v_j
	\nonumber\\\nonumber & 
	+\frac{e Q_\ell}{2} V_L \bar u_i \left[ \frac{2 p_i\cdot\epsilon^* + \cancel{\epsilon}^*\cancel{q}}{2p_i\cdot q}\cancel{\epsilon}_VP_L- \cancel{\epsilon}_V P_L \frac{2p_j\cdot \epsilon^* + \cancel{q} \cancel{\epsilon}^*}{2p_j\cdot q}\right]v_j
	\\\nonumber & 
	+\frac{e Q_\ell}{2} V_R \bar u_i \left[ \frac{2p_i\cdot\epsilon^* + \cancel{\epsilon}^*\cancel{q}}{2p_i\cdot q}\cancel{\epsilon}_V P_R - \cancel{\epsilon}_V P_R \frac{2p_j\cdot \epsilon^* + \cancel{q} \cancel{\epsilon}^*}{2p_j\cdot q}\right]v_j
	\\\nonumber & 
	+\frac{2 i e Q_\ell}{m_V} T_L \bar u_i \left[ \frac{2p_i\cdot\epsilon^* + \cancel{\epsilon}^*\cancel{q}}{2p_i\cdot q} \epsilon_V^\mu \sigma_{\mu\nu} p^\nu P_L - \epsilon_V^\mu \sigma_{\mu\nu}p^\nu P_L \frac{2p_j\cdot \epsilon^* + \cancel{q} \cancel{\epsilon}^*}{2p_j\cdot q}\right]v_j
	\\ & 
	+\frac{2 i e Q_\ell}{m_V} T_R \bar u_i \left[ \frac{2p_i\cdot\epsilon^* + \cancel{\epsilon}^*\cancel{q}}{2p_i\cdot q} \epsilon_V^\mu \sigma_{\mu\nu} p^\nu P_R - \epsilon_V^\mu \sigma_{\mu\nu}p^\nu P_R \frac{2p_j\cdot \epsilon^* + \cancel{q} \cancel{\epsilon}^*}{2p_j\cdot q}\right]v_j\,,
\end{align}
where the first three lines originate from initial state radiation and the last four from final state radiation. The coefficients are defined in Eqs.~\eqref{eq:VTL}, \eqref{eq:VTR} and \eqref{eq:SPA}.
The relevant scalar products are 
\begin{align}
	P\cdot q &= m_V^2 x_\gamma/2\,, &
	p^2 & = m_V^2\,, & 
	q^2 & = 0\,, & 
	p_i^2 & = y_i^2 m_V^2\,, &
	P\cdot p_{i} & = \frac{m_V^2}{2} x_{i}\,,  & 
	P\cdot q & = \frac{m_V^2}{2} x_\gamma \,, 
\end{align}
and the products of final state momenta are
\begin{align}
	p_i\cdot p_j & = \frac{m_V^2}{2} (1-y_i^2-y_j^2 -x_\gamma) \,, & 
	p_i\cdot q & = \frac{m_V^2}{2}  (1+y_j^2 -y_i^2 -x_j) \,, & 
\end{align}
where the variables $x$ and $y$ are defined as $x_{i}=2 E_{i}/m_V$ and  $y_i=m_i/m_V$, and the $x_i$ satisfy the relation $x_i+x_j+x_\gamma=2$. Similar expressions apply for $i\leftrightarrow j$.
Using \texttt{FeynCalc}~\cite{Mertig:1990an,Shtabovenko:2016sxi,Shtabovenko:2020gxv}, we calculate the summed and averaged squared matrix element
\begin{align}
\overline{|\mathcal{M}|^2}  & \equiv \frac13 \sum_{\rm spin,pol} |\mathcal{M}|^2\,.
\end{align}
The differential decay rate is then given by 
\begin{align}
	\frac{d\Gamma}{dx_i dx_\gamma} & = \frac{m_V}{256\pi^3}  \overline{|\mathcal{M}|^2}\,,
\end{align}
with 
\begin{align}
	&2y_a \leq x_a \leq 1+y_a^2 -y_b^2-y_c^2-2y_by_c \,, &
	x_{b}^-\leq  x_b \leq x_b^+\;,
\end{align}
\begin{equation}
\hspace*{-.22cm}
	x_b^\pm= \frac{1}{2(1-x_a + y_a^2) } \left[(2-x_a)(1+y_a^2+y_b^2-y_c^2-x_a) \pm \sqrt{x_a^2-4 y_a^2} \lambda^{1/2}(1+y_a^2-x_a,y_b^2,y_c^2)\right],
\end{equation}
where we choose $a=\gamma$, $b=i$ and $c=j$.
In the following we do not consider operators which are also constrained by the LFV 2-body decay, i.e.~we set $V_{L,R}=T_{L,R}=0$ and thus only the contributions from initial state radiation contribute. Furthermore, we assume a hierarchy among the final state lepton masses. The results slightly differ depending on which of the lepton masses is neglected, because the chirality of the massless lepton determines which operators interfere, {\it e.g.}~if the lepton is massless the relevant chirality is the one of the first lepton field $\bar  L, \bar e$ in the bilinear, while it is the second lepton field $L$, $e$ for massless antileptons.
We find, for the limiting case where either the lepton or antilepton is taken to be massless,
\begin{align}
	\frac{d\Gamma}{dx_\gamma} & = \frac{\alpha Q_q^2 m_V}{192\pi^2}
	\Big[ 
	\left(|A_L|^2 + |A_R|^2\right) g_A(x_\gamma,y) + I^\prime
	 g_{PA}(x_\gamma,y)
	\nonumber\\&
	+\left(
		|S_L+\tilde S_L x_\gamma|^2+|S_R+\tilde S_R x_\gamma|^2
		+ |P_L^\prime + i \tilde P_L^\prime x_\gamma |^2+|P_R^\prime + i\tilde P_R^\prime x_\gamma|^2
	\right) g_S(x_\gamma,y)
	\Big]\,,
\end{align}
where $y=y_i$ ($y=y_j$) denotes the non-zero (anti)lepton mass and the 
interference term is given by
\begin{equation}
    I^\prime  = \begin{cases}
+ \mathrm{Re}\big( A_L (P_L^{\prime}+i\tilde P_L^\prime x_\gamma)^* +  A_R (P_R^\prime+i \tilde P_R^\prime x_\gamma)^* \big) & \text{for }\, y=y_i\neq 0,\,  y_j= 0\,,\\
   - \mathrm{Re}\big( A_L (P_R^{\prime}+i\tilde P_R^\prime x_\gamma)^* + A_R (P_L^\prime+i \tilde P_L^\prime x_\gamma)^*\big) & \text{for }\, y_i=0,\, y = y_j\neq 0\,,\\
   \end{cases}
\end{equation}
The kinematic functions are given by
\begin{align}
	g_A(x,y) & =  \frac{x(1-x-y^2)^2( 2x^2-(6+y^2)x +4+5y^2)}{6(1-x)^3}
	\;,
	\nonumber\\
	g_S(x,y) & = \frac{x (1-x-y^2)^2}{2(1-x)}
	\;,
	\nonumber\\
	g_{PA}(x,y) & = \frac{y x (1-x-y^2)^2}{(1-x)^2}
	\;.
\end{align}
We find agreement with the results in Ref.~\cite{Li:2021phq} taking into account that the latter work studied final states with neutrinos instead of charged leptons. We also find agreement with Ref.~\cite{Hazard:2016fnc} for the scalar contribution up to a prefactor, but our result differs for the axial-vector contribution.

\section{Observables for indirect constraints}

In this appendix we collect the analytical expressions for the computation of the LFV observables we studied in this work and used to set indirect limits on the LFVQD. 
They are given in terms of the LEFT Wilson coefficients, with the exception of the LFV $Z$ decays that are given in the SMEFT. 
All these WCs are to be evaluated at the relevant scale, that is, setting $\mu$ to the mass of the decaying particle.

\label{app:obs}

\subsection{Radiative LFV decays: \texorpdfstring{$\ell_i \to \ell_j \gamma$}{li->lj gamma}}

Neglecting the mass of the final-state lepton,
the branching ratio for $\ell_i\to \ell_j \gamma$ reads~\cite{Kuno:1999jp}:
\begin{align}
{\rm BR}(\ell_i \to \ell_j  \gamma)=&~\frac{m_{\ell_i}^3}{4\pi\,  \Gamma_{\ell_i}}\Big(|C_{e\gamma,ji}|^2+|C_{e\gamma,ij}|^2\Big)\,,
\end{align}
where $\Gamma_{\ell_i}$ is the total width of the decaying lepton.
 
\subsection{3-body LFV decays: \texorpdfstring{$\ell_i \to \ell_j\ell_k\bar\ell_m$}{li->lj lk \bar lm}}

For $j=k=m$, {\it i.e.}~the decays $\mu\to ee\bar e$, $\tau\to ee\bar e$ and $\tau\to\mu\mu\bar \mu$, we have~\cite{Calibbi:2021pyh}
\begin{align}
{\rm BR}(\ell_i\to \ell_j \ell_j \bar \ell_j)  =&~\frac{m_{\ell_i}^5}{3 (16\pi)^3  \Gamma_{\ell_i}} 
\bigg\{16\big|C^{V,LL}_{ee,jijj}\big|^2+ 16 \big|C^{V,RR}_{ee,jijj}\big|^2 + 8 \big|C^{V,LR}_{ee,jijj}\big|^2 + 8\big|C^{V,LR}_{ee,jjji}\big|^2  \nonumber\\&
 +\big|C^{S,RR}_{ee,jijj}\big|^2 + \big|C^{S,RR}_{ee,jjij}\big|^2 
+ \frac{256 e^2}{m_{\ell_i}^2} \left(\log\frac{m_{\ell_i}^2}{m_{\ell_j}^2}-\frac{11}4\right)
\Big(\big|C_{e\gamma,ji}\big|^2 + \big|C_{e\gamma,ij}\big|^2\Big)\nonumber\\
&- \frac{64 e}{m_{\ell_i}}\, \Re\Big[\Big(2 C^{V,LL}_{ee,jijj} + C^{V,LR}_{ee,jijj}\Big) C^*_{e\gamma,ji} + \Big(2C^{V,RR}_{ee,jijj}+C^{V,LR}_{ee,jjji}\Big) C_{e\gamma,ji}\Big] \bigg\}\,.
\end{align}
Similarly, for $j\neq k=m$, that is, the decays $\tau\to e\mu\bar\mu$ and $\tau\to \mu e\bar e$, we find~\cite{Calibbi:2021pyh}
\begin{align}
{\rm BR}(\ell_i\to \ell_j \ell_k \bar \ell_k) =&~ \frac{m_{\ell_i}^5}{3 (16\pi)^3  \Gamma_{\ell_i}} 
\bigg\{8\big|C^{V,LL}_{ee,jikk}\big|^2+ 8 \big|C^{V,RR}_{ee,jikk}\big|^2 + 8 \big|C^{V,LR}_{ee,jikk}\big|^2 + 8\big|C^{V,LR}_{ee,kkij}\big|^2  \nonumber\\& +2\big|C^{S,RR}_{ee,jikk}\big|^2 + 2\big|C^{S,RR}_{ee,kkij}\big|^2 +
 \frac{256 e^2}{m_{\ell_i}^2} \left(\log\frac{m_{\ell_i}^2}{m_{\ell_k}^2}-3\right)
\Big(\big|C_{e\gamma,ji}\big|^2 + \big|C_{e\gamma,ij}\big|^2\Big)  \nonumber\\
&- \frac{64 e}{m_{\ell_i}}\, \Re\Big[\Big(C^{V,LL}_{ee,jikk} + C^{V,LR}_{ee,jikk}\Big) C^*_{e\gamma,ji} + \Big(C^{V,RR}_{ee,jikk}+C^{V,RL}_{ee,kkji}\Big) C_{e\gamma,ij}\Big] \bigg\}\,.
\end{align}
In the above expressions the masses of the lighter leptons have been all neglected. 

Finally, notice that the case $j=k \neq m$ corresponds to processes with $|\Delta L_e| =2$, {$|\Delta L_\mu|=|\Delta L_\tau| =1$} or  $|\Delta L_\mu| =2$, \mbox{$|\Delta L_e|=|\Delta L_\tau| =1$}  ($\tau\to e e \bar\mu$ and $\tau\to \mu\mu\bar e$) that are never relevant to constrain the hadronic LFV decays we are interested in.

\subsection{\texorpdfstring{$\mu\to e$}{mu->e} conversion in nuclei}
\label{app:meconv}

The conversion rate is defined as \mbox{$\Gamma(\mu N\to e N)/\Gamma_\text{capt}(N)$}, where $\Gamma_\text{capt}(N)$ is the capture rate of muons by the nucleus $N$~\cite{Kitano:2002mt}, and the $\mu N\to e N$ transition rate is given by is~\cite{Crivellin:2017rmk,Kitano:2002mt,Cirigliano:2009bz,Crivellin:2014cta}
\begin{align}
\Gamma(\mu N\to e N)=\frac{m_\mu^5}{4}&\Big|\frac{1}{m_\mu}C_{e\gamma,\mu e}^\ast \,D+4\Big(m_p C_{SR}^{(p)}\,S^{(p)}+ C_{VR}^{(p)}\,V^{(p)}+p\to n\Big)\Big|^2 \nonumber \\
+\,\frac{m_\mu^5}{4}&\Big|\frac{1}{m_\mu}C_{e\gamma,e\mu}\, D+4\Big(m_p C_{SL}^{(p)}\,S^{(p)}+ C_{VL}^{(p)}\,V^{(p)}+p\to n\Big)\Big|^2\,,
\end{align}
where 
\begin{align}
C_{VR}^{(N)}=&~\sum_{q=u,d,s}\Big(C_{qe,qq e\mu }^{V,LR}+C_{eq,e\mu qq}^{V,RR}\Big)f_{VN}^{(q)}\,,\\
C_{SR}^{(N)}=&~\sum_{q=u,d,s}\frac{C_{eq,\mu e qq}^{S,RR\ast}+C_{eq,\mu e qq}^{S,RL\ast}}{m_q} f_{SN}^{(q)}+\left[\frac{C_{eGG,\mu e}^\ast}{\alpha_s}-\frac{1}{12\pi}\sum_{q=c,b}\frac{C_{eq,\mu e qq}^{S,RR\ast}+C_{eq,\mu e qq}^{S,RL\ast}}{m_q}\right]f_{GN}\,,\\
C_{VL}^{(N)}=&~\sum_{q=u,d,s}\Big(C_{eq,e\mu qq}^{V,LR}+C_{eq,e\mu qq}^{V,LL}\Big)f_{VN}^{(q)}\,,\\
C_{SL}^{(N)}=&~\sum_{q=u,d,s}\frac{C_{eq,e\mu qq}^{S,RR}+C_{eq,e\mu qq}^{S,RL}}{m_q} f_{SN}^{(q)}+
\left[\frac{C_{eGG,e\mu}}{\alpha_s}-\frac{1}{12\pi}\sum_{q=c,b}\frac{C_{eq,e\mu qq}^{S,RR}+C_{eq,e\mu qq}^{S,RL}}{m_q}\right]f_{GN}\,,
\end{align}
with $N=p,n$. 
The nuclear vector form factors are determined from vector current conservation
\begin{align}
    f_{Vp}^{(u)} & = f_{Vn}^{(d)} =2\;, &
    f_{Vp}^{(d)} & = f_{Vn}^{(u)} =1\;, &
    f_{Vp}^{(s)} & = f_{Vn}^{(s)} =0\;,
\end{align}
and we follow Ref.~\cite{Crivellin:2017rmk} for the nuclear scalar form factors
\begin{align}
    f_{Sp}^{(u)} & = (20.8\pm 1.5)\times 10^{-3}\;,
    &
    f_{Sp}^{(d)} & = (41.1\pm2.8)\times 10^{-3}\;,
    &
    f_{SN}^{(s)} & = (53\pm27)\times 10^{-3}\;,
    \nonumber\\
    f_{Sn}^{(u)} & = (18.9\pm1.4)\times 10^{-3}\;,
    &
    f_{Sn}^{(d)} & = (45.1\pm 2.7)\times 10^{-3}\;.
\end{align}
The scalar form factors for up and down quarks are taken from Ref.~\cite{Hoferichter:2015dsa} and the ones for the strange quark have been obtained on the lattice~\cite{Junnarkar:2013ac}, which can also be determined using effective field theory~\cite{Alarcon:2011zs,Alarcon:2012nr}. See also Ref.~\cite{Cirigliano:2022ekw} for a recent calculation of next-to-leading order contributions.
Finally, the form factor $f_{GN}$ is related to the nuclear scalar form factors of light quarks
\begin{equation}
    f_{GN} = 1-\sum_{q=u,d,s} f_{SN}^{(q)}\;.
\end{equation}
The overlap integrals $D$, $S^{(p)}$, $S^{(n)}$, $V^{(p)}$, and $V^{(n)}$ for the different nuclei have been calculated in Ref.~\cite{Kitano:2002mt}. For a recent reassessment see Ref.~\cite{Heeck:2022wer}.

\subsection{Semileptonic LFV \texorpdfstring{$\tau$}{tau} decays: \texorpdfstring{$\tau \to \cal{P} \ell $, $\tau \to \cal{V} \ell $}{tau->P l, tau -> V l}}

Expressions for the $\tau \to \mathcal{V} \ell$ processes, with $\ell=e,\mu$ and $\mathcal{V}=\rho, \phi$ a vector meson, in terms of the LEFT Wilson coefficients can be found in Ref.~\cite{Aebischer:2018iyb}.
The branching ratio reads
\begin{align}
{\rm BR} (\tau \to \mathcal{V} \ell) = \frac{\sqrt{\lambda(m_\tau^2,m_\ell^2,m_{\mathcal{V}}^2)}}{16\pi\, m_\tau^3\,\Gamma_{\tau}} ~
\overline{\big|\mathcal M_{\tau \to \mathcal{V} \ell}\big|}^{\,2}\,,
\end{align}
where $\lambda(a,b,c)=a^2+b^2+c^2-2(ab+ac+bc)$. The squared amplitude is given by
\begin{align}
 \overline{\big|\mathcal M_{\tau \to \mathcal{V} \ell}\big|}^{\,2} =
 \overline{\big|\mathcal M^{\rm V}_{\tau \to \mathcal{V} \ell}\big|}^{\,2} +  \overline{\big|\mathcal M^{\rm T}_{\tau \to \mathcal{V} \ell}\big|}^{\,2}+ \mathcal I_{\,\tau \to \mathcal{V} \ell}\,,
\end{align}
where the first term comes from couplings of the meson to leptonic vector currents, the second one is due to tensor currents, and the last term is the interference of the two:
\begin{align}
 \overline{\big|\mathcal M^{\rm V}_{\tau\to \mathcal{V}\ell}\big|}^{\,2} &= 
~ \frac12\bigg[\Big(\big|\gVLV\big|^2+\big|\gVRV\big|^2\Big) \bigg(\frac{(m_\tau^2-m_\ell^2)^2}{m_\mathcal{V}^2} + m_\tau^2+m_\ell^2-2m_\mathcal{V}^2\bigg) \nonumber \\
 &-12 m_\tau m_\ell\, \Re\Big(\gVRV \big(\gVLV\big)^*\Big)
 \bigg]\,, \\
  \overline{\big|\mathcal M^{\rm T}_{\tau\to \mathcal{V}\ell}\big|}^{\,2} &= 
 ~\frac12\bigg[\Big(\big| \gTLV - \gTLVd \big|^2+\big| \gTRV + \gTRVd \big|^2\Big) \Big(2\big(m_\tau^2-m_\ell^2\big)^2 - m_\mathcal{V}^2(m_\tau^2+m_\ell^2) - m_\mathcal{V}^4\Big)\nonumber \\
 &  -12 m_\mathcal{V}^2m_\tau m_\ell\, \Re\Big(\big( \gTRV + \gTRVd \big)\big(\gTLV - \gTLVd \big)^*\Big)
 \bigg]\,,  \\
\mathcal I_{\,\tau\to \mathcal{V}\ell} ~&= 
~3 m_\tau (m_\tau^2-m_\ell^2-m_{\mathcal{V}}^2)\, \Re\Big(\gVLV \big(\gTRV +\gTRVd \big)^* + \gVRV\big(\gTLV -\gTLVd  \big)^*\Big) ~\nonumber\\
&+ 3 m_\ell\hspace{.05cm}(m_\ell^2-m_\tau^2-m_{\mathcal{V}}^2)\, \Re\Big(\gVRV \big(\gTRV + \gTRVd \big)^* + \gVLV\big(\gTLV -\gTLVd \big)^*\Big)\,.
\end{align}

In terms of the LEFT Wilson coefficients, the effective couplings appearing in the above expressions read in the case of decays into a $\rho$ meson
\begin{align}
\gVLV[\rho] = &~\frac{1}{2} m_\rho f_\rho \left( \frac{ C^{V,LL}_{eu,\ell\tau uu} -C^{V,LL}_{ed,\ell\tau dd} }{\sqrt{2}}  + \frac{ C^{V,LR}_{eu,\ell\tau uu} -C^{V,LR}_{ed,\ell\tau dd} }{\sqrt{2}} \right)\,, \\
\gVRV[\rho] = & ~\frac{1}{2} m_\rho f_\rho \left( \frac{ C^{V,RR}_{eu,\ell\tau uu} -C^{V,RR}_{ed,\ell\tau dd} }{\sqrt{2}}  + \frac{ C^{V,LR}_{ue, uu\ell\tau} -C^{V,LR}_{de,dd\ell\tau} }{\sqrt{2}} \right)\,,  \\
\gTLV[\rho] = &~f^T_\rho\,  \frac{ C^{T,RR\,*}_{eu,\tau\ell uu} -C^{T,RR\,*}_{ed,\tau\ell dd} }{\sqrt{2}} 	- \sqrt{2} e \,\frac{f_\rho}{m_\rho} C^{*}_{e\gamma,\tau\ell}\,,	\\
\gTRV[\rho] = &~f^T_\rho\,  \frac{ C^{T,RR}_{eu,\ell\tau uu} -C^{T,RR}_{ed,\ell\tau dd} }{\sqrt{2}} 	- \sqrt{2} e \,\frac{f_\rho}{m_\rho} C_{e\gamma,\ell\tau}\,,	 \\
\gTLVd[\rho] = &~-f^T_\rho\,  \frac{ C^{T,RR\,*}_{eu,\tau\ell uu} -C^{T,RR\,*}_{ed,\tau\ell dd} }{\sqrt{2}} \,,	\\
\gTRVd[\rho] = &~f^T_\rho\,  \frac{ C^{T,RR}_{eu,\ell\tau uu} -C^{T,RR}_{ed,\ell\tau dd} }{\sqrt{2}} \,,
\end{align}
where $f_\rho$ it the decay constant of the $\rho$ meson, and $f^T_\rho$ is the transverse decay constant.

The corresponding effective couplings for the case of the $\phi$ are
\begin{align}
\gVLV[\phi] = &~\frac{1}{2} m_\phi f_\phi \left( C^{V,LL}_{ed,\ell\tau ss} + C^{V,LR}_{ed,\ell\tau ss} \right)\,, \\
\gVRV[\phi] = &~\frac{1}{2} m_\phi f_\phi \left( C^{V,RR}_{ed,\ell\tau ss} + C^{V,LR}_{de, ss\ell\tau} \right)\,, \\
\gTLV[\phi] = &~f^T_\phi\,  C^{T,RR\,*}_{ed,\tau\ell ss} +   	\frac{2}{3} e \,\frac{f_\phi}{m_\phi} C^{*}_{e\gamma,\tau\ell}\,,	\\
\gTRV[\phi] =&~f^T_\phi\,  C^{T,RR}_{ed,\ell\tau ss} +   	\frac{2}{3} e \,\frac{f_\phi}{m_\phi} C_{e\gamma,\ell\tau}\,,	 \\
\gTLVd[\phi] = &~-f^T_\phi\,  C^{T,RR\,*}_{ed,\tau\ell ss} \,,	\\
\gTRVd[\phi] = &~f^T_\phi\,  C^{T,RR}_{ed,\ell\tau ss}\,.
\end{align}

As for the case of the decays into a vector, we can find the expressions for the decays into pseudoscalar mesons in Ref.~\cite{Aebischer:2018iyb}. The branching ratio for $\tau\to \mesonP\ell$, with $\ell=e,\mu$ is
\begin{align}
{\rm BR} (\tau \to \mathcal{P} \ell) = \frac{\sqrt{\lambda(m_\tau^2,m_\ell^2,m_{\mathcal{P}}^2)}}{16\pi\, m_\tau^3\,\Gamma_{\tau}} ~
\overline{\big|\mathcal M_{\tau\to \mesonP\ell}\big|}^{\,2}\,,
\end{align}
where
\begin{align}
 \overline{\big|\mathcal M_{\tau\to \mesonP\ell}\big|}^{\,2} =
 \frac12\big(m_\tau^2+m_\ell^2-m_{\mesonP}^2\big)\,\Big(\big|\gLP\big|^2 + \big|\gRP\big|^2\Big)
 +2m_\tau m_\ell \, \Re\Big(\gLP \big(\gRP\big)^*\Big)\,,
\end{align}
with
\begin{align}
\gLP = \gSLP - m_\ell \gVLP + m_\tau \gVRP\,,
\qquad
\gRP = \gSRP - m_\ell \gVRP + m_\tau \gVLP\,.
\end{align}
The effective couplings of $\pi^0$ to leptonic currents are
\begin{align}	
\gSLP[\pi] =&~ \frac{f_\pi m_\pi^2}{\sqrt{2}(m_u + m_d)} \, \left( \frac{ C^{S,RL\,*}_{eu,\tau\ell uu} -C^{S,RR\,*}_{eu,\tau\ell uu}}{2}  - \frac{ C^{S,RL\,*}_{ed,\tau\ell dd} -C^{S,RR\,*}_{ed,\tau\ell dd} }{2} \right)\,, \\
\gSRP[\pi] =&~ \frac{f_\pi m_\pi^2}{\sqrt{2}(m_u + m_d)} \, \left( \frac{ C^{S,RL}_{eu,\ell\tau uu} -C^{S,RR}_{eu,\ell\tau uu}}{2}  - \frac{ C^{S,RL}_{ed,\ell\tau dd} -C^{S,RR}_{ed,\ell\tau dd} }{2} \right)\,, \\
\gVLP[\pi] =&~ \frac{f_\pi}{\sqrt{2}} \, \left( \frac{ C^{V,LR}_{eu,\ell\tau uu} -C^{V,LR}_{ed,\ell\tau dd}}{2}  -  \frac{ C^{V,LL}_{eu,\ell\tau uu} -C^{V,LL}_{ed,\ell\tau dd} }{2} \right)\,, \\
\gVRP[\pi] =&~ \frac{f_\pi}{\sqrt{2}} \, \left( \frac{ C^{V,RR}_{eu,\ell\tau uu} -C^{V,RR}_{ed,\ell\tau dd}}{2}  -  \frac{ C^{V,LR}_{ue,uu\ell\tau} -C^{V,LR}_{de,dd\ell\tau} }{2} \right)\,.
\end{align}

%%%%%%%%%%%%%%%%%%%%%%%%%%%%%%%%%%%%%%%%%%%%%%%%%%%%

\subsection{LFV \texorpdfstring{$Z$}{Z} decays}

The branching ratios of LFV decays of the $Z$ boson are given by~\cite{Brignole:2004ah,Crivellin:2013hpa,Calibbi:2021pyh}
\begin{equation}
\mathrm{BR}\left( {Z \to \ell_i \ell_j } \right) =
\frac{m_Z}{12\pi \Gamma_Z} \left[
 \left| g^{ij}_{VR} \right|^2 + \left| g^{ij}_{VL} \right|^2  
 + \frac{m_Z^2}{2} \left(\left|  g^{ij}_{TR} \right|^2 +
 \left|  g^{ij}_{TL} \right|^2\right)
  \right],
  \label{eq:Zll}
\end{equation}
with
\begin{align}
  & g^{ij}_{VR} = \frac{e}{\sw\cw} \, [Z_{eR}]_{ij} \,,
   \quad
     g^{ij}_{VL} =\frac{e}{\sw\cw} \, [Z_{eL}]_{ij} \,,
   \\
&    g^{ij}_{TR} = \delta g^{ji\,*}_{TL} = -\frac{v}{\sqrt2 \Lambda^2}\,
    \Big(\sw C_{eB, ij} + \cw C_{eW, ij} \Big)\,,
\end{align}
where the LFV $Z$ couplings $Z_{eL},\,Z_{eR}$ are given in terms of WCs of lepton-Higgs operators in Eqs.~(\ref{eq:ZeL},\,\ref{eq:ZeR}).

%%%%%%%%%%%%%%%%%%%%%%%%%%%%%%%%%%%%%%%%

\bibliographystyle{JHEP} 
\bibliography{refs}% Produces the bibliography via BibTeX.

\providecommand{\href}[2]{#2}\begingroup\raggedright\begin{thebibliography}{100}

\bibitem{Calibbi:2017uvl}
L.~Calibbi and G.~Signorelli, \emph{{Charged Lepton Flavour Violation: An
  Experimental and Theoretical Introduction}},
  \href{https://doi.org/10.1393/ncr/i2018-10144-0}{\emph{Riv. Nuovo Cim.}
  {\bfseries 41} (2018) 71--174},
  [\href{https://arxiv.org/abs/1709.00294}{{\ttfamily 1709.00294}}].

\bibitem{Abi:2021gix}
{\scshape Muon g-2} collaboration, B.~Abi et~al., \emph{{Measurement of the
  Positive Muon Anomalous Magnetic Moment to 0.46 ppm}},
  \href{https://doi.org/10.1103/PhysRevLett.126.141801}{\emph{Phys. Rev. Lett.}
  {\bfseries 126} (2021) 141801},
  [\href{https://arxiv.org/abs/2104.03281}{{\ttfamily 2104.03281}}].

\bibitem{Bifani:2018zmi}
S.~Bifani, S.~Descotes-Genon, A.~Romero~Vidal and M.-H. Schune, \emph{{Review
  of Lepton Universality tests in $B$ decays}},
  \href{https://doi.org/10.1088/1361-6471/aaf5de}{\emph{J. Phys. G} {\bfseries
  46} (2019) 023001}, [\href{https://arxiv.org/abs/1809.06229}{{\ttfamily
  1809.06229}}].

\bibitem{London:2021lfn}
D.~London and J.~Matias, \emph{{$B$ Flavour Anomalies: 2021 Theoretical Status
  Report}},
  \href{https://doi.org/10.1146/annurev-nucl-102020-090209}{\emph{Ann. Rev.
  Nucl. Part. Sci.} {\bfseries 72} (2022) 37--68},
  [\href{https://arxiv.org/abs/2110.13270}{{\ttfamily 2110.13270}}].

\bibitem{Allwicher:2021jkr}
L.~Allwicher, L.~Di~Luzio, M.~Fedele, F.~Mescia and M.~Nardecchia, \emph{{What
  is the scale of new physics behind the muon g-2?}},
  \href{https://doi.org/10.1103/PhysRevD.104.055035}{\emph{Phys. Rev. D}
  {\bfseries 104} (2021) 055035},
  [\href{https://arxiv.org/abs/2105.13981}{{\ttfamily 2105.13981}}].

\bibitem{DiLuzio:2017chi}
L.~Di~Luzio and M.~Nardecchia, \emph{{What is the scale of new physics behind
  the $B$-flavour anomalies?}},
  \href{https://doi.org/10.1140/epjc/s10052-017-5118-9}{\emph{Eur. Phys. J. C}
  {\bfseries 77} (2017) 536},
  [\href{https://arxiv.org/abs/1706.01868}{{\ttfamily 1706.01868}}].

\bibitem{Glashow:2014iga}
S.~L. Glashow, D.~Guadagnoli and K.~Lane, \emph{{Lepton Flavor Violation in $B$
  Decays?}}, \href{https://doi.org/10.1103/PhysRevLett.114.091801}{\emph{Phys.
  Rev. Lett.} {\bfseries 114} (2015) 091801},
  [\href{https://arxiv.org/abs/1411.0565}{{\ttfamily 1411.0565}}].

\bibitem{Calibbi:2021qto}
L.~Calibbi, M.~L. L\'opez-Ib\'a\~nez, A.~Melis and O.~Vives,
  \emph{{Implications of the Muon g-2 result on the flavour structure of the
  lepton mass matrix}},
  \href{https://doi.org/10.1140/epjc/s10052-021-09741-1}{\emph{Eur. Phys. J. C}
  {\bfseries 81} (2021) 929},
  [\href{https://arxiv.org/abs/2104.03296}{{\ttfamily 2104.03296}}].

\bibitem{Isidori:2021gqe}
G.~Isidori, J.~Pag\`es and F.~Wilsch, \emph{{Flavour alignment of New Physics
  in light of the (g \ensuremath{-} 2)$_{\mu}$ anomaly}},
  \href{https://doi.org/10.1007/JHEP03(2022)011}{\emph{JHEP} {\bfseries 03}
  (2022) 011}, [\href{https://arxiv.org/abs/2111.13724}{{\ttfamily
  2111.13724}}].

\bibitem{Bigaran:2022kkv}
I.~Bigaran, T.~Felkl, C.~Hagedorn and M.~A. Schmidt, \emph{{Flavour anomalies
  meet flavour symmetry}},  \href{https://arxiv.org/abs/2207.06197}{{\ttfamily
  2207.06197}}.

\bibitem{Dorsner:2016wpm}
I.~Dor\v{s}ner, S.~Fajfer, A.~Greljo, J.~F. Kamenik and N.~Ko\v{s}nik,
  \emph{{Physics of leptoquarks in precision experiments and at particle
  colliders}}, \href{https://doi.org/10.1016/j.physrep.2016.06.001}{\emph{Phys.
  Rept.} {\bfseries 641} (2016) 1--68},
  [\href{https://arxiv.org/abs/1603.04993}{{\ttfamily 1603.04993}}].

\bibitem{Bhattacharya:2014wla}
B.~Bhattacharya, A.~Datta, D.~London and S.~Shivashankara, \emph{{Simultaneous
  Explanation of the $R_K$ and $R(D^{(*)})$ Puzzles}},
  \href{https://doi.org/10.1016/j.physletb.2015.02.011}{\emph{Phys. Lett. B}
  {\bfseries 742} (2015) 370--374},
  [\href{https://arxiv.org/abs/1412.7164}{{\ttfamily 1412.7164}}].

\bibitem{Calibbi:2015kma}
L.~Calibbi, A.~Crivellin and T.~Ota, \emph{{Effective Field Theory Approach to
  $b\to s\ell\ell^{(')}$, $B\to K^{(*)}\nu\overline{\nu}$ and $B\to
  D^{(*)}\tau\nu$ with Third Generation Couplings}},
  \href{https://doi.org/10.1103/PhysRevLett.115.181801}{\emph{Phys. Rev. Lett.}
  {\bfseries 115} (2015) 181801},
  [\href{https://arxiv.org/abs/1506.02661}{{\ttfamily 1506.02661}}].

\bibitem{Feruglio:2016gvd}
F.~Feruglio, P.~Paradisi and A.~Pattori, \emph{{Revisiting Lepton Flavor
  Universality in B Decays}},
  \href{https://doi.org/10.1103/PhysRevLett.118.011801}{\emph{Phys. Rev. Lett.}
  {\bfseries 118} (2017) 011801},
  [\href{https://arxiv.org/abs/1606.00524}{{\ttfamily 1606.00524}}].

\bibitem{Buttazzo:2017ixm}
D.~Buttazzo, A.~Greljo, G.~Isidori and D.~Marzocca, \emph{{B-physics anomalies:
  a guide to combined explanations}},
  \href{https://doi.org/10.1007/JHEP11(2017)044}{\emph{JHEP} {\bfseries 11}
  (2017) 044}, [\href{https://arxiv.org/abs/1706.07808}{{\ttfamily
  1706.07808}}].

\bibitem{BESIIInew}
{\scshape BESIII} collaboration, M.~Ablikim et~al., \emph{{Search for the
  lepton flavor violating decay $J/\psi\to e\mu$}},
  \href{https://arxiv.org/abs/2206.13956}{{\ttfamily 2206.13956}}.

\bibitem{Belle:2022cce}
{\scshape Belle} collaboration, S.~Patra et~al., \emph{{Search for charged
  lepton flavor violating decays of $\Upsilon(1S)$}},
  \href{https://doi.org/10.1007/JHEP05(2022)095}{\emph{JHEP} {\bfseries 05}
  (2022) 095}, [\href{https://arxiv.org/abs/2201.09620}{{\ttfamily
  2201.09620}}].

\bibitem{BESIII:2021slj}
{\scshape BESIII} collaboration, M.~Ablikim et~al., \emph{{Search for the
  charged lepton flavor violating decay $J/\psi\to e\tau$}},
  \href{https://doi.org/10.1103/PhysRevD.103.112007}{\emph{Phys. Rev. D}
  {\bfseries 103} (2021) 112007},
  [\href{https://arxiv.org/abs/2103.11540}{{\ttfamily 2103.11540}}].

\bibitem{BaBar:2010vxb}
{\scshape BaBar} collaboration, J.~P. Lees et~al., \emph{{Search for Charged
  Lepton Flavor Violation in Narrow Upsilon Decays}},
  \href{https://doi.org/10.1103/PhysRevLett.104.151802}{\emph{Phys. Rev. Lett.}
  {\bfseries 104} (2010) 151802},
  [\href{https://arxiv.org/abs/1001.1883}{{\ttfamily 1001.1883}}].

\bibitem{BES:2004jiw}
{\scshape BES} collaboration, M.~Ablikim et~al., \emph{{Search for the lepton
  flavor violation processes $J/\psi \to \mu \tau$ and $e \tau$}},
  \href{https://doi.org/10.1016/j.physletb.2004.08.005}{\emph{Phys. Lett. B}
  {\bfseries 598} (2004) 172--177},
  [\href{https://arxiv.org/abs/hep-ex/0406018}{{\ttfamily hep-ex/0406018}}].

\bibitem{BESIII:2020nme}
{\scshape BESIII} collaboration, M.~Ablikim et~al., \emph{{Future Physics
  Programme of BESIII}},
  \href{https://doi.org/10.1088/1674-1137/44/4/040001}{\emph{Chin. Phys. C}
  {\bfseries 44} (2020) 040001},
  [\href{https://arxiv.org/abs/1912.05983}{{\ttfamily 1912.05983}}].

\bibitem{Barnyakov:2020vob}
{\scshape Super Charm-Tau Factory} collaboration, A.~Y. Barnyakov, \emph{{The
  project of the Super Charm-Tau Factory in Novosibirsk}},
  \href{https://doi.org/10.1088/1742-6596/1561/1/012004}{\emph{J. Phys. Conf.
  Ser.} {\bfseries 1561} (2020) 012004}.

\bibitem{Zhou:2021rgi}
{\scshape STCF Working Group} collaboration, X.~Zhou, \emph{{Experimental
  Program at Super Tau-Charm Facility}},
  \href{https://doi.org/10.22323/1.385.0007}{\emph{PoS} {\bfseries CHARM2020}
  (2021) 007}.

\bibitem{Lyu:2021tlb}
{\scshape STCF Working Group} collaboration, X.-R. Lyu, \emph{{Physics Program
  of the Super Tau-Charm Factory}},
  \href{https://doi.org/10.22323/1.391.0060}{\emph{PoS} {\bfseries BEAUTY2020}
  (2021) 060}.

\bibitem{Kou:2018nap}
{\scshape Belle-II} collaboration, W.~Altmannshofer et~al., \emph{{The Belle II
  Physics Book}}, \href{https://doi.org/10.1093/ptep/ptz106}{\emph{PTEP}
  {\bfseries 2019} (2019) 123C01},
  [\href{https://arxiv.org/abs/1808.10567}{{\ttfamily 1808.10567}}].

\bibitem{Nussinov:2000nm}
S.~Nussinov, R.~D. Peccei and X.~M. Zhang, \emph{{On unitarity based relations
  between various lepton family violating processes}},
  \href{https://doi.org/10.1103/PhysRevD.63.016003}{\emph{Phys. Rev. D}
  {\bfseries 63} (2001) 016003},
  [\href{https://arxiv.org/abs/hep-ph/0004153}{{\ttfamily hep-ph/0004153}}].

\bibitem{Calibbi:2021pyh}
L.~Calibbi, X.~Marcano and J.~Roy, \emph{{Z lepton flavour violation as a probe
  for new physics at future $e^+e^-$ colliders}},
  \href{https://doi.org/10.1140/epjc/s10052-021-09777-3}{\emph{Eur. Phys. J. C}
  {\bfseries 81} (2021) 1054},
  [\href{https://arxiv.org/abs/2107.10273}{{\ttfamily 2107.10273}}].

\bibitem{TheMEG:2016wtm}
{\scshape MEG} collaboration, A.~M. Baldini et~al., \emph{{Search for the
  lepton flavour violating decay $\mu ^+ \rightarrow \mathrm {e}^+ \gamma $
  with the full dataset of the MEG experiment}},
  \href{https://doi.org/10.1140/epjc/s10052-016-4271-x}{\emph{Eur. Phys. J. C}
  {\bfseries 76} (2016) 434},
  [\href{https://arxiv.org/abs/1605.05081}{{\ttfamily 1605.05081}}].

\bibitem{Baldini:2018nnn}
{\scshape MEG II} collaboration, A.~Baldini et~al., \emph{{The design of the
  MEG II experiment}},
  \href{https://doi.org/10.1140/epjc/s10052-018-5845-6}{\emph{Eur. Phys. J. C}
  {\bfseries 78} (2018) 380},
  [\href{https://arxiv.org/abs/1801.04688}{{\ttfamily 1801.04688}}].

\bibitem{Bellgardt:1987du}
{\scshape SINDRUM} collaboration, U.~Bellgardt et~al., \emph{{Search for the
  Decay $\mu^+\to e^+e^+e^-$}},
  \href{https://doi.org/10.1016/0550-3213(88)90462-2}{\emph{Nucl. Phys. B}
  {\bfseries 299} (1988) 1--6}.

\bibitem{Blondel:2013ia}
A.~Blondel et~al., \emph{{Research Proposal for an Experiment to Search for the
  Decay $\mu \to eee$}},  \href{https://arxiv.org/abs/1301.6113}{{\ttfamily
  1301.6113}}.

\bibitem{Bertl:2006up}
{\scshape SINDRUM II} collaboration, W.~H. Bertl et~al., \emph{{A Search for
  muon to electron conversion in muonic gold}},
  \href{https://doi.org/10.1140/epjc/s2006-02582-x}{\emph{Eur. Phys. J. C}
  {\bfseries 47} (2006) 337--346}.

\bibitem{Kuno:2013mha}
{\scshape COMET} collaboration, Y.~Kuno, \emph{{A search for muon-to-electron
  conversion at J-PARC: The COMET experiment}},
  \href{https://doi.org/10.1093/ptep/pts089}{\emph{PTEP} {\bfseries 2013}
  (2013) 022C01}.

\bibitem{Bartoszek:2014mya}
{\scshape Mu2e} collaboration, L.~Bartoszek et~al., \emph{{Mu2e Technical
  Design Report}},  \href{https://arxiv.org/abs/1501.05241}{{\ttfamily
  1501.05241}}.

\bibitem{ATLAS:2022uhq}
{\scshape ATLAS} collaboration, {ATLAS Collaboration}, \emph{{Search for the
  charged-lepton-flavor-violating decay $Z\rightarrow e\mu$ in $pp$ collisions
  at $\sqrt{s}=13$ TeV with the ATLAS detector}},
  \href{https://arxiv.org/abs/2204.10783}{{\ttfamily 2204.10783}}.

\bibitem{Dam:2018rfz}
M.~Dam, \emph{{Tau-lepton Physics at the FCC-ee circular e$^+$e$^-$ Collider}},
  \href{https://doi.org/10.21468/SciPostPhysProc.1.041}{\emph{SciPost Phys.
  Proc.} {\bfseries 1} (2019) 041},
  [\href{https://arxiv.org/abs/1811.09408}{{\ttfamily 1811.09408}}].

\bibitem{Aubert:2009ag}
{\scshape BaBar} collaboration, B.~Aubert et~al., \emph{{Searches for Lepton
  Flavor Violation in the Decays $\tau^\pm \to e^\pm\gamma$ and $\tau^\pm \to
  \mu^\pm\gamma$}},
  \href{https://doi.org/10.1103/PhysRevLett.104.021802}{\emph{Phys. Rev. Lett.}
  {\bfseries 104} (2010) 021802},
  [\href{https://arxiv.org/abs/0908.2381}{{\ttfamily 0908.2381}}].

\bibitem{Banerjee:2022xuw}
S.~Banerjee et~al., \emph{{Snowmass 2021 White Paper: Charged lepton flavor
  violation in the tau sector}},
  \href{https://arxiv.org/abs/2203.14919}{{\ttfamily 2203.14919}}.

\bibitem{Hayasaka:2010np}
K.~Hayasaka et~al., \emph{{Search for Lepton Flavor Violating Tau Decays into
  Three Leptons with 719 Million Produced Tau+Tau- Pairs}},
  \href{https://doi.org/10.1016/j.physletb.2010.03.037}{\emph{Phys. Lett. B}
  {\bfseries 687} (2010) 139--143},
  [\href{https://arxiv.org/abs/1001.3221}{{\ttfamily 1001.3221}}].

\bibitem{Miyazaki:2007jp}
{\scshape Belle} collaboration, Y.~Miyazaki et~al., \emph{{Search for lepton
  flavor violating $\tau^-$ decays into $\ell^- \eta$, $\ell^- \eta'$ and
  $\ell^- \pi^0$}},
  \href{https://doi.org/10.1016/j.physletb.2007.03.027}{\emph{Phys. Lett. B}
  {\bfseries 648} (2007) 341--350},
  [\href{https://arxiv.org/abs/hep-ex/0703009}{{\ttfamily hep-ex/0703009}}].

\bibitem{Miyazaki:2011xe}
{\scshape Belle} collaboration, Y.~Miyazaki et~al., \emph{{Search for
  Lepton-Flavor-Violating tau Decays into a Lepton and a Vector Meson}},
  \href{https://doi.org/10.1016/j.physletb.2011.04.011}{\emph{Phys. Lett. B}
  {\bfseries 699} (2011) 251--257},
  [\href{https://arxiv.org/abs/1101.0755}{{\ttfamily 1101.0755}}].

\bibitem{newATLAS}
{\scshape ATLAS} collaboration, G.~Aad et~al., \emph{{Search for
  lepton-flavor-violation in $Z$-boson decays with $\tau$-leptons with the
  ATLAS detector}},
  \href{https://doi.org/10.1103/PhysRevLett.127.271801}{\emph{Phys. Rev. Lett.}
  {\bfseries 127} (2022) 271801},
  [\href{https://arxiv.org/abs/2105.12491}{{\ttfamily 2105.12491}}].

\bibitem{Belle:2021ysv}
{\scshape Belle} collaboration, A.~Abdesselam et~al., \emph{{Search for
  lepton-flavor-violating tau decays to $\ell\gamma$ modes at Belle}},
  \href{https://doi.org/10.1007/JHEP10(2021)019}{\emph{JHEP} {\bfseries 10}
  (2021) 019}, [\href{https://arxiv.org/abs/2103.12994}{{\ttfamily
  2103.12994}}].

\bibitem{Aubert:2006cz}
{\scshape BaBar} collaboration, B.~Aubert et~al., \emph{{Search for Lepton
  Flavor Violating Decays $\tau^\pm \to \ell^\pm \pi^0$, $\ell^\pm \eta$,
  $\ell^\pm \eta^\prime$}},
  \href{https://doi.org/10.1103/PhysRevLett.98.061803}{\emph{Phys. Rev. Lett.}
  {\bfseries 98} (2007) 061803},
  [\href{https://arxiv.org/abs/hep-ex/0610067}{{\ttfamily hep-ex/0610067}}].

\bibitem{Jenkins:2017jig}
E.~E. Jenkins, A.~V. Manohar and P.~Stoffer, \emph{{Low-Energy Effective Field
  Theory below the Electroweak Scale: Operators and Matching}},
  \href{https://doi.org/10.1007/JHEP03(2018)016}{\emph{JHEP} {\bfseries 03}
  (2018) 016}, [\href{https://arxiv.org/abs/1709.04486}{{\ttfamily
  1709.04486}}].

\bibitem{Buchmuller:1985jz}
W.~Buchmuller and D.~Wyler, \emph{{Effective Lagrangian Analysis of New
  Interactions and Flavor Conservation}},
  \href{https://doi.org/10.1016/0550-3213(86)90262-2}{\emph{Nucl. Phys. B}
  {\bfseries 268} (1986) 621--653}.

\bibitem{Grzadkowski:2010es}
B.~Grzadkowski, M.~Iskrzynski, M.~Misiak and J.~Rosiek, \emph{{Dimension-Six
  Terms in the Standard Model Lagrangian}},
  \href{https://doi.org/10.1007/JHEP10(2010)085}{\emph{JHEP} {\bfseries 10}
  (2010) 085}, [\href{https://arxiv.org/abs/1008.4884}{{\ttfamily 1008.4884}}].

\bibitem{Brivio:2017vri}
I.~Brivio and M.~Trott, \emph{{The Standard Model as an Effective Field
  Theory}}, \href{https://doi.org/10.1016/j.physrep.2018.11.002}{\emph{Phys.
  Rept.} {\bfseries 793} (2019) 1--98},
  [\href{https://arxiv.org/abs/1706.08945}{{\ttfamily 1706.08945}}].

\bibitem{Crivellin:2017rmk}
A.~Crivellin, S.~Davidson, G.~M. Pruna and A.~Signer,
  \emph{{Renormalisation-group improved analysis of $\mu\to e$ processes in a
  systematic effective-field-theory approach}},
  \href{https://doi.org/10.1007/JHEP05(2017)117}{\emph{JHEP} {\bfseries 05}
  (2017) 117}, [\href{https://arxiv.org/abs/1702.03020}{{\ttfamily
  1702.03020}}].

\bibitem{Jenkins:2013zja}
E.~E. Jenkins, A.~V. Manohar and M.~Trott, \emph{{Renormalization Group
  Evolution of the Standard Model Dimension Six Operators I: Formalism and
  lambda Dependence}},
  \href{https://doi.org/10.1007/JHEP10(2013)087}{\emph{JHEP} {\bfseries 10}
  (2013) 087}, [\href{https://arxiv.org/abs/1308.2627}{{\ttfamily 1308.2627}}].

\bibitem{Jenkins:2013wua}
E.~E. Jenkins, A.~V. Manohar and M.~Trott, \emph{{Renormalization Group
  Evolution of the Standard Model Dimension Six Operators II: Yukawa
  Dependence}}, \href{https://doi.org/10.1007/JHEP01(2014)035}{\emph{JHEP}
  {\bfseries 01} (2014) 035},
  [\href{https://arxiv.org/abs/1310.4838}{{\ttfamily 1310.4838}}].

\bibitem{Alonso:2013hga}
R.~Alonso, E.~E. Jenkins, A.~V. Manohar and M.~Trott, \emph{{Renormalization
  Group Evolution of the Standard Model Dimension Six Operators III: Gauge
  Coupling Dependence and Phenomenology}},
  \href{https://doi.org/10.1007/JHEP04(2014)159}{\emph{JHEP} {\bfseries 04}
  (2014) 159}, [\href{https://arxiv.org/abs/1312.2014}{{\ttfamily 1312.2014}}].

\bibitem{Jenkins:2017dyc}
E.~E. Jenkins, A.~V. Manohar and P.~Stoffer, \emph{{Low-Energy Effective Field
  Theory below the Electroweak Scale: Anomalous Dimensions}},
  \href{https://doi.org/10.1007/JHEP01(2018)084}{\emph{JHEP} {\bfseries 01}
  (2018) 084}, [\href{https://arxiv.org/abs/1711.05270}{{\ttfamily
  1711.05270}}].

\bibitem{Gutsche:2009vp}
T.~Gutsche, J.~C. Helo, S.~Kovalenko and V.~E. Lyubovitskij, \emph{{On lepton
  flavor violating decays of vector mesons}},
  \href{https://doi.org/10.1103/PhysRevD.81.037702}{\emph{Phys. Rev. D}
  {\bfseries 81} (2010) 037702},
  [\href{https://arxiv.org/abs/0912.4562}{{\ttfamily 0912.4562}}].

\bibitem{Carpentier:2010ue}
M.~Carpentier and S.~Davidson, \emph{{Constraints on two-lepton, two quark
  operators}}, \href{https://doi.org/10.1140/epjc/s10052-010-1482-4}{\emph{Eur.
  Phys. J. C} {\bfseries 70} (2010) 1071--1090},
  [\href{https://arxiv.org/abs/1008.0280}{{\ttfamily 1008.0280}}].

\bibitem{Crivellin:2013hpa}
A.~Crivellin, S.~Najjari and J.~Rosiek, \emph{{Lepton Flavor Violation in the
  Standard Model with general Dimension-Six Operators}},
  \href{https://doi.org/10.1007/JHEP04(2014)167}{\emph{JHEP} {\bfseries 04}
  (2014) 167}, [\href{https://arxiv.org/abs/1312.0634}{{\ttfamily 1312.0634}}].

\bibitem{Cai:2015poa}
Y.~Cai and M.~A. Schmidt, \emph{{A Case Study of the Sensitivity to LFV
  Operators with Precision Measurements and the LHC}},
  \href{https://doi.org/10.1007/JHEP02(2016)176}{\emph{JHEP} {\bfseries 02}
  (2016) 176}, [\href{https://arxiv.org/abs/1510.02486}{{\ttfamily
  1510.02486}}].

\bibitem{Hazard:2016fnc}
D.~E. Hazard and A.~A. Petrov, \emph{{Lepton flavor violating quarkonium
  decays}}, \href{https://doi.org/10.1103/PhysRevD.94.074023}{\emph{Phys. Rev.}
  {\bfseries D94} (2016) 074023},
  [\href{https://arxiv.org/abs/1607.00815}{{\ttfamily 1607.00815}}].

\bibitem{Hazard:2017udp}
D.~E. Hazard and A.~A. Petrov, \emph{{Radiative lepton flavor violating B, D,
  and K decays}}, \href{https://doi.org/10.1103/PhysRevD.98.015027}{\emph{Phys.
  Rev. D} {\bfseries 98} (2018) 015027},
  [\href{https://arxiv.org/abs/1711.05314}{{\ttfamily 1711.05314}}].

\bibitem{Davidson:2018rqt}
S.~Davidson and A.~Saporta, \emph{{Constraints on $2\ell 2q$ operators from
  $\mu - e$ flavour-changing meson decays}},
  \href{https://doi.org/10.1103/PhysRevD.99.015032}{\emph{Phys. Rev. D}
  {\bfseries 99} (2019) 015032},
  [\href{https://arxiv.org/abs/1807.10288}{{\ttfamily 1807.10288}}].

\bibitem{Dib:2018rpy}
C.~O. Dib, T.~Gutsche, S.~G. Kovalenko, V.~E. Lyubovitskij and I.~Schmidt,
  \emph{{Bounds on lepton flavor violating physics and decays of neutral mesons
  from $\tau(\mu) \to 3 \ell, \ell \gamma\gamma$-decays}},
  \href{https://doi.org/10.1103/PhysRevD.99.035020}{\emph{Phys. Rev. D}
  {\bfseries 99} (2019) 035020},
  [\href{https://arxiv.org/abs/1812.02638}{{\ttfamily 1812.02638}}].

\bibitem{Angelescu:2020uug}
A.~Angelescu, D.~A. Faroughy and O.~Sumensari, \emph{{Lepton Flavor Violation
  and Dilepton Tails at the LHC}},
  \href{https://doi.org/10.1140/epjc/s10052-020-8210-5}{\emph{Eur. Phys. J. C}
  {\bfseries 80} (2020) 641},
  [\href{https://arxiv.org/abs/2002.05684}{{\ttfamily 2002.05684}}].

\bibitem{Gonzalez:2021tqc}
M.~Gonz\'alez, S.~Kovalenko, N.~A. Neill and J.~Vignatti, \emph{{RGE effects on
  the LFV scale from meson decays}},
  \href{https://doi.org/10.1140/epjc/s10052-022-10206-2}{\emph{Eur. Phys. J. C}
  {\bfseries 82} (2022) 312},
  [\href{https://arxiv.org/abs/2101.11692}{{\ttfamily 2101.11692}}].

\bibitem{Cirigliano:2021img}
V.~Cirigliano, K.~Fuyuto, C.~Lee, E.~Mereghetti and B.~Yan, \emph{{Charged
  Lepton Flavor Violation at the EIC}},
  \href{https://doi.org/10.1007/JHEP03(2021)256}{\emph{JHEP} {\bfseries 03}
  (2021) 256}, [\href{https://arxiv.org/abs/2102.06176}{{\ttfamily
  2102.06176}}].

\bibitem{Hoferichter:2022mna}
M.~Hoferichter, J.~Men\'endez and F.~No\"el, \emph{{Improved limits on
  lepton-flavor-violating decays of light pseudoscalars via spin-dependent
  $\mu\to e$ conversion in nuclei}},
  \href{https://arxiv.org/abs/2204.06005}{{\ttfamily 2204.06005}}.

\bibitem{Arun:2022uqn}
M.~T. Arun, P.~Lamba and S.~K. Vempati, \emph{{Restricting $q^2 l^2$ operators
  from $\pi^0 \rightarrow \mu e$}},
  \href{https://arxiv.org/abs/2204.06948}{{\ttfamily 2204.06948}}.

\bibitem{Liao:2020zyx}
Y.~Liao, X.-D. Ma and Q.-Y. Wang, \emph{{Extending low energy effective field
  theory with a complete set of dimension-7 operators}},
  \href{https://doi.org/10.1007/JHEP08(2020)162}{\emph{JHEP} {\bfseries 08}
  (2020) 162}, [\href{https://arxiv.org/abs/2005.08013}{{\ttfamily
  2005.08013}}].

\bibitem{Abada:2015zea}
A.~Abada, D.~Be\v{c}irevi\'c, M.~Lucente and O.~Sumensari, \emph{{Lepton flavor
  violating decays of vector quarkonia and of the $Z$ boson}},
  \href{https://doi.org/10.1103/PhysRevD.91.113013}{\emph{Phys. Rev. D}
  {\bfseries 91} (2015) 113013},
  [\href{https://arxiv.org/abs/1503.04159}{{\ttfamily 1503.04159}}].

\bibitem{Mertig:1990an}
R.~Mertig, M.~Bohm and A.~Denner, \emph{{FEYN CALC: Computer algebraic
  calculation of Feynman amplitudes}},
  \href{https://doi.org/10.1016/0010-4655(91)90130-D}{\emph{Comput. Phys.
  Commun.} {\bfseries 64} (1991) 345--359}.

\bibitem{Shtabovenko:2016sxi}
V.~Shtabovenko, R.~Mertig and F.~Orellana, \emph{{New Developments in FeynCalc
  9.0}}, \href{https://doi.org/10.1016/j.cpc.2016.06.008}{\emph{Comput. Phys.
  Commun.} {\bfseries 207} (2016) 432--444},
  [\href{https://arxiv.org/abs/1601.01167}{{\ttfamily 1601.01167}}].

\bibitem{Shtabovenko:2020gxv}
V.~Shtabovenko, R.~Mertig and F.~Orellana, \emph{{FeynCalc 9.3: New features
  and improvements}},
  \href{https://doi.org/10.1016/j.cpc.2020.107478}{\emph{Comput. Phys. Commun.}
  {\bfseries 256} (2020) 107478},
  [\href{https://arxiv.org/abs/2001.04407}{{\ttfamily 2001.04407}}].

\bibitem{Li:2021phq}
T.~Li, X.-D. Ma, M.~A. Schmidt and R.-J. Zhang, \emph{{{Implication of
  $J/\psi\rightarrow(\gamma+)$invisible for the effective field theories of
  neutrino and dark matter}}},
  \href{https://doi.org/10.1103/PhysRevD.104.035024}{\emph{Phys. Rev. D}
  {\bfseries 104} (2021) 035024},
  [\href{https://arxiv.org/abs/2104.01780}{{\ttfamily 2104.01780}}].

\bibitem{Barger:1987nn}
V.~D. Barger and R.~J.~N. Phillips, \emph{{Collider Physics}}.
\newblock 1987.

\bibitem{Cheng:2013fba}
H.-Y. Cheng, C.-K. Chua, K.-C. Yang and Z.-Q. Zhang, \emph{{Revisiting
  charmless hadronic B decays to scalar mesons}},
  \href{https://doi.org/10.1103/PhysRevD.87.114001}{\emph{Phys. Rev. D}
  {\bfseries 87} (2013) 114001},
  [\href{https://arxiv.org/abs/1303.4403}{{\ttfamily 1303.4403}}].

\bibitem{Workman:2022ynf}
{\scshape Particle Data Group} collaboration, R.~L. Workman and Others,
  \emph{{Review of Particle Physics}}, {\emph{PTEP} {\bfseries 2022} (2022)
  083C01}.

\bibitem{Straub:2018kue}
D.~M. Straub, \emph{{flavio: a Python package for flavour and precision
  phenomenology in the Standard Model and beyond}},
  \href{https://arxiv.org/abs/1810.08132}{{\ttfamily 1810.08132}}.

\bibitem{Aebischer:2018bkb}
J.~Aebischer, J.~Kumar and D.~M. Straub, \emph{{Wilson: a Python package for
  the running and matching of Wilson coefficients above and below the
  electroweak scale}},
  \href{https://doi.org/10.1140/epjc/s10052-018-6492-7}{\emph{Eur. Phys. J. C}
  {\bfseries 78} (2018) 1026},
  [\href{https://arxiv.org/abs/1804.05033}{{\ttfamily 1804.05033}}].

\bibitem{Godfrey:2015vda}
S.~Godfrey and H.~E. Logan, \emph{{Probe of new light Higgs bosons from
  bottomonium $\chi_{b0}$ decay}},
  \href{https://doi.org/10.1103/PhysRevD.93.055014}{\emph{Phys. Rev. D}
  {\bfseries 93} (2016) 055014},
  [\href{https://arxiv.org/abs/1510.04659}{{\ttfamily 1510.04659}}].

\bibitem{Godfrey:2015dia}
S.~Godfrey and K.~Moats, \emph{{Bottomonium Mesons and Strategies for their
  Observation}}, \href{https://doi.org/10.1103/PhysRevD.92.054034}{\emph{Phys.
  Rev. D} {\bfseries 92} (2015) 054034},
  [\href{https://arxiv.org/abs/1507.00024}{{\ttfamily 1507.00024}}].

\bibitem{ParticleDataGroup:2020ssz}
{\scshape Particle Data Group} collaboration, P.~A. Zyla et~al., \emph{{Review
  of Particle Physics}},
  \href{https://doi.org/10.1093/ptep/ptaa104}{\emph{PTEP} {\bfseries 2020}
  (2020) 083C01}.

\bibitem{Hatton:2020qhk}
{\scshape HPQCD} collaboration, D.~Hatton, C.~T.~H. Davies, B.~Galloway,
  J.~Koponen, G.~P. Lepage and A.~T. Lytle, \emph{{Charmonium properties from
  lattice $QCD$+QED : Hyperfine splitting, $J/\psi$ leptonic width, charm quark
  mass, and $a^c_\mu$}},
  \href{https://doi.org/10.1103/PhysRevD.102.054511}{\emph{Phys. Rev. D}
  {\bfseries 102} (2020) 054511},
  [\href{https://arxiv.org/abs/2005.01845}{{\ttfamily 2005.01845}}].

\bibitem{Hatton:2020vzp}
{\scshape HPQCD} collaboration, D.~Hatton, C.~T.~H. Davies, G.~P. Lepage and
  A.~T. Lytle, \emph{{Renormalization of the tensor current in lattice QCD and
  the $J/\psi$ tensor decay constant}},
  \href{https://doi.org/10.1103/PhysRevD.102.094509}{\emph{Phys. Rev. D}
  {\bfseries 102} (2020) 094509},
  [\href{https://arxiv.org/abs/2008.02024}{{\ttfamily 2008.02024}}].

\bibitem{Hatton:2021dvg}
D.~Hatton, C.~T.~H. Davies, J.~Koponen, G.~P. Lepage and A.~T. Lytle,
  \emph{{Bottomonium precision tests from full lattice QCD: Hyperfine
  splitting, $\Upsilon$ leptonic width, and b quark contribution to $e^+e^-
  \rightarrow$ hadrons}},
  \href{https://doi.org/10.1103/PhysRevD.103.054512}{\emph{Phys. Rev. D}
  {\bfseries 103} (2021) 054512},
  [\href{https://arxiv.org/abs/2101.08103}{{\ttfamily 2101.08103}}].

\bibitem{Colquhoun:2014ica}
B.~Colquhoun, R.~J. Dowdall, C.~T.~H. Davies, K.~Hornbostel and G.~P. Lepage,
  \emph{{$\Upsilon$ and $\Upsilon^{\prime}$ Leptonic Widths, $a_{\mu}^b$ and
  $m_b$ from full lattice QCD}},
  \href{https://doi.org/10.1103/PhysRevD.91.074514}{\emph{Phys. Rev. D}
  {\bfseries 91} (2015) 074514},
  [\href{https://arxiv.org/abs/1408.5768}{{\ttfamily 1408.5768}}].

\bibitem{Chung:2020zqc}
H.~S. Chung, \emph{{$\overline {MS}$ renormalization of $S$-wave quarkonium
  wavefunctions at the origin}},
  \href{https://doi.org/10.1007/JHEP12(2020)065}{\emph{JHEP} {\bfseries 12}
  (2020) 065}, [\href{https://arxiv.org/abs/2007.01737}{{\ttfamily
  2007.01737}}].

\bibitem{Becirevic:2013bsa}
D.~Be\v{c}irevi\'c, G.~Duplan\v{c}i\'c, B.~Klajn, B.~Meli\'c and F.~Sanfilippo,
  \emph{{Lattice QCD and QCD sum rule determination of the decay constants of
  $\eta_c$, J/$\psi$ and $h_c$ states}},
  \href{https://doi.org/10.1016/j.nuclphysb.2014.03.024}{\emph{Nucl. Phys. B}
  {\bfseries 883} (2014) 306--327},
  [\href{https://arxiv.org/abs/1312.2858}{{\ttfamily 1312.2858}}].

\bibitem{Khodjamirian:2015dda}
A.~Khodjamirian, T.~Mannel and A.~A. Petrov, \emph{{Direct probes of
  flavor-changing neutral currents in $e^+ e^-$-collisions}},
  \href{https://doi.org/10.1007/JHEP11(2015)142}{\emph{JHEP} {\bfseries 11}
  (2015) 142}, [\href{https://arxiv.org/abs/1509.07123}{{\ttfamily
  1509.07123}}].

\bibitem{Appelquist:1974zd}
T.~Appelquist and H.~D. Politzer, \emph{{Orthocharmonium and e+ e-
  Annihilation}}, \href{https://doi.org/10.1103/PhysRevLett.34.43}{\emph{Phys.
  Rev. Lett.} {\bfseries 34} (1975) 43}.

\bibitem{DeRujula:1974rkb}
A.~De~Rujula and S.~L. Glashow, \emph{{Is Bound Charm Found?}},
  \href{https://doi.org/10.1103/PhysRevLett.34.46}{\emph{Phys. Rev. Lett.}
  {\bfseries 34} (1975) 46--49}.

\bibitem{Kuhn:1979bb}
J.~H. Kuhn, J.~Kaplan and E.~G.~O. Safiani, \emph{{Electromagnetic Annihilation
  of e+ e- Into Quarkonium States with Even Charge Conjugation}},
  \href{https://doi.org/10.1016/0550-3213(79)90055-5}{\emph{Nucl. Phys. B}
  {\bfseries 157} (1979) 125--144}.

\bibitem{Keung:1980ev}
W.-Y. Keung, \emph{{Off Resonance Production of Heavy Vector Quarkonium States
  in $e^+ e^-$ Annihilation}},
  \href{https://doi.org/10.1103/PhysRevD.23.2072}{\emph{Phys. Rev. D}
  {\bfseries 23} (1981) 2072}.

\bibitem{Berger:1980ni}
E.~L. Berger and D.~L. Jones, \emph{{Inelastic Photoproduction of J/psi and
  Upsilon by Gluons}},
  \href{https://doi.org/10.1103/PhysRevD.23.1521}{\emph{Phys. Rev. D}
  {\bfseries 23} (1981) 1521--1530}.

\bibitem{Clavelli:1982hp}
L.~Clavelli, \emph{{Associated Heavy Vector Meson Production in $e^+ e^-$
  Annihilation}}, \href{https://doi.org/10.1103/PhysRevD.26.1610}{\emph{Phys.
  Rev. D} {\bfseries 26} (1982) 1610}.

\bibitem{Clavelli:2001zi}
L.~Clavelli, T.~Gajdosik and I.~Perevalova, \emph{{Parity and time reversal in
  J / psi decay}},
  \href{https://doi.org/10.1016/S0370-2693(01)01359-4}{\emph{Phys. Lett. B}
  {\bfseries 523} (2001) 249--254},
  [\href{https://arxiv.org/abs/hep-ph/0110076}{{\ttfamily hep-ph/0110076}}].

\bibitem{Clavelli:2001gb}
L.~Clavelli, P.~Coulter and T.~Gajdosik, \emph{{Quarkonium polarization in
  lepton annihilation}},
  \href{https://doi.org/10.1016/S0370-2693(01)01516-7}{\emph{Phys. Lett. B}
  {\bfseries 526} (2002) 360--364},
  [\href{https://arxiv.org/abs/hep-ph/0111250}{{\ttfamily hep-ph/0111250}}].

\bibitem{Kitano:2002mt}
R.~Kitano, M.~Koike and Y.~Okada, \emph{{Detailed calculation of lepton flavor
  violating muon electron conversion rate for various nuclei}},
  \href{https://doi.org/10.1103/PhysRevD.76.059902}{\emph{Phys. Rev. D}
  {\bfseries 66} (2002) 096002},
  [\href{https://arxiv.org/abs/hep-ph/0203110}{{\ttfamily hep-ph/0203110}}].

\bibitem{Cirigliano:2009bz}
V.~Cirigliano, R.~Kitano, Y.~Okada and P.~Tuzon, \emph{{On the model
  discriminating power of $\mu \to e$ conversion in nuclei}},
  \href{https://doi.org/10.1103/PhysRevD.80.013002}{\emph{Phys. Rev. D}
  {\bfseries 80} (2009) 013002},
  [\href{https://arxiv.org/abs/0904.0957}{{\ttfamily 0904.0957}}].

\bibitem{Crivellin:2014cta}
A.~Crivellin, M.~Hoferichter and M.~Procura, \emph{{Improved predictions for
  $\mu\to e$ conversion in nuclei and Higgs-induced lepton flavor violation}},
  \href{https://doi.org/10.1103/PhysRevD.89.093024}{\emph{Phys. Rev. D}
  {\bfseries 89} (2014) 093024},
  [\href{https://arxiv.org/abs/1404.7134}{{\ttfamily 1404.7134}}].

\bibitem{Katana}
{PVC (Research Infrastructure), UNSW Sydney}, \emph{Katana},  2010.
\newblock 10.26190/669X-A286.

\bibitem{Kuno:1999jp}
Y.~Kuno and Y.~Okada, \emph{{Muon decay and physics beyond the standard
  model}}, \href{https://doi.org/10.1103/RevModPhys.73.151}{\emph{Rev. Mod.
  Phys.} {\bfseries 73} (2001) 151--202},
  [\href{https://arxiv.org/abs/hep-ph/9909265}{{\ttfamily hep-ph/9909265}}].

\bibitem{Hoferichter:2015dsa}
M.~Hoferichter, J.~Ruiz~de Elvira, B.~Kubis and U.-G. Mei\ss{}ner,
  \emph{{High-Precision Determination of the Pion-Nucleon \ensuremath{\sigma}
  Term from Roy-Steiner Equations}},
  \href{https://doi.org/10.1103/PhysRevLett.115.092301}{\emph{Phys. Rev. Lett.}
  {\bfseries 115} (2015) 092301},
  [\href{https://arxiv.org/abs/1506.04142}{{\ttfamily 1506.04142}}].

\bibitem{Junnarkar:2013ac}
P.~Junnarkar and A.~Walker-Loud, \emph{{Scalar strange content of the nucleon
  from lattice QCD}},
  \href{https://doi.org/10.1103/PhysRevD.87.114510}{\emph{Phys. Rev. D}
  {\bfseries 87} (2013) 114510},
  [\href{https://arxiv.org/abs/1301.1114}{{\ttfamily 1301.1114}}].

\bibitem{Alarcon:2011zs}
J.~M. Alarcon, J.~Martin~Camalich and J.~A. Oller, \emph{{The chiral
  representation of the $\pi N$ scattering amplitude and the pion-nucleon sigma
  term}}, \href{https://doi.org/10.1103/PhysRevD.85.051503}{\emph{Phys. Rev. D}
  {\bfseries 85} (2012) 051503},
  [\href{https://arxiv.org/abs/1110.3797}{{\ttfamily 1110.3797}}].

\bibitem{Alarcon:2012nr}
J.~M. Alarcon, L.~S. Geng, J.~Martin~Camalich and J.~A. Oller, \emph{{The
  strangeness content of the nucleon from effective field theory and
  phenomenology}},
  \href{https://doi.org/10.1016/j.physletb.2014.01.065}{\emph{Phys. Lett. B}
  {\bfseries 730} (2014) 342--346},
  [\href{https://arxiv.org/abs/1209.2870}{{\ttfamily 1209.2870}}].

\bibitem{Cirigliano:2022ekw}
V.~Cirigliano, K.~Fuyuto, M.~J. Ramsey-Musolf and E.~Rule,
  \emph{{Next-to-leading order scalar contributions to $\mu\rightarrow e$
  conversion}}, \href{https://doi.org/10.1103/PhysRevC.105.055504}{\emph{Phys.
  Rev. C} {\bfseries 105} (2022) 055504},
  [\href{https://arxiv.org/abs/2203.09547}{{\ttfamily 2203.09547}}].

\bibitem{Heeck:2022wer}
J.~Heeck, R.~Szafron and Y.~Uesaka, \emph{{Isotope dependence of
  muon-to-electron conversion}},
  \href{https://doi.org/10.1016/j.nuclphysb.2022.115833}{\emph{Nucl. Phys. B}
  {\bfseries 980} (2022) 115833},
  [\href{https://arxiv.org/abs/2203.00702}{{\ttfamily 2203.00702}}].

\bibitem{Aebischer:2018iyb}
J.~Aebischer, J.~Kumar, P.~Stangl and D.~M. Straub, \emph{{A Global Likelihood
  for Precision Constraints and Flavour Anomalies}},
  \href{https://doi.org/10.1140/epjc/s10052-019-6977-z}{\emph{Eur. Phys. J. C}
  {\bfseries 79} (2019) 509},
  [\href{https://arxiv.org/abs/1810.07698}{{\ttfamily 1810.07698}}].

\bibitem{Brignole:2004ah}
A.~Brignole and A.~Rossi, \emph{{Anatomy and phenomenology of mu-tau lepton
  flavor violation in the MSSM}},
  \href{https://doi.org/10.1016/j.nuclphysb.2004.08.037}{\emph{Nucl. Phys. B}
  {\bfseries 701} (2004) 3--53},
  [\href{https://arxiv.org/abs/hep-ph/0404211}{{\ttfamily hep-ph/0404211}}].

\end{thebibliography}\endgroup

\end{document}